\pdfoutput=1
\documentclass[12pt, a4paper]{article}
\usepackage{jheppub}
\usepackage[usenames, dvipsnames]{xcolor}
\usepackage{graphicx}
\usepackage{microtype}
\usepackage{amssymb} 
\usepackage{amsmath}
\usepackage{framed}
\usepackage{mathtools}
\usepackage{amsfonts}    
\usepackage{dsfont}
\usepackage{pdfpages}
\usepackage{verbatim}
\usepackage{braket}
\usepackage{bm}
\usepackage{tensor}
\usepackage{upgreek}
\usepackage{subcaption}
\usepackage{simpler-wick}
\usepackage{enumitem}
\usepackage[hang, flushmargin]{footmisc}

\usepackage{tikz}
\usetikzlibrary{decorations.markings}
\usetikzlibrary{shapes.misc}
\usetikzlibrary{arrows}
\usetikzlibrary{fit, chains}
\usetikzlibrary{patterns}

\usepackage{listings}
\usepackage{color}

\usepackage[framemethod=default]{mdframed}
\usepackage{etoolbox}

\DeclareMathOperator*{\SumInt}{%
\mathchoice%
  {\ooalign{$\displaystyle\sum$\cr\hidewidth$\displaystyle\int$\hidewidth\cr}}
  {\ooalign{\raisebox{.14\height}{\scalebox{.7}{$\textstyle\sum$}}\cr\hidewidth$\textstyle\int$\hidewidth\cr}}
  {\ooalign{\raisebox{.2\height}{\scalebox{.6}{$\scriptstyle\sum$}}\cr$\scriptstyle\int$\cr}}
  {\ooalign{\raisebox{.2\height}{\scalebox{.6}{$\scriptstyle\sum$}}\cr$\scriptstyle\int$\cr}}
}

\newenvironment{FramedBox}[1][]{%
    \begin{mdframed}[
        skipabove=7pt,
        skipbelow=7pt,
        rightline=true,
        leftline=true,
        topline=true,
        bottomline=true,
        backgroundcolor=gray!10,
        linecolor=gray!10,
        innerleftmargin=10pt,
        innerrightmargin=10pt,
        innertopmargin=10pt,
        innerbottommargin=10pt,
        leftmargin=0cm,
        rightmargin=0cm,
        linewidth=1pt,
        #1
    ]%
    \ignorespaces%
}{%
    \end{mdframed}%
}

\usepackage[framemethod=default]{mdframed}
\newmdenv[skipabove=7pt,
skipbelow=7pt,
rightline=false,
leftline=false,
topline=false,
bottomline=false,
backgroundcolor=gray!10,
linecolor=gray,
innerleftmargin=5pt,
innerrightmargin=5pt,
innertopmargin=5pt,
innerbottommargin=5pt,
leftmargin=0cm,
rightmargin=0cm,
linewidth=4pt]{eBox}

\usepackage{tikz}
\usetikzlibrary{decorations.pathreplacing, decorations.markings,calc,shapes.misc,decorations.pathmorphing,patterns.meta, math, shadings, backgrounds}

\definecolor{rosewood}{rgb}{0.4, 0.0, 0.04}
\definecolor{pyblue}{RGB}{31, 119, 180}
\colorlet{lightpyblue}{pyblue!30!white}
\definecolor{pyorange}{RGB}{255, 127, 14}
\colorlet{lightpyorange}{pyorange!30!white}
\definecolor{pygreen}{RGB}{44, 160, 44}
\colorlet{lightpygreen}{pygreen!30!white}
\definecolor{pyred}{RGB}{214, 39, 40}
\colorlet{lightpyred}{pyred!30!white}
\definecolor{lightgray}{gray}{0.9}

\usepackage{hyperref}
\hypersetup{
    pdfencoding=unicode,
    colorlinks=true,
    urlcolor=rosewood,
    linkcolor=pyblue,
    citecolor=pygreen,
    pdftitle={This is the title},
    pdfauthor={Denis Werth},
    pdfdisplaydoctitle=true,
    pdfstartview=FitH,
    linktocpage=true
}

\def \d {\mathrm{d}}

\def \x {\bm{x}}

\def \k {\bm{k}}
\def \p {\bm{p}}
\def \m {\bm{m}}

\def \a {{\sf{a}}}
\def \b {{\sf{b}}}

\def \C {\mathcal{C}}
\def \D {\mathcal{D}}
\def \E {\mathcal{E}}
\def \F {\mathcal{F}}
\def \K {\mathcal{K}}
\def \L {\mathcal{L}}
\def \G {\mathcal{G}}
\def \I {\mathcal{I}} 
\def \M {\mathcal{M}}
\def \N {\mathcal{N}}
\def \O {\mathcal{O}}
\def \P {\mathcal{P}}
\def \T {\mathcal{T}}

\def \S {\mathcal{S}}
\def \H {\mathcal{H}}
\def \V {\mathcal{V}}

\def \aa {{\sf{a}}}

\def \Im {\mathrm{Im}}
\def \Re {\mathrm{Re}}

\def \SO {\mathrm{SO}}
\def \dS {\mathrm{dS}}

\def \EAdS {\mathrm{EAdS}}
\def \Span {\mathrm{Span}}
\def \vol {\mathrm{vol}}
\def \KLF {\mathrm{KLF}}

\title{Kontorovich-Lebedev-Fourier Space for de Sitter Correlators}

\author{Nathan Belrhali$^{\rm K}$}
\author{Arthur Poisson$^{\rm K}$}
\author{S\'ebastien Renaux-Petel$^{\rm K}$}
\author{Denis Werth$^{\rm K, L, F}$}

\affiliation{$^{\rm K}$Institut d'Astrophysique de Paris, CNRS, Sorbonne Universit\'e, FR-75014, France \\
$^{\rm L}$Max-Planck-Institut f\"ur Physik, Werner-Heisenberg-Institut, Garching bei M\"unchen, D-85748, Germany \\
$^{\rm F}$Max Planck-IAS-NTU Center for Particle Physics, Cosmology and Geometry}

\abstract{In this work, we build a novel frequency-momentum space for $(d+1)$-dimensional de Sitter (dS) correlators from first principles.
This construction follows directly from the decomposition into unitary irreducible representations (UIRs) of the spacetime isometry group $\SO(1,d+1)$. While the spatial momentum space is given by the standard $d$-dimensional Fourier transform, the frequency space arises from diagonalising the quadratic Casimir operator, leading to the $(d+1)$-dimensional Kontorovich-Lebedev-Fourier (KLF) transform.
We show that square-integrable functions decompose only along the principal series, whereas more general functions can receive discrete contributions from other UIRs. Applying this framework to the bulk CFT two-point function reproduces its Källén-Lehmann representation. Using the path integral formulation, we derive the Feynman rules for in-in perturbation theory in KLF space, leading to the introduction of KLF-space correlators, which are simply related to late-time correlation functions through a reduction formula. Furthermore, the KLF-space formulation sheds light on the simple mathematical structure of perturbative computations. In particular, the propagators take the form of simple rational functions, and tree-level diagrams can be written as spectral integrals over known meromorphic functions, as demonstrated in the example of the single-exchange four-point function. At the loop level, we show, through the example of the self-energy correction to the scalar propagator, that the group-theoretical nature of the construction allows the momentum integral to be recast as an orthogonality relation among $\mathrm{SO}(1,d+1)$ Clebsch-Gordan coefficients.}

\begin{document}

\setcounter{tocdepth}{3}
\maketitle
\setcounter{page}{2}

\newpage
\section{Introduction}
Theoretical physics is often about properly understanding kinematics. For instance, scattering amplitudes in Minkowski spacetime quantum field theory (QFT) are better understood when expressed in terms of energy and momentum variables rather than in terms of space and time. Since it allows one to take advantage of the spacetime translation invariance, the theoretical building blocks, such as the Feynman propagator, take a very simple form, making perturbative computations straightforward to organise. It is also the natural language to reveal non-perturbative constraints from causality and unitarity~\cite{Froissart:1961ux,Martin:1962rt,Bros:1965kbd,Martin:1965jj,Paulos:2017fhb,Bellazzini:2020cot}, as they often arise in the complex plane of Mandelstam variables (see~\cite{Correia:2020xtr,Kruczenski:2022lot} for modern reviews).
Another example is provided by the revival of the conformal bootstrap~\cite{Rattazzi:2008pe,Rychkov:2009ij,Caracciolo:2009bx,Rattazzi:2010gj}, in which a thorough understanding of the kinematic building blocks, such as conformal blocks and partial waves~\cite{Dolan:2000ut,Dolan:2003hv,Costa:2011dw}, is essential. The latter play a great role in the non-perturbative understanding of QFT in Anti-de Sitter (AdS) spacetime, where the boundary correlation functions satisfy the same axioms as those of a lower dimensional Lorentzian conformal field theory (CFT)~\cite{Callan:1989em,Paulos:2016fap,Carmi:2018qzm}.
\vskip 4pt
The success of these two descriptions of QFT defined on a given spacetime shares a common origin, which lies in the high degree of symmetry of the background. More precisely, all the features mentioned above arise when expanding either the fields in flat spacetime, or the boundary correlators in AdS, in a harmonic basis appropriate to the spacetime symmetry group.\footnote{See e.g.~\cite{Mack:1974jjo,Mack:1974sa,Dobrev:1975ru,Dobrev:1977qv} for harmonic analysis on the conformal group and~\cite{Dolan:2003hv,Hogervorst:2017sfd,Simmons-Duffin:2017nub,Karateev:2018oml} for more recent work.} Unfortunately, we do not live in any of these spacetimes. The relevant background for cosmological applications is rather approached by de Sitter (dS) spacetime, which describes an expanding Universe with constant expansion rate. As such, it can describe both the early inflationary Universe and its late-time accelerated expansion.
\vskip 4pt
Although dS is also a maximally symmetric space, we still lack a proper understanding of its kinematics. While the isometry group also acts on the boundary correlators as the conformal group in one lower dimension, the scaling dimensions of the operators involved in the harmonic decomposition do not always fall into its unitary irreducible representations (UIR), 
which prevents a convergent operator product expansion (OPE) \cite{Hogervorst:2021uvp,DiPietro:2021sjt,SalehiVaziri:2024joi}.\footnote{This statement is due to the absence of a bulk state-boundary operator correspondence, and ultimately to the space-like nature of the dS boundary. Instead, see e.g. \cite{Paulos:2016fap,SalehiVaziri:2024joi} for the statement of this correspondence in AdS.} Therefore, although conformal invariance can still be used to bootstrap late-time observables at the perturbative level by mean of boundary differential equations~\cite{Maldacena:2011nz,Bzowski:2011ab,Mata:2012bx,Bzowski:2013sza,Kundu:2014gxa,Kundu:2015xta,Ghosh:2014kba,Shukla:2016bnu,Arkani-Hamed:2018kmz,Baumann:2019oyu,Baumann:2020dch,Pimentel:2022fsc,Jazayeri:2022kjy,Qin:2022fbv,Qin:2023ejc,Aoki:2024uyi,Qin:2025xct}, its constraining power at a non-perturbative level is not guaranteed (nevertheless, see~\cite{Hogervorst:2021uvp} for a recent attempt).
\vskip 4pt
At a more primitive level, we even lack a proper momentum space to organise perturbative computations, as we do in flat spacetime. 
The absence of time translation invariance means that energy is not conserved, rendering the usual Fourier frequency space ineffective.
The traditional method to compute late-time correlation functions is therefore to use the Schwinger-Keldysch (or in-in) formalism~\cite{Schwinger:1960qe,Feynman:1963fq,Keldysh:1964ud,Maldacena:2002vr,Weinberg:2005vy} in spatial Fourier space and conformal time. In this framework, even the simplest Feynman diagrams require the evaluation of nested time integrals, which quickly become daunting (see e.g.~\cite{Qin:2023bjk,Qin:2023nhv,Xianyu:2023ytd,Liu:2024xyi,Fan:2025scu} for recent progress). For correlators in the Bunch-Davies vacuum, a common alternative is to perform computations in the Euclidean version of spacetime, which is simply the sphere, and analytically continue the result to dS~\cite{Higuchi:1986wu,Bros:1990cu,Bros:1995js,Marolf:2010zp,Marolf:2010nz,Higuchi:2010xt}. In this way, fields and correlators can be expanded in the $(d+1)$-dimensional spherical harmonics basis, providing a dS momentum space. However, although this has proved useful to evaluate IR-finite loop corrections to 
propagators~\cite{Marolf:2010nz,Marolf:2010zp,Chen:2016hrz,Chakraborty:2023qbp,Chakraborty:2023eoq,Chakraborty:2025myb,Loparco:2025azm}, the evaluation of higher-point correlation functions appears impossible in that context. 
\vskip 4pt
Due to this issue, one can instead make use of another Wick rotation that maps the Poincar\'e patch of dS to the one of Euclidean Anti-de Sitter (EAdS)~\cite{Sleight:2019mgd,Sleight:2019hfp,Sleight:2020obc,Sleight:2021plv,DiPietro:2021sjt,Loparco:2023rug}. A convenient approach is then to take maximal advantage of the scale invariance in the bulk by diagonalising the dilatation generator, resulting in the Mellin-Barnes (MB) formalism~\cite{Sleight:2019mgd,Sleight:2019hfp,Sleight:2020obc,Sleight:2021plv}. However, like for radial quantisation in CFT, this is incompatible with staying in spatial Fourier space, because dilatation and translations do not commute.
The corresponding ``CFT-basis'' $\ket{\mu,\x}$ is expressed in terms of the scaling dimension $\Delta=\frac{d}{2}+i\mu$ and a spatial point $\x$ on the conformal boundary 
\cite{Sun:2021thf}. Thus, it is especially convenient when working in the embedding-space formalism where $\x$ can be identified with a point on the projective null cone \cite{Dirac:1936fq,Bros:2001yw,Cacciatori:2007in,Costa:2011dw,Rychkov:2016iqz,Karateev:2017jgd,Baumann:2019oyu,Moschella:2024kvk,SalehiVaziri:2024joi}.
\vskip 4pt
In this work, we formulate QFT in dS by using an alternative basis $\ket{\mu,\k}$ where the spatial translation generators are diagonalised together with the Casimir operator, paralleling the Wigner classification of the Poincar\'e group~\cite{Bargmann:1948ck}. In this picture, the Mellin transform is traded for the Kontorovich-Lebedev transform, resulting in the introduction of what we call the Kontorovich-Lebedev-Fourier (KLF) space, labeled by a frequency $\mu$ and a spatial momentum $\k$.
\vskip 4pt
Let us highlight some features of our construction:
\begin{itemize}
\item The building blocks are the harmonic functions $\Phi_{\k}^{\mu}(z, \x)$ \eqref{eq: harmonic function expression}, which provide the overlap between momentum states and their position-space duals, in particular between the Euclidean time $z$ and the frequency $\mu$.
    \item The principal series has a privileged status: only the harmonic functions with real $\mu$ enter the Kontorovitch-Lebedev transform and its inverse:
    \begin{equation}
    \begin{aligned}
    f(z,\x) &= \int_{\KLF}\d\mu\,\N_{\mu}\frac{\d^d\k}{(2\pi)^d} \, \Phi^{\mu}_{\k}(z,\x)f^{\mu}_{\k}\,, \\
    f^{\mu}_{\k} & = \int_\EAdS\frac{\d z\d^d\x}{(H z)^{d+1}}\left[\Phi^{\mu}_{\k}(z,\x)\right]^*f(z,\x)\,.
    \nonumber
    \end{aligned}
\end{equation}
    They are enough to decompose square-integrable functions, from which more general results can be simply deduced by analytical continuation to complex $\mu$. In particular, the contributions from other UIRs can be read from the non-analyticities in
the complex $\mu$-plane of the KLF modes $f^{\mu}_{\k}$, see Sec.~\ref{sec: Role of the other Series}.
    \item Like in Minkowski momentum space, the KLF version of the time-ordered Feynman propagator \eqref{eq: KLF (anti-)time ordered} has a simple rational form with poles located at the on-shell mass of the propagating degree of freedom.
\item The KLF Wightman function of a local operator $\O$ reduces to the corresponding K\"all\'en-Lehmann spectral density: $ \G^{\O}_{\pm}(\mu)=\rho_\O(\mu)/\N_{\mu}$.
\item The computation of correlation functions can be systematically organised using KLF Feynman rules, giving rise to spectral integrals over frequencies, with key ingredients the meromorphic vertex functions $\mathcal{I}^{\mu_1\ldots\mu_n}_{k_1\ldots k_n}$ \eqref{eq: def vertex function}.
\item A reduction formula \eqref{eq: general late time subdiagrams} straightforwardly gives cosmological correlators in terms of KLF amputated diagrams. Like in Minkowski, these building blocks do not involve time, but only frequencies and momenta.
\item The introduction of dS $3\mu$ symbols \eqref{eq: 3mu symbol} together with group-theoretical arguments enables to easily compute the loop momentum integral of the self-energy correction to the scalar propagator. 
\end{itemize}

\paragraph{Outline.} The outline of the paper is as follows: In Sec.~\ref{section-dS}, we review de Sitter geometry and the representation theory of its isometry group. Sec.~\ref{sec: poincare slicing} is devoted to the construction of the KLF momentum space, paying attention to the role of non-principal series. Sec.~\ref{sec:correlators} describes the perturbative formulation of QFT in KLF space, and Sec.~\ref{sec:examples} puts this into practice with the computations of cosmological correlators and loop corrections to propagators. We state our conclusions and discuss future directions in Sec.~\ref{sec:conclusion}. We collect important technical details in a collection of appendices.

\newpage
\section{De Sitter Space}
\label{section-dS}

In this section, we set up our notations and review de Sitter geometry and the representation theory of its isometry group.

\subsection{De Sitter Geometry}\label{sec: de Sitter}

De Sitter spacetime in $d+1$ dimensions, denoted $\dS_{d+1}$, can be realised as a hyperboloid embedded in $(d+2)$-dimensional Minkowski spacetime $\mathbb{M}^{1, d+1}$. It is defined as the set of points at a fixed Lorentzian distance from the origin
\begin{equation}
\label{eq: dS def}
    \eta_{AB} X^A X^B = -(X^0)^2 + (X^1)^2 + \ldots + (X^{d+1})^2 = R_{\dS}^2 \,,
\end{equation}
with $A, B = 0, 1, \ldots, d+1$. Here, $R_{\dS} \equiv H^{-1}$ is the curvature radius, equal to the inverse of the constant Hubble parameter. The spacetime $\dS_{d+1}$ satisfies Einstein's equations in vacuum with a positive cosmological constant given by $\Lambda \equiv \tfrac{1}{2}d(d-1)H^2$. We now introduce several coordinate systems that can be used to parametrise this geometry.

\begin{figure}[t!]
    \hspace*{-2cm}
    \centering
    \begin{subfigure}{.45\textwidth}
        \centering
        \includegraphics[width=1\linewidth]{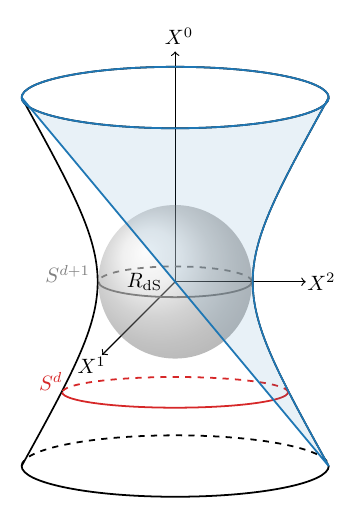}
    \end{subfigure}
    \begin{subfigure}{.45\textwidth}
        \centering
        \raisebox{0.6cm}{\includegraphics[width=1.3\linewidth]{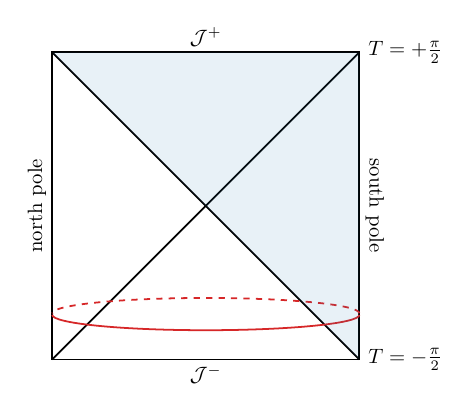}}
    \end{subfigure}
    \caption{{\it Left}: Illustration of $(d+1)$-dimensional de Sitter spacetime $\dS_{d+1}$, for $d=1$, represented as a hyperboloid in $\mathbb{M}^{1, d+1}$. The Euclidean $(d+1)$-dimensional sphere with radius $R_\dS$ is also shown in \textcolor{gray}{gray} for comparison. The \textcolor{pyred}{global coordinates} cover the entire spacetime and correspond to a foliation by $d$-dimensional spheres $S^d$, whereas the \textcolor{pyblue}{Poincar\'e coordinates} describe an expanding patch covering only half of the spacetime. {\it Right}: Penrose diagram of $\dS_{d+1}$ in conformal coordinates. The \textcolor{pyblue}{blue region} represents the Poincar\'e patch as the causal future of a south pole observer sitting in the asymptotic past.} 
\label{fig: dS geometry}
\end{figure}

\paragraph{Global coordinates.} The most natural coordinates are global coordinates. As the name suggests, they cover the entire $\dS_{d+1}$ spacetime. These coordinates are obtained by analytically continuing the standard $(d+1)$-dimensional spherical coordinates, allowing the azimuthal angle to take imaginary values:
\begin{equation}
\label{eq: embedding coordinate split for global}
    X^0 = R_\dS \sinh(\rho)\,, \quad X^a = R_\dS \cosh(\rho)\, \omega^a \,,
\end{equation}
where $\rho \in \mathbb{R}$ is the global time, $a = 1, \ldots, d+1$ and $\omega^a \in S^d\subset \mathbb{R}^{d+1}$ is a unit vector ($\omega^a\omega_a=1$) spanning a $d$-dimensional sphere. The latter can be parametrised as
\begin{equation}
    \begin{aligned}
        \omega^1 &= \cos\theta_1 \,, \\
        \omega^2 &= \sin\theta_1 \, \cos\theta_2 \,, \\
        & \,\,\, \vdots \\
        \omega^{d} &= \sin\theta_1 \ldots \sin\theta_{d-1} \, \sin\theta_d \,, \\
        \omega^{d+1} &= \sin\theta_1 \ldots \sin\theta_{d-1} \, \sin\theta_d \,,
    \end{aligned}
\end{equation}
with $0\leq \theta_i < \pi$ for $1\leq i \leq d-1$ and $0\leq \theta_d < 2\pi$, which yields the standard line element of the unit $d$-dimensional sphere $S^d$:
\begin{equation}
    \d\Omega_d^2 \equiv \sum_{a=1}^{d+1}(\d\omega^a)^2 = \d\theta_1^2 + \sin^2\theta_1 \d\theta_2^2 + \ldots + \sin^2\theta_1 \ldots \sin^2\theta_{d-1} \d\theta_d^2 \,.
\end{equation}
The induced $\dS_{d+1}$ metric in global coordinates is therefore given by
\begin{equation}
\label{eq: dS metric global}
    \d s^2 = R_\dS^2 (-\d\rho^2 + \cosh^2\rho\, \d\Omega_d^2) \,.
\end{equation}
Notice that we recover the $(d+1)$-dimensional sphere $S^{d+1}$ metric by setting $\rho = i(\theta-\tfrac{\pi}{2})$ with $\theta\in [0, \pi)$. At the level of embedding space, de Sitter and the sphere are related by the Wick rotation $X^0\to iX^0$. As illustrated in Fig.~\ref{fig: dS geometry}, the metric~\eqref{eq: dS metric global} describes a foliation of $d$-dimensional spheres that begin with infinite radius at $\rho=-\infty$, contract to a minimal radius $R_\dS$ at $\rho=0$, where it coincides with the equilateral plane of the Euclidean sphere, and expand again to infinite size at $\rho=+\infty$. From Eq.~\eqref{eq: embedding coordinate split for global}, it follows that the global coordinates correspond to a $1+(d+1)$ decomposition of the embedding coordinates $X^A$.

\paragraph{Conformal coordinates.} Global time can be compactified via the change of variable $\cosh(\rho) = 1/\cos(T)$ so that we have $-\pi/2<T<\pi/2$. This leads to the metric
\begin{equation}
\label{eq: dS metric conformal}
    \d s^2 = R_\dS^2 \, \frac{-\d T^2 + \d\Omega_d^2}{\cos^2(T)}\,,
\end{equation}
which makes manifest that global $\dS_{d+1}$ is conformal to a compact cylinder. The form~\eqref{eq: dS metric conformal} is particularly useful for understanding the causal structure of de Sitter spacetime, as illustrated by the corresponding Penrose diagram in Fig.~\ref{fig: dS geometry}. In particular, no single observer can access the entire spacetime.

\paragraph{Poincar\'e coordinates.} An alternative splitting choice is to adopt the following $(1+d+1)$ decomposition, which foliates de Sitter spacetime using flat slices:
\begin{equation}
    X^0 = R_\dS \, \frac{\tau^2-\x^2-1}{2\tau} \,, \quad X^i = -R_\dS\,\frac{x^i}{\tau}\,, \quad X^{d+1} = R_\dS\,\frac{\x^2-\tau^2-1}{2\tau} \,,
\end{equation}
where $i=1, \ldots, d$. The Poincar\'e coordinates $(\tau, \x)$, with $\tau<0$ denoting conformal time and $x^i\in \mathbb{R}^d$ being spatial coordinates, cover only half of de Sitter space, which is therefore not geodesically complete. For concreteness, we focus on the expanding patch defined by $X^0+X^{d+1}\geq 0$, as illustrated in Fig.~\ref{fig: dS geometry}. This region corresponds to the causal future of an observer located at some point in the asymptotic past. The induced metric in these coordinates is
\begin{equation}\label{eq: dS metric poincare}
    \d s^2 = \frac{-\d\tau^2 + \d\x^2}{(H\tau)^2} \;.
\end{equation}
This parametrisation is relevant in cosmology since the metric~\eqref{eq: dS metric poincare} corresponds to a FLRW spacetime with a constant expansion rate $H$ and a scale factor $a(\tau) = -1/(H\tau)$. Constant-$\tau$ spatial hypersurfaces are $d$-dimensional Euclidean planes $\mathbb{R}^d$, which are manifestly symmetric under the action of the Euclidean group $\mathbb{E}_d \equiv \mathbb{T}_d \rtimes \SO(d)$, composed of translations $\mathbb{T}_d$ and rotations $\SO(d)$ in $d$ dimensions. Poincar\'e and global coordinates are related via the map
\begin{equation}
    \tau = \frac{-1}{\sinh(\rho) + \cosh(\rho)\,\omega^{d+1}} \,, \quad x^i = \frac{\omega^i}{\tanh(\rho) + \omega^{d+1}} \,.
\end{equation}
The Poincar\'e expanding patch in global coordinates is the region covering $\omega^{d+1}+\tanh(\rho) \geq 0$.

\paragraph{Two-point invariants.} When considering two-point correlation functions, an important quantity is the two-point invariant $\sigma^{\dS}$ between the points $X_1$ and $X_2$, defined as:
\begin{equation}\label{eq: def two-point invariant dS}
    \sigma^{\dS} = H^2 X_1^AX_2^A\eta_{A B}\,.
\end{equation}
This is related to the so-called chordal distance $d(X_1,X_2)$ as:
\begin{equation}
    d(X_1,X_2) = (X_1^A-X_2^A)(X_1^B-X_2^B)\eta_{AB} = \frac{2}{H^2}(1-\sigma^{\dS})\;.
    \label{def-chordal-distance-dS}
\end{equation}
As a result, $\sigma^{\dS}=1$ corresponds to two points separated by a null geodesic, while $\sigma^{\dS}>1\;(\textrm{resp.}<1)$ corresponds to time-like (resp. space-like) separations. In Poincar\'e coordinates, one has 
\begin{equation}
\sigma^{\dS}= \frac{\tau_1^2+\tau^2_2-|\x_{12}|^2}{2\tau_1\tau_2}\;,\, \textrm{with} \quad   \x_{12}=\x_2-\x_1\,.
\end{equation}

\subsection{Representation Theory of $\SO(1, d+1)$}
We now wish to formulate quantum field theory on the de Sitter geometry introduced above. The corresponding Hilbert space decomposes as a direct sum of invariant subspaces corresponding to unitary irreducible representations (UIRs) of the de Sitter isometry group. Before turning to specific decompositions of the symmetry generators, let us first review how elementary particles in $\dS_{d+1}$ can be organised into UIRs of the de Sitter group in a coordinate-independent manner, see~\cite{Thomas1941, Newton1950, Dixmier1961, Hirai1962, Takahashi1963}. Further details can be found in the reviews~\cite{Basile:2016aen, Sun:2021thf} and references therein.
\paragraph{Symmetries of de Sitter.} From its definition~\eqref{eq: dS def}, de Sitter spacetime $\dS_{d+1}$ is manifestly invariant under $\SO(1, d+1)$, which is isomorphic to the Euclidean conformal group in $\mathbb{R}^d$. As such, it has $\tfrac{1}{2}(d+2)(d+1)$ Killing vectors. A standard basis for the corresponding Lie algebra is $\{J_{AB}\}$ where
\begin{equation}
    J_{AB} \equiv X_A \frac{\partial}{\partial X^B} - X_B \frac{\partial}{\partial X^A}\,,
\end{equation}
is the vector $(1/2, 1/2)$ representation, with $A, B = 0, 1, \ldots, d+1$. The generators are antisymmetric $J_{AB} = -J_{BA}$ and verify the following commutation relation
\begin{equation}
\label{eq: Lorentz algebra}
    [J_{AB}, J_{CD}] = i\,(\eta_{BC}J_{AD} - \eta_{AC} J_{BD} + \eta_{AD} J_{BC} - \eta_{BD} J_{AC}) \,,
\end{equation}
which is the $(d+2)$-dimensional Lorentz algebra.

\paragraph{Unitary irreducible representations.} According to Wigner's theorem~\cite{Wigner1948}, particles in $\dS_{d+1}$ can be identified as UIRs of the de Sitter group. Since $\SO(1, d+1)$ is non-compact, these representations are infinite-dimensional. To determine the Hilbert space, a standard procedure is to identify the largest set of commuting Hermitian operators among the generators and diagonalise them in a common orthogonal basis. As such, the Hilbert space $\H$ can be decomposed as a direct sum over UIRs
\begin{equation}
    \H = \bigoplus_\alpha \V_{(\alpha)} \,,
\end{equation}
where $\alpha$ is an abstract (for now unspecified) label running over UIRs of $\SO(1, d+1)$, and $\V_{(\alpha)}$ is the corresponding invariant subspace. By Schur's lemma, any operator $\C$ that commutes with all elements of the group must act as a multiple of the identity when restricted to $\V_{(\alpha)}$ \cite{Tung:1985iqd,Georgi:2000vve,Isaev:2018xcg}. This operator and its action on any vector $\ket{\alpha}\in \V_{(\alpha)}$ therefore takes the form
\begin{equation}
    \C = \bigoplus_\alpha M^2_{(\alpha)} \mathds{1}^{(\alpha)} \,, \quad \C \ket{\alpha} = M_{(\alpha)}^2 \ket{\alpha}\,.
\end{equation}
For a Lie group, an example is given by the quadratic Casimir operator, which for $\dS_{d+1}$ is given by
\begin{equation}
\label{eq: Casimir}
    \C^{\SO(1, d+1)}  = -\frac{1}{2}J^{AB} J_{AB} \,.
\end{equation}
Infinite-dimensional representations of $\SO(1, d+1)$ are labeled by an {\it a priori} complex conformal dimension $\Delta$ and a representation of the subgroup $\SO(d)$. When restricting to representations realised by scalar fields or traceless-symmetric tensor fields in spacetime, this representation is characterised by an integer spin $S$, with $S=0$ corresponding to the trivial representation. The quadratic Casimir eigenvalues are parametrised by $M_{(\alpha)}^2 = M_{(\Delta, S)}^2 = \Delta(d-\Delta) + S(S+d-2)$. From now on, we will only consider scalar representations by setting $S=0$. It is then convenient to introduce the parameter $\mu$ through $\Delta \equiv \tfrac{d}{2}+i \mu$. With this parametrisation, the associated Casimir eigenvalues become
\begin{equation}
\label{eq: Casimir eigenvalues}
    M_{(\alpha)}^2 = M_{\mu}^2 = \mu^2 + \frac{d^2}{4} \,.
\end{equation}
The form~\eqref{eq: Casimir eigenvalues} implies a $\mathbb{Z}_2$ shadow symmetry $\mu \leftrightarrow -\mu$, making manifest the isomorphism between the representations $(\mu)$ and $(-\mu)$. As we shall see, the parameter $\mu$ is related to the mass $m$ of the corresponding bulk scalar field realisation through $(m/H)^2 = M_{\mu}^2$.

\vskip 4pt
To ensure unitarity of the representations of $\SO(1, d+1)$, the generators of the Lie algebra must be represented by Hermitian operators on the Hilbert space, i.e.~$J_{AB}^\dagger = J_{AB}$. This requirement enforces the Casimir eigenvalues $M_{\mu}^2$ to be real, implying that the parameter $\mu$ must be either real or purely imaginary. Discarding the ever-present trivial representation, the possible scalar UIRs of $\SO(1, d+1)$ fall into three categories:

\begin{itemize}
    \item The \textbf{scalar principal series $\P_{\mu}$}: $\mu \in \mathbb{R}$. The bulk theory whose single-particle Hilbert space realises this representation corresponds to a free scalar field with mass that is heavy in Hubble units, $m > \tfrac{d}{2}H$. In the flat-space limit $R_\dS\to \infty$, these representations reduce to the familiar massive particle representations of the Poincar\'e group.

    \item The \textbf{scalar complementary series $\C_{\mu}$}: $\nu \equiv i\mu \in [-\tfrac{d}{2}, \tfrac{d}{2}]$. The corresponding bulk fields represent light particles, with $m\leq \tfrac{d}{2}H$. These representations include the massless UIR at their endpoints, $\nu =\pm\tfrac{d}{2}$.
    
    \item The \textbf{type-I exceptional series $\E_{\mu}$}: $\nu \equiv i\mu = \tfrac{d}{2} + k$ with $k\in \mathbb{Z}_{\geq 0}$. These representations correspond to shift symmetric bulk scalar fields in $\dS_{d+1}$, see e.g.~\cite{Bonifacio:2018zex}, with $k$ referred to as the level of the state. The discrete nature of these representations is reminiscent of the phenomenon of partial masslessness in (A)dS~\cite{Deser:1983tm, Deser:1983mm, Higuchi:1986py, Brink:2000ag, Zinoviev:2001dt}. These are tachyonic states with $m^2 = -k(k+d)H^2$. For example, the $k=1$ representation is realised by the galileon~\cite{Luty:2003vm, Nicolis:2008in}.
\end{itemize}

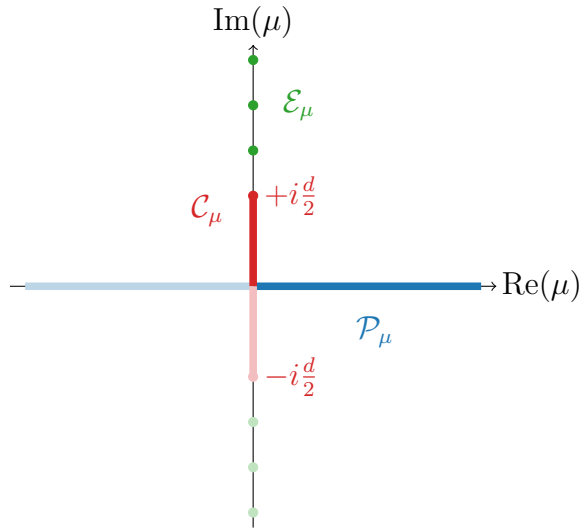
\begin{figure}[h!]
    \centering
    	\begin{tikzpicture}[scale = 2]
		\draw[black, ->] (-1.6,0) -- (1.6,0) coordinate (xaxis);
		\draw[black, ->] (0,-1.6) -- (0,1.6) coordinate (yaxis);
		\node at (1.9, 0) {$\text{Re}(\mu)$};
		\node at (0, 1.75) {$\text{Im}(\mu)$};

		\node at (0.8, -0.3) {\textcolor{pyblue}{$\mathcal{P}_{\mu}$}};
		\draw[-, pyblue, line width=1mm] (0, 0) -- (1.5, 0);
		\draw[-, lightpyblue, line width=1mm] (0, 0) -- (-1.5, 0);

		\node at (-0.3, 0.5) {\textcolor{pyred}{$\mathcal{C}_{\mu}$}};
		\draw[-, pyred, line width=1mm] (0, 0) -- (0, 0.6);
		\draw[-, lightpyred, line width=1mm] (0, 0) -- (0, -0.6);
		
		\draw[pyred, fill = pyred] (0, 0.6) circle (.03cm);
		\node at (0.25, 0.6) {$\textcolor{pyred}{+i\tfrac{d}{2}}$};
		\draw[lightpyred, fill = lightpyred] (0, -0.6) circle (.03cm);
		\node at (0.25, -0.6) {$\textcolor{pyred}{-i\tfrac{d}{2}}$};

		\node at (0.3, 1.2) {\textcolor{pygreen}{$\mathcal{E}_{\mu}$}};
		\draw[pygreen, fill = pygreen] (0, 0.9) circle (.03cm);
		\draw[pygreen, fill = pygreen] (0, 1.2) circle (.03cm);
		\draw[pygreen, fill = pygreen] (0, 1.5) circle (.03cm);
		
		\draw[lightpygreen, fill = lightpygreen] (0, -0.9) circle (.03cm);
		\draw[lightpygreen, fill = lightpygreen] (0, -1.2) circle (.03cm);
		\draw[lightpygreen, fill = lightpygreen] (0, -1.5) circle (.03cm);
	\end{tikzpicture}
    \caption{Scalar unitary irreducible representations of $\dS_{d+1}$ in the complex $\mu$-plane, illustrating the \textcolor{pyblue}{scalar principal series}, the scalar \textcolor{pyred}{complementary series} and the \textcolor{pygreen}{type-I exceptional series}. The shadow symmetry appears as a manifest $\mathbb{Z}_2$ invariance under $\mu \leftrightarrow -\mu$.}
\label{fig: de Sitter UIRs}
\end{figure}

These representations are depicted in Fig.~\ref{fig: de Sitter UIRs}.

\paragraph{Invariant subspace basis.} After charting the Hilbert space $\H$ into UIRs of the de Sitter group, one must still identify an explicit basis for the invariant subspaces $\V_{\mu}$. A convenient strategy is to select a subgroup $G\subset \SO(1, d+1)$ and further decompose $\V_{\mu}$ into a direct sum of the UIRs of $G$, as
\begin{equation}
    \V_{\mu} = \bigoplus_{\ell, m, \ldots} \V_{\mu}^{\ell, m, \ldots}\,,
\end{equation}
where the indices $\ell, m, \ldots$, which label UIRs of $G$, play the role of {\it quantum numbers}. In what follows, we will show that choosing $G$ as the isometry group of the spatial hypersurfaces corresponding to the Poincar\'e slicing naturally leads to a {\it momentum basis}, conjugate to the position-space variables. This construction provides the de Sitter counterpart of the energy-momentum Fourier space in $(d+1)$-dimensional Minkowski spacetime $\mathbb{M}^{1, d}$.

\newpage
\section{Kontorovich-Lebedev-Fourier Momentum Space}
\label{sec: poincare slicing}

In this section, we describe our approach to construct a natural momentum-space representation of the Hilbert space, dual to position-space states, by exploiting the symmetries of the Poincar\'e slicing. This construction naturally gives rise to a {\it Kontorovitch-Lebedev-Fourier} (KLF) transform in $d+1$ dimensions, which diagonalises the dynamical and kinematical structure of the bulk de Sitter theory.

\vskip 4pt
We begin by leveraging the invariance of Cauchy slices of constant conformal time under the $d$-dimensional Euclidean group to build a complete, orthonormal basis for the Hilbert space, Sec.~\ref{symmetry}. Then, using the decomposition of the de Sitter generators into the Euclidean conformal algebra, we solve the corresponding quadratic Casimir differential equation to determine the harmonic function connecting the position and momentum representations, Sec.~\ref{sec: Harmonic functions and Position Space}. We describe how (Euclidean) square-integrable functions can be decomposed into harmonic functions of the principal series through the KLF transform, Sec.~\ref{KLF-space}. We then show how more general functions whose spectral decomposition have support on other UIRs can be naturally described in our formalism, Sec.~\ref{sec: Role of the other Series}

\subsection{Momentum Space from Symmetry Generators}
\label{symmetry}

The isomorphism between the $\SO(1, d+1)$ Lie algebra~\eqref{eq: Lorentz algebra} and the $d$-dimensional Euclidean conformal algebra is made manifest through the map
\begin{equation}
\label{eq: de Sitter generator conformal split}
    D = J_{0, d+1} \,, \quad M_{ij} = J_{ij} \,, \quad P_i = J_{d+1, i}+J_{0, i} \,, \quad K_i = J_{d+1, i} - J_{0, i} \,,
\end{equation}
where the newly defined generators $D, P_i, K_i$ and $M_{ij} = -M_{ji}$, with $i, j = 1, \ldots d$, correspond to dilatation, spatial translations, special conformal transformations (sometimes referred to as de Sitter boosts), and rotations, respectively. From Eq.~\eqref{eq: Lorentz algebra} and~\eqref{eq: de Sitter generator conformal split}, it follows that the generators satisfy the  commutation relations of the conformal algebra:
\begin{equation}
\label{eq: Lie algebra poincare}
    \begin{aligned}
        [D, P_i] &= iP_i \,, \quad [D, K_i] = -iK_i \,, [K_i, P_j] = 2i\,(\delta_{ij}D - M_{ij}) \,, \\
        [M_{ij}, P_k] &= i\,(\delta_{jk}P_i - \delta_{ik} P_j) \,, \quad [M_{ij}, K_k] = i\,(\delta_{jk}K_i - \delta_{ik} K_j) \,, \\
        [M_{ij}, M_{k\ell}] &= i\,(\delta_{jk}M_{i\ell} - \delta_{ik}M_{j\ell} + \delta_{i\ell}M_{jk} - \delta_{j\ell}M_{ik}) \,,
    \end{aligned}
\end{equation}
together with $[P_i, P_j] = [K_i, K_j]=[D, M_{ij}] = 0$. Throughout, we adopt the convention that all generators are Hermitian, so that the corresponding group elements are unitary. The quadratic Casimir operator~\eqref{eq: Casimir} is given by
\begin{equation}
    \C^{\SO(1, d+1)} = D(D-id) - P_i K^i - \frac{1}{2}M_{ij}^2 \,,
    \label{dS-Casimir}
\end{equation}
where $-\tfrac{1}{2}M_{ij}^2 \equiv -\tfrac{1}{2}M_{ij}M^{ij}$ is the quadratic Casimir operator of $\SO(d)$. In the Poincar\'e coordinates, a local bulk scalar operator $\O(\tau, \x)$ in de Sitter transforms under the $\SO(1, d+1)$ isometries as the following commutators
\begin{equation}
\label{eq: Poincare killing vectors}
    \begin{aligned}
        [D, \O(\tau, \x)] &= \D\O(\tau, \x) \equiv -i\,(\tau\partial_\tau + x^i \partial_i) \O(\tau, \x) \,, \\
        [P_i, \O(\tau, \x)] &= \P_i\O(\tau, \x) \equiv -i\,\partial_i \O(\tau, \x) \,, \\
        [K_i, \O(\tau, \x)] &= \K_i\O(\tau, \x) \equiv -i\,[2x_i \tau\partial_\tau + 2x_ix^j\partial_j + (\tau^2-\x^2)\partial_i]\O(\tau, \x) \,, \\
        [M_{ij}, \O(\tau, \x)] &= \M_{ij}\O(\tau, \x) \equiv -i\,(x_i\partial_j - x_j \partial_i) \O(\tau, \x) \,,
    \end{aligned}
\end{equation}
where $\D, \P_i, \K_i$ and $\M_{ij}$ are the corresponding Killing vector differential operators.

\vskip 4pt
The intrinsic $d$-dimensional nature of the algebra~\eqref{eq: Lie algebra poincare} is reminiscent of the conformal invariance of the late-time boundary ($\tau \to 0$) of $\dS_{d+1}$. However, the constant-$\tau$ Cauchy slices are invariant only under spatial rotations and translations. As announced, we construct our basis by diagonalising the Casimir and the spatial translations, yielding the following momentum basis for the Poincar\'e-sliced Hilbert space\footnote{Since the Casimir operator is unbounded, its eigenstates are typically not part of the Hilbert space $\H$. As reviewed in appendix~\ref{sec: Rigged Hilbert Space}, they can be seen as linear forms acting on elements of $\H$. For simplicity, we ignore these subtleties in the bulk of the paper.}
\begin{equation}
\label{eq: Hilbert space Poincare}
    \H \equiv \Span\left(\ket{\mu, \k}\right) \,,
\end{equation}
where the basis elements $\ket{\mu, \k}$ satisfy the following properties
\begin{equation}
    \C^{\SO(1, d+1)}\ket{\mu, \k} = M_{\mu}^2 \ket{\mu, \k} \,, \quad P_i \ket{\mu, \k} = k_i \ket{\mu, \k} \,,
\end{equation}
with $M_{\mu}^2$ defined in~\eqref{eq: Casimir eigenvalues}. They are identified with single-particle states that can be constructed from the Bunch-Davies vacuum $\ket{\Omega}$ by acting with the appropriate creation operator:
\begin{equation}
    \ket{\mu, \k} \equiv a_{\k}^{(\mu)\dagger} \ket{\Omega} \,.
\end{equation}
Since $\ket{\Omega}$ is invariant under $\SO(1,d+1)$, it is annihilated by every $J_{A B}$ of the Lie algebra
\begin{equation}
    J_{A B}\ket{\Omega} = 0\;.
\end{equation}
The basis states form a complete and orthonormal basis of the Hilbert space
\begin{equation}
\label{eq: orthonormality momentum states}
    \braket{\mu, \k | \mu', \k'} = \frac{(2\pi)^d |\mu|}{\N_{\mu}} \, \delta\left(M_{\mu}^2 - M_{\mu'}^2\right) \, \delta^{(d)}(\k - \k') \,,
\end{equation}
and
\begin{equation}
\label{eq: completeness momentum states}
    \mathds{1} = \int_{\S}\d\mu\N_{{\mu}}\frac{\d^d\k}{(2\pi)^d}\ket{\mu,\k}\bra{\mu,\k} \,.
\end{equation}
The $\mu$-integration in principle runs over the full spectrum $\mathcal{S}$ of the Casimir operator, i.e.~$\P_{\mu}, \C_{\mu}$ and $\E_{\mu}$, the latter for which the integral turns into a discrete sum. Equivalently, one can also see \eqref{eq: orthonormality momentum states} as a direct consequence of the canonical commutation relations
\begin{equation}
    [a_{\k}^{\mu}, a_{\k'}^{\mu'\dagger}] = \frac{(2\pi)^d|\mu|}{\N_{\mu}} \,  \delta\left(M_{\mu}^2 - M_{\mu'}^2\right) \, \delta^{(d)}(\k-\k') \,.
\end{equation}
In what follows, it will be enough to compute $\N_{\mu}$ for the principal series, see \eqref{Nmu} below.

\paragraph{Comparison with Minkowski.} 

Here we comment on a simple conceptual difference between Minkowski and dS that is worth highlighting. For Minkowski, the Casimir of the Poincar\'e isometry group $\mathbb{R}^{d+1} \rtimes \mathrm{SO}(d+1)$ is simply $\C=-P_\mu P^\mu=-P_0 P^0+P ^i P^i$, where $P^\mu$ correspond to spacetime translations, which form an abelian subalgebra of the full isometry group. As a result, one can diagonalise at the same time the Casimir, the spatial translations $P_i$, and the time-translation operator $P_0$, with eigenvalues that are related through $M^2=\omega^2-\k^2$. There is no such equivalent for the dS Casimir~\eqref{dS-Casimir}, whose building blocks do not all commute with one another. Hence, one can diagonalise the Casimir and spatial translations, as we do in our construction, but there is no dispersion relation between the frequency $\mu$ and the spatial momentum $\k$. Saying it otherwise, the de-Sitter mass-shell condition simply states that the Casimir eigenvalue $M_{\mu}$ coincides with the mass of a particle of the spectrum.

\begin{table}[h!]
    \centering
    \begin{tabular}{|c|c|c|}
        \hline
         & $\ket{\mu,\x}\equiv e^{-i\boldsymbol{P}\cdot\x}\ket{\mu,\x=0}$ & $\ket{\mu,\k}$\\
        \hline
        Casimir Operator $\C$ & $\frac{d^2}{4}+\mu^2$ & $\frac{d^2}{4}+\mu^2$\\
        \hline
        Spatial Translation Operator $P_i$ & $-i\partial_i$ & $k_i$\\
        \hline
        Dilatation Operator $D$ & $\frac{d}{2}+i\mu -ix^j\partial_j$ & $i\left(k_i\partial_{k^i} + \tfrac{d}{2}\right) $\\
        \hline
    \end{tabular}
    \caption{Action of the Casimir, translations and dilatation operators on the CFT-basis $\ket{\mu,\x}$ and the basis $\ket{\mu,\k}$ used in this paper.}
    \label{tab:action of generator}
\end{table}

\paragraph{Comparison with the CFT basis.} In Ref.~\cite{Hogervorst:2021uvp}, the basis $\ket{\mu,\k}$ was introduced but quickly abandoned for the CFT basis mentioned in the introduction. In the latter, one diagonalises the dilatation, with eigenvalue $\Delta=\frac{d}{2}+i \mu$, and then acts with the translation operator to obtain the CFT position basis $\ket{\mu,\x}$ \cite{Sun:2021thf}. This is related to the basis $\ket{\mu,\k}$ by the following tilted Fourier transform \cite{Hogervorst:2021uvp}:
\begin{equation}
    \ket{\mu,\x} = \int\frac{\d^d\k}{(2\pi)^d}e^{i\k\cdot\x}k^{i\mu}\ket{\mu,\k}\;,
\end{equation}
and we compare in Table~\ref{tab:action of generator} how the Casimir, translations and dilatation operators act on each basis. Note that $\ket{\mu,\x}$ is not a dilatation eigenstate for $\x \neq 0$.

\subsection{Harmonic Functions and Position Space}\label{sec: Harmonic functions and Position Space}

In order to obtain the transformation properties of the states $\ket{\mu, \k}$, and to construct the integral transform that connects the KLF momentum space to position space, we define the following wavefuntions:
\begin{equation}
\label{eq: harmonic function}
     \Phi^\mu_{\k,\O}(\tau,\x)\equiv\braket{\Omega | \O(\tau, \x) | \mu, \k}\,,
\end{equation}
where $\O(\tau,\x)$ is any local operator in the theory. Physically, this object represents the transition amplitude from the Bunch-Davies vacuum of the theory to a specific field configuration at some time $\tau$, given by $\O(\tau, \x)$, projected onto a single-particle state.
\vskip 4pt
Before deriving the explicit expression for $\Phi_{\k,\O}^{\mu}(\tau, \x)$, let us first perform a Wick rotation to Euclidean signature. In order to define the Bunch-Davies vacuum state, we need an appropriate $i\epsilon$-prescription in the infinite past.\footnote{The $i\epsilon$-prescription prepares the Hartle-Hawking state of the theory in the far past, as shown in~\cite{Higuchi:2010xt}.} Because of this, the time variable we use acquires a non-zero imaginary value and the theory must be studied on the tilted conformal time axes $\tau_c =\tau e^{-i \a \epsilon}$, with $\a=\pm 1$, on which the Casimir operator is not self-adjoint. This can be solved by analytically continuing the conformal time to the imaginary axis. To this aim, it is convenient to introduce the variable $z\in \mathbb{C}$ as a phase shift of the complexified conformal time $\tau_c$:
\begin{equation}
    \tau_c = z\, e^{\frac{i\a\pi}{2}}\;.
\end{equation}
The Euclidean signature is defined when $z\in \mathbb{R}_+$ and the tilted axis is recovered when:
\begin{equation}\label{eq: z Lorentzian}
    z=- \tau e^{i\, \a \left(\frac{\pi}{2}-\epsilon\right)}\;.
\end{equation}
This analytic continuation maps the geometry of $\dS_{d+1}$ to the one of $\EAdS_{d+1}$, with $z$ acting as the radial variable with conformal boundary located at $z=0$. The conformal time integrals are deformed to 
\begin{equation}\label{eq: Wick rotation to EAdS}
    \int_{-\infty_\a}^0\d\tau f(\tau) = e^{-\frac{i\a\pi}{2}}\int_0^\infty\d z f(e^{ \frac{i\a\pi}{2}}z) \,,
\end{equation}
where $-\infty_\pm \equiv -\infty(1\mp i\epsilon)$. As shown in~\eqref{eq: Lorentzian two-point invariant off-diag} and~\eqref{eq: Lorentzian two-point invariant diag}, this analytic continuation procedure can be repeated at the level of the two-point invariant~\eqref{eq: def two-point invariant dS}. We stress that we do not need to perform the Wick rotation for the Hubble radius. This is because we are interested in dS correlation functions, and the analytic continuation to EAdS should be seen as a technical step.

\vskip 4pt
Using~\eqref{eq: Poincare killing vectors}, the Killing vector differential expression for the Casimir operator is given by the Laplace-Beltrami operator in $\dS_{d+1}$
\begin{equation}
    \begin{aligned}
        \C : H^{-2}\Box_\dS &= -\left(\tau^2\partial_\tau^2 - (d-1)\tau\partial_\tau - \tau^2\partial_i^2\right) \\
        &= -\left(z^2\partial_z^2 - (d-1)z\partial_z + z^2\partial_i^2\right) \,.
    \end{aligned}
\end{equation}
Then, the definition \eqref{eq: harmonic function} together with the hermiticity of the generators provides us with a set of differential equations for the wavefunction $\Phi^\mu_{\k,\O}$:
\begin{equation}
\label{eq: diff eq Poincare}
    \begin{aligned}
        \left(z^2 \partial_z^2 - (d-1)z\partial_z + z^2 \partial_i^2\right) \Phi_{\k,\O}^{\mu}(z, \x) &= -M_{\mu}^2 \Phi_{\k,\O}^{\mu}(z, \x) \,, \\
        \partial_i \Phi_{\k,\O}^{\mu}(z, \x) &= -ik_i \Phi_{\k,\O}^{\mu}(z, \x) \,.
    \end{aligned}
\end{equation}
This can be seen in the following way for the momentum, and similarly for the Casimir:
\begin{equation}
    \begin{aligned}
        -i k_i\Phi^{\mu}_{\k,\O}(z,\x) &= -i\bra{\Omega}\O(z,\x)P_i\ket{\mu,\k} \\
        & = -i \bra{\Omega}[\O(z,\x),P_i]\ket{\mu,\k}\\
        & = \partial_i\Phi^{\mu}_{\k,\O}(z,\x)\;.
    \end{aligned}
\end{equation}
Since these equations are independent of the nature of $\O$, we can write the wavefuntion~\eqref{eq: harmonic function} as:
\begin{equation}
    \Phi^\mu_{\k,\O}(z,\x) = c_\O(\mu)\Phi^\mu_{\k}(z,\x)\;,
\end{equation}
where the harmonic function $\Phi^\mu_{\k}(z,\x)$ satisfies the system~\eqref{eq: diff eq Poincare} with Bunch-Davies initial conditions and $c_\O(\mu)$ is a theory-dependent quantity encoding the physics of the operator $\O(z,\x)$. Equation~\eqref{eq: harmonic function} together with the resolution of the identity~\eqref{eq: completeness momentum states} can be used to derive the Euclidean K\"all\'en-Lehmann decomposition of the two-point function of the operator $\O(z,\x)$~\cite{Loparco:2023rug}:
\begin{equation}
    \bra{\Omega}\O(z_1,\x_1)\O(z_2,\x_2)\ket{\Omega} = \int_{\S}\d\mu\;\frac{\rho_\O(\mu)}{H^2}\frac{\d^d\k}{(2\pi)^d}\Phi^\mu_{\k}(z_1,\x_1)\left[\Phi^\mu(z_2,\x_2)\right]^*\;,
\end{equation}
where $\rho_\O(\mu) = H^2\N_\mu |c_\O(\mu)|^2$ is called the spectral density of $\O$.\footnote{If $\O$ has mass dimension $\delta_\O$, the dimension of the spectral density is $\delta_\rho=2\delta_\O-d+1$.} For reasons that will become clear later, this only makes sense when the two operators $\O(z_{1,2},\x_{1,2})$ do not live on the same copy of EAdS, which corresponds to the Wightman two-point function. The Lorentzian version can be obtained by consistently applying the Wick rotation as explained in appendix~\ref{sec: propagators in real space}.

\vskip 4pt
Notice that the differential operators in \eqref{eq: diff eq Poincare} are self-adjoint, i.e.~$\braket{\D f, g} = \braket{f, \D g}$, with respect to the natural $L^2(\EAdS_{d+1})$-inner product
\begin{equation}\label{eq: Natural L2 Inner Product}
    \braket{f, g} \equiv \int_{\dS_{d+1}} f^*(X^E)g(X^E) \, \vol_g = \int_\EAdS\frac{\d z \d^d\x}{(H z)^{d+1}} \, f^*(z, \x) g(z, \x) \,,
\end{equation}
with $\vol_g \equiv \d^{d+2}X^E \, \delta(X^E\cdot X^E + R_\EAdS^2)$, where we have specified to Poincar\'e coordinates. This can be checked explicitly after successive integrations by parts.

\vskip 4pt
Keeping only the solution that decays at infinity $z\to\infty$, which is equivalent to requiring the absence of states with negative energy, the set of equations~\eqref{eq: diff eq Poincare} fixes the functional dependence of $\Phi$:
\begin{FramedBox}
\begin{equation}
\label{eq: harmonic function expression}
    \Phi_{\k}^{\mu}(z, \x) = \frac{H^{\frac{d+1}{2}}}{\sqrt{\pi}} \, e^{-i \k \cdot \x} z^{\frac{d}{2}} K_{i\mu}(kz)\,,
\end{equation}
\end{FramedBox}
where $k\equiv |\k|$, $K_{i\mu}$ is the modified Bessel function of the second kind (Macdonald function), and the pre-factor has been chosen for later convenience. For real values of $\mu$, the functions $K_{i\mu}$ are the kernels of the Kontorovich-Lebedev integral transformation reviewed in App.~\ref{sec: Rigged Hilbert Space}. This will be central in what follows. 
The harmonic functions~\eqref{eq: harmonic function expression} verify the following property under complex conjugation
\begin{equation}
    \left[\Phi_{\k}^{\mu}(z, \x)\right]^* = \Phi_{-\k}^{\mu}(z, \x) \,,
\label{complex-conjugate-Harmonic-function}    
\end{equation}
which can be interpreted as a CPT symmetry,\footnote{See e.g.~\cite{Goodhew:2024eup,Thavanesan:2025kyc,Thavanesan:2025ibm} for constraints from CPT on cosmological correlators.} and the shadow symmetry yields
\begin{equation}
    \Phi_{\k}^{\mu}(z, \x) = \Phi_{\k}^{-\mu}(z, \x) \,,
\end{equation}
for $\mu \in \mathbb{C}$ and $z\in\mathbb{C}\backslash(-\infty, 0]$. Analytic continuation to the Lorentzian signature is recovered when one evaluates $z$ on the tilted axis~\eqref{eq: z Lorentzian}:
\begin{equation}\label{eq: Lorentzian harmonic functions}
    \Phi^{\mu}_{\k}\left(-\tau e^{i \a \left( \frac{\pi}{2}-\epsilon \right)},\x\right) = H^{\frac{d+1}{2}}e^{-i\k\cdot\x}u^{-\a}_k\left(\tau (1-i \a\epsilon),\mu\right)\;,
\end{equation}
where the $u^\a_k(\tau,\mu)$ are the familiar mode functions of a massive field of mass $m=HM_\mu$ in momentum space: 
\begin{equation}\label{eq: Lorentzian mode functions} 
\begin{aligned}
     u^+_k(\tau,\mu)&=\frac{\sqrt{\pi}}{2}e^{-\frac{\pi\mu}{2}-\frac{i\pi(d-2)}{4}}(-\tau)^{\frac{d}{2}}H_{i\mu}^{(1)}(-k\tau)\\
     u^-_k(\tau,\mu) &= \frac{\sqrt{\pi}}{2}e^{+\frac{\pi\mu}{2}+\frac{i\pi(d-2)}{4}}(-\tau)^{\frac{d}{2}}H_{i\mu}^{(2)}(-k\tau)\;,
\end{aligned}
\end{equation}
and $H_{i\mu}^{(1, 2)}$ are the Hankel functions. Notice that the $i\epsilon$ prescription is not needed in order to evaluate expression~\eqref{eq: Lorentzian harmonic functions} as the mode functions are analytic for $\tau\in\mathbb{R}_-$. However, we will see that the Lorentzian correlation functions have branch cuts when two operators are within each other's light cones, and keeping the $\epsilon$'s explicit will be useful. The Wick rotation~\eqref{eq: Lorentzian harmonic functions} and the relation~\eqref{completeness-orthogonality-harmonic-principal} can be combined to show the completenesses relation among Hankel functions used in~\cite{Grafe:2026qsm,Lee:2025kgs}.

\paragraph{Action of generators.} The definition \eqref{eq: harmonic function} of the wavefunctions, together with the functional form \eqref{eq: harmonic function expression} of their Euclidean versions, and the commutation relations \eqref{eq: Poincare killing vectors}, dictate how the $\SO(1, d+1)$ generators~\eqref{eq: de Sitter generator conformal split} act on the (scalar) single-particle states $\ket{\mu, \k}$. Explicitly, these actions are given by\footnote{Our minus sign and factor $i$ convention differ from~\cite{Hogervorst:2021uvp} because we take the generators to be Hermitian.}
\begin{equation}
\label{eq: generator action on Poincare momentum states}
\begin{aligned}
    D \ket{\mu, \k} &= i\left(k_i\partial_{k^i} + \tfrac{d}{2}\right) \ket{\mu, \k} \,, \\
    P_i \ket{\mu, \k} &= k_i \ket{\mu, \k} \,, \\
    K_i \ket{\mu, \k} &= \left[k_i \delta_{j\ell}\partial_{k^j}\partial_{k^\ell} - 2\left(2k_j \partial_{k^j} + d\right)\partial_{k^i} - \tfrac{\mu^2}{k^2}k_i\right] \ket{\mu, \k} \,, \\
    M_{ij} \ket{\mu, \k} &= -i \left(k_i \partial_{k^j} - k_j \partial_{k^i} \right) \ket{\mu, \k} \,.
\end{aligned}
\end{equation}
For instance, using~\eqref{eq: Poincare killing vectors}, one obtains
\begin{equation}
    \begin{aligned}
        \D\Phi_{\k,\O}^{\mu}(z, \x) &= -i \left(z\partial_z + x^i\partial_{x^i}\right) \Phi_{\k,\O}^{\mu}(z, \x) \\
        &= -i \left(\tfrac{d}{2} - i \k\cdot \x\right)\Phi_{\k,\O}^{\mu}(z, \x) - i c_\O(\mu) e^{-i \k\cdot \x} z^{d/2} k^i\partial_{k^i} K_{i\mu}(kz) \\
        &= -i \left(k^i \partial_{k^i} + \tfrac{d}{2}\right)\Phi_{\k,\O}^{\mu}(z, \x) \\
        &= -i \left(k^i \partial_{k^i} + \tfrac{d}{2}\right) \braket{\Omega|\O(z, \x)|\mu, \k} \,.
    \end{aligned}
\end{equation}
The symmetry of $K_{i\mu}(kz)$ under $k\leftrightarrow z$ was used to write $z\partial_z K_{i\mu}(kz) = k^i\partial_{k^i} K_{i\mu}(kz)$. Additionally, the dilatation acts on a bulk operator as the commutator
\begin{equation}
    \D\Phi_{\k,\O}^{\mu}(z, \x) = \braket{\Omega| [D, \O(z, \x)]|\mu, \k} = -\braket{\Omega|\O(z, \x)D|\mu, \k} \,,
\end{equation}
where we used the invariance of the Bunch-Davies vacuum under de Sitter isometries, i.e.~$D\ket{\Omega} = 0$. Since the generators are Hermitian, it follows that $\bra{\Omega}D = 0$. Identifying both expressions, we recover the first relation in~\eqref{eq: generator action on Poincare momentum states}. Deriving the remaining identities is similar and straightforward. In the same way, one can derive the transformation properties of the position-space states $\ket{z, \x}$ that we now introduce.

\paragraph{Back to position space.}  The harmonic functions allow to define the position-space states to which the momentum-space basis elements $\ket{\mu, \k}$ are dual as:
\begin{equation}\label{eq: def position-space states}
\begin{aligned}
    \ket{z, \x}_{\S} &\equiv \int_{\S}\d\mu \N_{\mu} \frac{\d^d\k}{(2\pi)^d} \left[\Phi_{\k}^{\mu}(z, \x)\right]^* \ket{\mu, \k}\;,
\end{aligned}
\end{equation}
A consequence of~\eqref{eq: def position-space states} is that the harmonic functions provide the overlap between momentum states and their position-space duals
\begin{equation}
    \;_\S\braket{z, \x| \mu, \k} =  \braket{\mu, \k | z, \x}^*_\S = \Phi_{\k}^{\mu}(z, \x) \,.
\end{equation}
As the states $\ket{\mu,\k}$ represent the propagation of some unlocalised particle of mass $M_{\mu}$ and momentum $\k$, the dual state $\ket{z,\x}_\S$ stands for a localised quanta in a superposition of Casimir and momentum eigenstates. In the next sub-section, we will show that these position states form a complete basis of the space of square-integrable functions when the $\mu$ integral is restricted from the full spectrum $\S$ to the principal series $\P$.

\subsection{Kontorovich-Lebedev-Fourier Space} 
\label{KLF-space}

In the above section, we showed that the differential operators in the eigenvalue equations \eqref{eq: diff eq Poincare} for the harmonic functions are self-adjoint with respect to the natural $L^2(\EAdS_{d+1})$-inner product \eqref{eq: Natural L2 Inner Product}. When dealing with self-ajoint bounded operators on a compact domain, the Sturm-Liouville theory guarantees that the corresponding discrete set of eigenfunctions form an orthogonal basis of the associated Hilbert space. For non-compact operators, as relevant here, the spectrum acquires a continuous part, but, more importantly, 
the proper mathematical structure becomes the rigged Hilbert space (also known as the Gelfand triple), and only a subset of the eigenfunctions enter into the corresponding expansion.
This rather technical procedure is carried out in App.~\ref{sec: Rigged Hilbert Space} (see also \cite{Belrhali:2026ktb} for a summary), where we show that any square integrable function on $\EAdS_{d+1}$, $f(z,\x)\in L^2(\EAdS_{d+1})$ can be decomposed along the principal series harmonic functions:
\begin{equation}\label{eq: KLF decomposition}
    f(z,\x) = \int_{\KLF}\d\mu\,\N_{\mu}\frac{\d^d\k}{(2\pi)^d} \, \Phi^{\mu}_{\k}(z,\x)f^{\mu}_{\k}\,,
\end{equation}
where, importantly, the $\mu$-integration range spans the real axis only:
\begin{equation}
    \int_{\KLF} \d\mu\,\N_{\mu}\frac{\d^d\k}{(2\pi)^d} \equiv \int_{\mathbb{R}} \d\mu\,\N_{\mu}\int_{\mathbb{R}^d}\frac{\d^d\k}{(2\pi)^d}\,,
\end{equation}
and
\begin{equation}
    \N_{\mu} = \frac{\mu}{\pi}\sinh(\pi\mu) 
    \label{Nmu}
\end{equation}
is often called the de Sitter density of (principal series) states. The temporal component of the transformation~\eqref{eq: KLF decomposition} can be recognised as the the Kontorovich-Lebedev (KL) transform, introduced in App.~\ref{sec: Rigged Hilbert Space}. Since the spatial part is the usual $d$-dimensional Fourier transform, we name this (restricted) momentum space the {\it Kontorovich-Lebedev-Fourier} (KLF) space.

\vskip 4pt
Equivalently to the decomposition \eqref{eq: KLF decomposition}, the function $f(z,\x)$ can be seen as a state living on the principal series subspace
\begin{equation}
\label{eq: definition L2 functions}
    f(z,\x) = \braket{z,\x|f}\,, \quad \text{where} \quad \ket{f}\in \bigcup_{\mu\in \mathbb{R}}\P_{\mu}\,,
\end{equation}
and we defined $\ket{z,\x}$ as the restriction of the position state \eqref{eq: def position-space states} to a superposition of principal series states only:
\begin{equation}\label{eq: restricted position states}
    \ket{z,\x} = \int_\KLF\d\mu\N_{\mu}\frac{\d^d\k}{(2\pi)^d}\left[\Phi^{\mu}_{\k}(z,\x)\right]^*\ket{\mu,\k}\;.
\end{equation}
As advertised before, the restricted position states \eqref{eq: restricted position states} constitute an orthonormal and complete basis of the principal series subspace. Indeed, the completeness and orthogonality relations of the principal series harmonic functions \eqref{completeness-orthogonality-harmonic-principal} implies
\begin{equation}
    \braket{z,\x|z',\x'}
     = (H z)^{d+1}\delta(z-z')\delta^{(d)}(\x-\x')\;,
\end{equation}
and 
\begin{equation}\label{eq: KLF resolution of indentity}
\begin{aligned}
    \mathds{1}_\P&=\int_\EAdS\frac{\d z\d^d\x}{(H z)^{d+1}}\ket{z,\x}\bra{z,\x} \\
    &= \int_\KLF\d\mu\N_{\mu}\frac{\d^d\k}{(2\pi)^d}\ket{\mu,k}\bra{\mu,\k}\;.
\end{aligned}
\end{equation}
Consequently, the general KLF decomposition \eqref{eq: KLF decomposition} can be easily recovered by inserting a resolution of identity \eqref{eq: KLF resolution of indentity} in \eqref{eq: definition L2 functions}. Similarly, the momentum-space representation of the $f^{\mu}_{\k} = \braket{\mu,\k|f}$ in the integral \eqref{eq: KLF decomposition} is given by the following inverse formula
\begin{equation}
\label{eq: def KLF transform}
    f^{\mu}_{\k} = \int_\EAdS\frac{\d z\d^d\x}{(H z)^{d+1}}\left[\Phi^{\mu}_{\k}(z,\x)\right]^*f(z,\x)\,.
\end{equation}
A crucial consequence is that the principal series provides a complete representation only within the {\it subspace} of the full Hilbert space $\H$ consisting of square-integrable $L^2(\EAdS_{d+1})$ functions. For completeness, let us mention the existence of the Parseval identity, i.e.~for any pair of square integrable functions $f(z,\x), g(z,\x) \in L^2(\EAdS_{d+1})$, the scalar product is preserved by the KLF transformation:
\begin{equation}
    \int_{\KLF}\d\mu\,\N_{\mu}\frac{\d^d\k}{(2\pi)^d}\left[f^{\mu}_{\k}\right]^*g^{\mu}_{\k} = \int_{\EAdS}\frac{\d z\,\d^d\x}{(H z)^{d+1}}f(z,\x)^*g(z,\x) \,.
\end{equation}
This identity also holds for more general functions provided that the momentum-space integral also includes contributions from the complementary and exceptional series, as we now discuss.

\subsection{Role of the Non-Principal Series}\label{sec: Role of the other Series}
Motivated by the path-integral formalism, we paid special attention to the space $L^2\left(\EAdS_{d+1}\right)$. However, the objects we encounter in physical situations can be more general functions and other UIRs can appear in their spectral decomposition. As we show in this section, these contributions can be read from the non-analyticities in the $\mu$ complex plane of the KLF modes. Moreover, since the latter are meromorphic functions in every physical example, the non-principal UIRs only contribute as isolated points in the spectrum and can be taken into account in the decomposition~\eqref{eq: KLF decomposition} by promoting $f^{\mu}_{\k}$ to a distribution. In order to achieve this, we generalise the orthogonality
relation among the harmonic functions:
\begin{equation}
    \int_{\EAdS}\frac{\d z\d^d\x}{(H z)^{d+1}}\left[\Phi^{\mu}_{\k}(z,\x)\right]^*\Phi^{\alpha}_{\k'}(z,\x)=\frac{(2\pi)^d}{\N_\mu}\delta^{(d)}(\k-\k')\hat{\delta}_{\alpha}(\mu)\;,
\end{equation}
where for any $\alpha\in\mathbb{C}$, $\hat{\delta}_{\alpha}(\mu)$ acts on KLF space as:
\begin{equation}\label{eq: definition delta hat}
    \hat{\delta}_\alpha[f]\equiv\int\displaylimits^{\infty}_{-\infty}\d\mu\;\hat{\delta}_\alpha(\mu)f^{\mu}_{\k}=\frac{1}{2}\left(f^{(\alpha)}_{\k}+f^{(-\alpha)}_{\k}\right)\;.
\end{equation}
Notice that for real values of $\alpha$, \eqref{eq: definition delta hat} reduces to the definition of the usual symmetrised delta function. For convenience, we introduce the following notation for the KLF generalised delta function:
\begin{equation}
    \delta\left(^{\mu_1\mu_2}_{\k_1\k_2}\right)\equiv \frac{(2\pi)^d}{\N_{\mu_1}}\hat{\delta}_{\mu_2}(\mu_1)\delta^{(d)}(\k_1-\k_2)\;.
\label{KLF-delta-function}    
\end{equation}
\vskip 4pt
\paragraph{Generic case.}
Let us now explain how non-principal contributions are naturally taken into account in our formalism. A generic smooth function $f(z,\x)$ defined on EAdS can fail to be square-integrable either at $z\to 0$ or $z\to \infty$. Since the large $z$ behavior of the harmonic function is independent of the frequency, it cannot impact the structure of the KLF mode and does not lead to the appearance of new UIRs in the spectrum. We thus consider functions that are UV square-integrable in what follows. However, if the function fails to be square-integrable in the IR, this necessarily yields contributions from the complementary series. To see this, let us consider the following series expansion:\footnote{IR logarithmic behaviours are also taken into account as they can be obtained by differentiation: $z^\Delta\log^n(z) = \frac{\d^n}{\d\Delta^n}z^\Delta$. If the KLF modes of the monomials are meromorphic, the differentiation cannot induce a branch cut.}
\begin{equation}\label{eq: generic function f}
    f(z,\x)=\sum_{n=0}^\infty z^{\Delta+n}f_n(\x)\;,
\end{equation}
where $\Delta$ is the leading power near $z=0$ and we assume the spatial coefficients $f_n(\x)$ to be square-integrable.
The function $f(z,\x)$ is square-integrable only
if $\text{Re}(\Delta)>\frac{d}{2}$. In this case, the KLF transform \eqref{eq: def KLF transform} is well defined and can be evaluated exactly:\footnote{The evaluation of the time integral can be performed by using eq. $6.561$, $16$ of \cite{gradshteyn2007}.} 
\begin{equation}\label{eq: general KLF modes Principal series}
    f^{(\mu),\P}_{\k}=\frac{H^{-\frac{d+1}{2}}}{4\sqrt{\pi}}\sum_{n=0}^{\infty}\hat{f}_n(\k)\left(\frac{2}{k}\right)^{\Delta-\frac{d}{2}+n}\Gamma\left(\frac{\Delta-\frac{d}{2}+n\pm i\mu}{2}\right)\;,
\end{equation}
where $\hat{f}(\k)$ is the usual Fourier transform and we add the subscript ${\cal P}$ to recall that $\mu$ is real. Now, in the $\mu$ complex plane, this is a meromorphic function with two series of poles in the lower and upper half-planes:
\begin{equation}
    \mu_\pm^{\ell}=\pm i\left(\ell+\Delta-\frac{d}{2}\right)\;,\quad \ell\in \mathbb{N}\;,
\end{equation}
where the residues are expressed as the following sums of $\lfloor\frac{\ell}{2}+1\rfloor$ terms: 
\begin{equation}
\begin{aligned}
    &\textrm{Res}\left(f^{(\mu),\P}_{\k},\mu^\ell_{\pm}\right) = \mp \frac{iH^{-\frac{d+1}{2}}}{2\sqrt{\pi}}\S^\ell_\Delta\;,\\
   &\S^\ell_\Delta =\sum_{j\textrm{ s.t. }\ell-j\in2\mathbb{N}}\frac{(-1)^{\frac{\ell-j}{2}}\hat{f}_j(\k)}{\left(\frac{\ell-j}{2}\right)!}&\left(\frac{2}{k}\right)^{\Delta-\frac{d}{2}+j}\Gamma\left(\Delta-\frac{d}{2}+\frac{\ell-j}{2}\right)\;.
\end{aligned}
\end{equation}
The direct consequence of the integrability condition is that the KLF modes are analytic in the strip $| \Im(\mu)|\leq \text{Re}(\Delta)-\frac{d}{2}$. Since the function \eqref{eq: generic function f} is quite general, we expect this statement to hold for a very large class of functions. In particular, we do not know any example of a function whose KLF mode features a cut in the $\mu$ plane. The corresponding analytic structure is pictured in the left panel of figure \ref{fig: KLF modes analytical structure}. In the regime of square-integrability, $f(z,\x)$ can be written as:
\begin{equation}\label{eq: general f principal integrql expansionKLF}
\begin{aligned}
    f(z,\x)=i\frac{H^{\frac{d+1}{2}}z^\frac{d}{2}}{\sqrt{\pi}}\int\frac{\d^d\k}{(2\pi)^d}e^{-i\k\cdot\x}\int\displaylimits^\infty_{-\infty}\d\mu\;\mu f^{(\mu),\P}_{\k} I_{i\mu}(k z)\;,
\end{aligned}
\end{equation}
where we split $\N_\mu K_{i \mu}(z)=\frac{i}{2}\mu(I_{i \mu}(z)-I_{-i \mu}(z))$ to use the decay of $I_{i \mu}$ in the lower $\mu$ half-plane, the two contributions being equal due to the shadow-symmetry of $f^{(\mu),\P}_{\k}$.
The spectral integral can then be performed by collecting the residues $\mu_-^{\ell}$:
\begin{equation}\label{eq: general f series expansionKLF}
\begin{aligned}
    f(z,\x) & = 2\sqrt{\pi}H^{\frac{d+1}{2}} z^{\frac{d}{2}} \sum_{\ell=0}^{\infty}\int\frac{\d^d\k}{(2\pi)^d}e^{-i\k\cdot\x}\mu^{\ell}_-\text{Res}\left( f^{(\mu),\P}_{\k},\mu_-^{\ell}\right)I_{i\mu_-^{\ell}}(k z)\;.
\end{aligned}
\end{equation}
It is not obvious that this complicated sum reproduces a simple power-law dependence, but this can be checked using the following Neumann-type expansion (see $10.44.4$ of~\cite{NIST:DLMF}):\footnote{This kind of resummation formulas can be systematically obtained as the residues expansion of the inverse KL transform of known functions.} 
\begin{equation}
    \left(\frac{z}{2}\right)^\alpha=\sum_{\ell=0}^\infty\frac{(-1)^\ell}{\ell!}(\alpha+2\ell)\Gamma(\alpha+\ell)I_{\alpha+2\ell}(z)\;,\quad\alpha\in \mathbb{C}\setminus\{0,-1,-2,\ldots\}\;.
\end{equation}
\vskip 4pt
When $\Delta-\frac{d}{2}\notin -\mathbb{N}$, it is clear that the expansion \eqref{eq: general f series expansionKLF} remains valid in the regime $\textrm{Re}(\Delta)<\frac{d}{2}$, where the function $f(z,\x)$ fails to be square-integrable. In that case, the integral over the principal series \eqref{eq: general f principal integrql expansionKLF} does not coincide with the residues expansion \eqref{eq: general f series expansionKLF} and extra terms are needed in the spectral decomposition. To see that, one can look at the analytical structure of the KLF modes \eqref{eq: general KLF modes Principal series} in this regime, which is pictured in the right panel of figure \ref{fig: KLF modes analytical structure}. We can see that the two series of poles cross the real axis, and the previous region of analyticity $| \Im(\mu)|\leq | \text{Re}(\Delta)-\frac{d}{2}|$ is now populated by poles coming from both Gamma factors. Consequently, the integral \eqref{eq: general f principal integrql expansionKLF} misses some $\mu_-^{\ell}$ residues and gets extra contributions from some $\mu_+^{\ell}$ residues. If $\frac{d}{2}-\textrm{Re}(\Delta)\notin \mathbb{N}$, we have:
\begin{equation}
\begin{aligned}
    \int\displaylimits^\infty_{-\infty}\d\mu\;\mu f^{(\mu),\P}_{\k} I_{i\mu}(k z) =& -2\pi i\sum_{\ell=\lceil\frac{d}{2}-\textrm{Re}(\Delta)\rceil}^\infty\mu^{0,\ell}_-\text{Res}\left( f^{(\mu),\P}_{\k},\mu_-^{\ell}\right)I_{i\mu_-^{\ell}}(k z)\\
    &- 2\pi i\sum_{\ell=0}^{\lfloor\frac{d}{2}-\textrm{Re}(\Delta)\rfloor}\mu^{\ell}_+\text{Res}\left( f^{(\mu),\P}_{\k},\mu_+^{\ell}\right)I_{i\mu_+^{\ell}}(k z)\;.
\end{aligned}
\end{equation}
In the integer case, some $\mu_\pm^{\ell}$ poles collide, resulting in a second-order singularity. Moreover, the pole for $\ell=\frac{d}{2}-\Delta$, being located on the real axis, requires the integral to be regularised by taking its principal value. In order to avoid dealing with these complications, we consider the non-integer case in what follows.

\begin{figure}[h!]
\centering
\begin{subfigure}[h!]{0.4\textwidth}
    \hspace{-0.25cm}
    	\begin{tikzpicture}[scale = 2]
        \draw[pyblue,thick] (-1.6,0.4) -- (1.6,0.4);
        \draw[pyred,thick] (-1.6,-0.4) -- (1.6,-0.4);
        \fill[lightpygreen] (-1.6,0.4) -- (1.6,0.4) -- (1.6,-0.4) -- (-1.6,-0.4) -- cycle;
        \draw[pyorange,<->] (-0.4,0.4) -- (-0.4,0);
        \node at (-1,0.2) {\textcolor{pyorange}{$\Re(\Delta)-\frac{d}{2}$}};
        \draw[pyorange,<->] (-0.4,-0.4) -- (-0.4,0);
        \node at (-1,-0.2) {\textcolor{pyorange}{$\Re(\Delta)-\frac{d}{2}$}};

        \draw[black, ->] (-1.6,0) -- (1.6,0) coordinate (xaxis);
		\draw[black, ->] (0,-1.6) -- (0,1.6) coordinate (yaxis);
		\node at (1.9, 0) {$\text{Re}(\mu)$};
		\node at (0, 1.75) {$\text{Im}(\mu)$};
        \node at (-0.9,1.2) {$\boxed{\Re(\Delta)>\frac{d}{2}}$};
        
        \draw[pyblue, fill = pyblue] (0.2, 0.4) circle (.03cm);
		\draw[pyblue, fill = pyblue] (0.2, 0.7) circle (.03cm);
		\draw[pyblue, fill = pyblue] (0.2, 1.) circle (.03cm);
		\draw[pyblue, fill = pyblue] (0.2, 1.3) circle (.03cm);
        \node at (0.5, 1.) {\textcolor{pyblue}{$\mu^{\ell}_+$}};

        \draw[pyred, fill = pyred] (-0.2, -0.4) circle (.03cm);
		\draw[pyred, fill = pyred] (-0.2, -0.7) circle (.03cm);
		\draw[pyred, fill = pyred] (-0.2, -1.0) circle (.03cm);
		\draw[pyred, fill = pyred] (-0.2, -1.3) circle (.03cm);
        \node at (-0.5, -1.) {\textcolor{pyred}{$\mu^{\ell}_-$}};

        \draw[xshift=0,pyblue!80!black,decoration={markings,mark=between positions 0.1 and 1 step 0.2 with \arrow{>}},postaction={decorate}] (-1.6,0) -- (1.6,0) arc (0:-180:1.6);
    \end{tikzpicture}
\end{subfigure}
\hfill
\begin{subfigure}[h!]{0.4\textwidth}
        \hspace{-1.5cm}
    	\begin{tikzpicture}[scale = 2]
        \draw[pyred,thick] (-1.6,0.4) -- (1.6,0.4);
        \draw[pyblue,thick] (-1.6,-0.4) -- (1.6,-0.4);
        \fill[lightpygreen] (-1.6,0.4) -- (1.6,0.4) -- (1.6,-0.4) -- (-1.6,-0.4) -- cycle;
        \draw[pyorange,<->] (-0.4,0.4) -- (-0.4,0);
        \node at (-1,0.2) {\textcolor{pyorange}{$\frac{d}{2}-\Re(\Delta)$}};
        \draw[pyorange,<->] (-0.4,-0.4) -- (-0.4,0);
        \node at (-1,-0.2) {\textcolor{pyorange}{$\frac{d}{2}-\Re(\Delta)$}};

        \draw[black, ->] (-1.6,0) -- (1.6,0) coordinate (xaxis);
		\draw[black, ->] (0,-1.6) -- (0,1.6) coordinate (yaxis);
		\node at (1.9, 0) {$\text{Re}(\mu)$};
		\node at (0, 1.75) {$\text{Im}(\mu)$};
        \node at (-0.9,1.2) {$\boxed{\Re(\Delta)<\frac{d}{2}}$};
        
        \draw[pyblue, fill = pyblue] (0.2, -0.4) circle (.03cm);
		\draw[pyblue, fill = pyblue] (0.2, -0.1) circle (.03cm);
		\draw[pyblue, fill = pyblue] (0.2, 0.2) circle (.03cm);
		\draw[pyblue, fill = pyblue] (0.2, 0.5) circle (.03cm);
        \draw[pyblue, fill = pyblue] (0.2, 0.8) circle (.03cm);
        \draw[pyblue, fill = pyblue] (0.2, 1.1) circle (.03cm);
        \draw[pyblue, fill = pyblue] (0.2, 1.4) circle (.03cm);
        \node at (0.5, 1.) {\textcolor{pyblue}{$\mu^{\ell}_+$}};

        \draw[pyred, fill = pyred] (-0.2, 0.4) circle (.03cm);
		\draw[pyred, fill = pyred] (-0.2, 0.1) circle (.03cm);
		\draw[pyred, fill = pyred] (-0.2, -0.2) circle (.03cm);
		\draw[pyred, fill = pyred] (-0.2, -0.5) circle (.03cm);
        \draw[pyred, fill = pyred] (-0.2, -0.8) circle (.03cm);
        \draw[pyred, fill = pyred] (-0.2, -1.1) circle (.03cm);
        \draw[pyred, fill = pyred] (-0.2, -1.4) circle (.03cm);
        \node at (-0.5, -1.) {\textcolor{pyred}{$\mu^{\ell}_-$}};

        \draw[xshift=0,pyblue!80!black,decoration={markings,mark=between positions 0.1 and 1 step 0.2 with \arrow{>}},postaction={decorate}] (-1.6,0) -- (1.6,0) arc (0:-180:1.6);
	\end{tikzpicture}
\end{subfigure}
\caption{Analytic structure in the $\mu$ complex plane of \eqref{eq: general KLF modes Principal series}, in the generic situation where $\Im(\Delta) \neq 0$. \textbf{Left:} case where the function \eqref{eq: generic function f} is square-integrable as its leading fall-off is greater than $\frac{d}{2}$. The function has two distinct series of poles separated by a region of analyticity. \textbf{Right:} case where the function \eqref{eq: generic function f} is not square-integrable as its leading fall-off is smaller than $\frac{d}{2}$. The two series of poles are now intertwined in the former analyticity region.}
\label{fig: KLF modes analytical structure}
\end{figure}
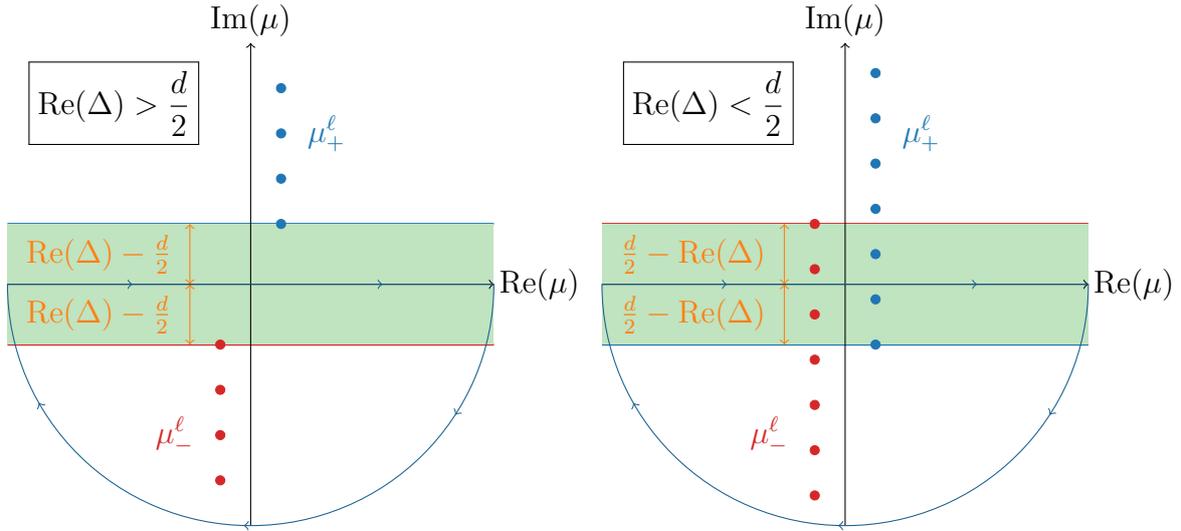 

In order to recover the right expansion, one should add the missing contributions from the $\mu_-^{\ell}$ poles and subtract those coming from the $\mu_+^{\ell}$ ones, yielding the following non-principal contribution to the spectral decomposition:
\begin{equation}\label{eq: non-principal contributions}
\begin{aligned}
    &2\pi i\int\frac{\d^d\k}{(2\pi)^d}\frac{H^{\frac{d+1}{2}}}{\sqrt{\pi}}z^{\frac{d}{2}}e^{-i\k\cdot\x}\sum_{a=\pm}\sum_{\ell=0}^{\lfloor\frac{d}{2}-\Re(\Delta)\rfloor} a\, i\mu_a^{\ell}\text{Res}\left(f^{(\mu),\P}_{\k},\mu_a^{\ell}\right)I_{i\mu_a^{\ell}}(k z)\\
    &= 4\pi i\int\frac{\d^d\k}{(2\pi)^d}\sum_{\ell=0}^{\lfloor\frac{d}{2}-\Re(\Delta)\rfloor}\N_{\mu_+^{\ell}}\text{Res}\left(f^{(\mu),\P}_{\k},\mu_+^{\ell}\right)\Phi^{\mu_+^{\ell}}_{\k}(z,\x)\\
    &= 4\pi i \int_{\KLF}\sum_{\ell=0}^{\lfloor\frac{d}{2}-\Re(\Delta)\rfloor}\hat{\delta}_{\mu_+^{\ell}}(\mu)\text{Res}\left(f^{(\mu),\P}_{\k},\mu_+^{\ell}\right)\Phi^{\mu}_{\k}(z,\x)\;,
    \end{aligned}
\end{equation}
where we made use of the shadow anti-symmetry of the residue and again of $\N_\mu K_{i \mu}(z)=\frac{i}{2}\mu(I_{i \mu}(z)-I_{-i \mu}(z))$. 

\vskip 4pt
A few comments are in order. In physical applications coming from a unitary theory, one expects these contributions to fall in $\SO(1,d+1)$ UIRs. This corresponds to the following behaviours for the possible fall-offs: 
\begin{equation}
\begin{aligned} 
& \textrm{Unitarity:}\\
       & \bullet \Re(\Delta)> \frac{d}{2},\,\, \,\,\,\,\,\, \,\,\,\, \,\Im(\Delta)\, \textrm{arbitrary}, & \textrm{principal}\\
       & \bullet 0 \leq \Re(\Delta)< \frac{d}{2}, \,\,\Im(\Delta)=0, & \textrm{principal} +\textrm{complementary} \\
       & \bullet \Re(\Delta) \in \mathbb{N}^{-},\,\, \,\,\,\,\,\, \,\,\Im(\Delta)=0,\quad &\textrm{principal} +\textrm{(complementary)} +\textrm{exceptional}
       \nonumber
         \end{aligned} 
\end{equation}
where we listed the ever-present principal series contribution, and the possible non-principal ones according to each case (in the last situation, complementary series contributions arise or not depending on the values of $d$). As advertised before, the non-principal contributions are fully determined by the non analyticities of the principal series contribution. Since for our class of functions, the latter is meromorphic in the $\mu$ plane, it can only constitute a discrete set. 
\vskip 4pt
To summarise, any generic function of the form \eqref{eq: generic function f} can be expanded into KLF modes, simply by promoting them to the sum of a meromorphic function and a set of distributional terms encoding the contributions outside the principal series:
\begin{FramedBox}
\begin{equation}
    f^{\mu}_{\k,0} = f^{(\mu),\P}_{\k,0} + 4\pi i\Theta\left(\frac{d}{2}-\Re(\Delta)\right)\sum_{\ell=0}^{\lfloor\frac{d}{2}-\Re(\Delta)\rfloor}\hat{\delta}_{\mu_+^{\ell}}(\mu)\text{Res}\left(f^{(\mu),\P}_{\k},\mu_+^{\ell}\right)\;.
\end{equation}
\end{FramedBox}

\vskip 4pt
\paragraph{An example: scalar conformal two-point function.}
For concreteness, let us look at the example of a conformal two point function~\cite{Hogervorst:2021uvp,Loparco:2023rug}. In a dS CFT, the two-point function of a local scalar operator $\O$ with scaling dimension $\Delta\geq \frac{d-1}{2}$ is completely fixed by conformal invariance to be: 
\begin{equation}
    \braket{\O(X_1)\O(X_2)} \equiv\left(\frac{R_{\dS}^2}{d(X_1,X_2)}\right)^\Delta= \frac{1}{2^\Delta(1-\sigma^{\dS})^\Delta}\;,
    \label{dS-CFT}
\end{equation}
where $X_{1,2} \equiv (\tau_{1,2},\x_{1,2})$ denotes two points in $\dS$, and the chordal distance $d(X_1,X_2)$ and two-point invariant $\sigma^\dS$ are defined in Eqs.~\eqref{eq: def two-point invariant dS}-\eqref{def-chordal-distance-dS}. As explained in appendix~\ref{sec: propagators in real space}, one should regulate the branch cut at time-like separations by means of the appropriate $i\epsilon$ prescription. Our interest here lies in the Wightman two-point function, whose Euclidean counterpart simply follows from \eqref{dS-CFT} and the Wick rotation \eqref{eq: Lorentzian two-point invariant off-diag} as\footnote{There is no need to keep track of the $i \epsilon$ here as \eqref{EAdS-CFT} $1-\sigma^E$ is always positive.} 
\begin{equation}
    \braket{\O(X_1^E)\O(X_2^E)} = \frac{1}{2^\Delta(1-\sigma^E)^\Delta}\;.
    \label{EAdS-CFT}
\end{equation}
\vskip 4pt
In the regime where this is a square-integrable function, its KLF transform with respect to one of the two variables converges. Since the leading fall-off of this function is given by the scaling dimension of the operator $\Delta$, we know from the above analysis that this is the case when $\Delta>\frac{d}{2}$. As explained in appendix \ref{sec: CFT details}, the KLF transform in $X_1^E$ can be evaluated by direct computation and it takes the following form: 
\begin{equation}\label{eq: KLF tranform CFT 2points}
\begin{aligned}
    \int_{\EAdS}\frac{\d z_1\d^d\x_1}{(H z)^{d+1}}\left[\Phi^{\mu}_{\k}(z_1,\x_1)\right]^*\braket{\O(X_1^E)\O(X_2^E)} = \frac{\rho^{\P}_\O(\mu)}{H^2 \N_{\mu}}\left[\Phi^{\mu}_{\k}(z_2,\x_2)\right]^*\;,
\end{aligned}
\end{equation}
where the spectral density $\rho^{\P}_\O(\mu)$ is given by: 
\begin{equation}
    \rho^{\P}_\O(\mu) = \frac{\N_\mu}{H^{d-1}}\underbrace{\frac{2^{-2\Delta+d+1}\pi^{\frac{d+1}{2}}}{\Gamma(\Delta)\Gamma\left(\Delta-\frac{d}{2}+\frac{1}{2}\right)}}_{c_\Delta}\Gamma\left(\Delta-\frac{d}{2}\pm i\mu\right)\;.
\end{equation}
As above, this has two distinct series of poles separated by an analyticity region (see figure \ref{fig: KLF modes analytical structure}). Because of the simple form of the function we do not need the late-time expansion and the poles are labeled by a single integer $n$:
\begin{equation}
    \mu^n_\pm=\pm i\left(\Delta-\frac{d}{2}+n\right)\;.
\end{equation}
Now, let us insert this in the KLF decomposition~\eqref{eq: KLF decomposition}: 
\begin{equation}\label{eq: KL CFT example integral}
    \braket{\O(X_1^E)\O(X_2^E)}_{\Delta>\frac{d}{2}}= \int\displaylimits^{\infty}_{-\infty}\d\mu\rho^\P_\O(\mu)W_\mu(\sigma^E)\;,
\end{equation}
where the function $W_\mu(\sigma^E)$ coming from the evaluation of the momentum integral is defined in \eqref{eq:definition of W}. For the function \eqref{eq: generic function f}, it was convenient, in order to evaluate the integral, to split $K_{i \mu}$ into the two modified Bessel functions $I_{\pm i\mu}$. Here, since we already performed the spatial momentum integration, the same procedure amounts to split the function $W_\mu(\sigma^E)$ in two EAdS bulk-to-bulk propagators $\Pi^{\EAdS}_{\mu}(\sigma^E)$ according to the connection formula \eqref{eq: splittin W in AdS}. This function, defined in \eqref{PiEAdsS}, both decays and is analytic in the $\mu$ lower half-plane. One can thus evaluate the integral \eqref{eq: KL CFT example integral} by closing the contour and collecting the residues at the poles $\mu_n^-$:
\begin{equation}
\begin{aligned}
   \braket{\O(X_1^E)\O(X_2^E)}=-2\pi\sum_{n=0}^\infty\mu_-^n\textrm{Res}\left(\frac{\rho^\P_\O(\mu)}{\N_\mu},\mu^n_-\right)\Pi^{\EAdS}_{\mu_-^n}(\sigma^E)\;.
\label{CFT-expansion}    
\end{aligned}
\end{equation}
An explicit expression is provided in equation \eqref{eq: G-+ CFT Series App}. 
\vskip 4pt
This is the analogue for the CFT of Eq.~\eqref{eq: general f series expansionKLF} for the test function \eqref{eq: generic function f}.
Like for this case, Eq.~\eqref{CFT-expansion} remains valid when $\Delta<\frac{d}{2}$ for $\frac{d}{2}-\Delta\notin \mathbb{N}$, and the principal series integral should be corrected by adding and subtracting the appropriate pole contributions. The integer case has the same pathology as above. This is the dS counterpart of the singularities one encounters by taking the Fourier transform of a conformal two-point function in flat spacetime, resulting in the appearance of a scale anomaly~\cite{Bzowski:2013sza}. 
\vskip 4pt
Finally, the extension of the KLF expansion takes the same form as in the generic case:
\begin{equation}\label{eq: full spectral decomposition}
\begin{aligned}
   \braket{\O(X_1^E)\O(X_2^E)}_{\Delta<\frac{d}{2}}&= \int\displaylimits^{\infty}_{-\infty}\d\mu\rho^\P_\O(\mu)W_\mu(\sigma^E)\\
    &+4 i\pi  \sum_{n=0}^{\lfloor\frac{d}{2}-\Delta\rfloor}\text{Res}\left(\rho^P_\O(\mu)W_\mu(\sigma^E),\mu=\mu^n_+\right)\;.
\end{aligned}
\end{equation}
An explicit expression can be found in equation~\eqref{eq: explicit expression full spectral decomp CFT}.
\vskip 4pt
In the case of a unitary CFT, i.e. $\Delta>\frac{d-1}{2}$, there is only a single contribution from the complementary series. Using our generalised delta function \eqref{eq: definition delta hat}, this can be rewritten as a KLF decomposition: 
\begin{equation}
    \braket{\O(X_1^E)\O(X_2^E)}=\int_{\KLF}\frac{\rho_\O(\mu)}{H^2\N_{\mu}}\Phi^{\mu}_{\k}(X_1^E)\left[\Phi^{\mu}_{\k}(X_2^E)\right]^*\;,
\end{equation}
where the KLF spectral density is given by the sum of a smooth function and a distributional part
\begin{equation}\label{eq: CFT spectral density}
    \rho_\O(\mu) = \rho^\P_\O(\mu) + 4 i\pi\, \Theta\left(\frac{d}{2}-\Delta\right)\hat{\delta}_{\mu^0_+}(\mu)\text{Res}\left(\rho^\P_\O(\mu),\mu = \mu^0_+\right)\;.
    \end{equation}
This reproduces the result of \cite{Loparco:2023rug} for the K\"all\'en-Lehmann decomposition of the CFT Wightman function.

\section{Cosmological Correlators in Kontorovich-Lebedev-Fourier Space}
\label{sec:correlators}

As it is now clear that KLF space is the dS counterpart of the $(d+1)$-dimensional energy-momentum space in Minkowski, let us follow this example and turn the kinematical statements we obtained into handy prescriptions to formulate perturbation theory in any QFT. To take a maximal advantage of our functional approach, we make use of the path-integral formalism. 
\subsection{KLF Space Correlators}
\begin{figure}[h!]
\centering
\begin{subfigure}[h!]{0.4\textwidth}
    \hspace{-0.25cm}
    	\begin{tikzpicture}[scale = 2]
        \draw[black, ->] (-1.6,0) -- (1.6,0) coordinate (xaxis);
		\draw[black, ->] (1.,-1.6) -- (1.,1.6) coordinate (yaxis);
		\node at (1.9, 0) {$\text{Re}(\tau)$};
		\node at (1., 1.75) {$\text{Im}(\tau)$};

        \draw[pyorange,<->] (-0.9,0.3) -- (-0.9,0) node [above right] {$+i\epsilon$};
        \draw[pyorange,<->] (-0.9,-0.3) -- (-0.9,0) node [below left] {$-i\epsilon$};

        \draw[pyblue!60,thick,decoration={markings,mark=between positions 0.1 and 1 step 0.28 with \arrow{>}},postaction={decorate}] (-1.6,0.3) -- (0.7,0.3) arc (90:-90:0.3) (0.7,-0.3) -- (-1.6,-0.3);

        \draw[pyred, fill = pyred] (-1.3, 0.3) circle (.03cm) node [above] {$\O_1^+$};
		\draw[pyred, fill = pyred] (-0.7, -0.3) circle (.03cm) node [below] {$\O_2^-$};
		\draw[pyred, fill = pyred] (-0.1, 0.3) circle (.03cm) node [above] {$\O_3^+$};
		\draw[pyred, fill = pyred] (0.5, -0.3) circle (.03cm) node [below] {$\O_4^-$};
    \end{tikzpicture}
\end{subfigure}
\hfill
\begin{subfigure}[h!]{0.4\textwidth}
    \hspace{-1.5cm}
    	\begin{tikzpicture}[scale = 2]
        \draw[black, ->] (-1.6,0) -- (1.6,0) coordinate (xaxis);
		\draw[black, ->] (1.,-1.6) -- (1.,1.6) coordinate (yaxis);
		\node at (1.9, 0) {$\text{Re}(\tau)$};
		\node at (1., 1.75) {$\text{Im}(\tau)$};

        \draw[gray,<->] (-0.9,0.3) -- (-0.9,0) node [above right] {$+i\epsilon$};
        \draw[gray,<->] (-0.9,-0.3) -- (-0.9,0) node [below left] {$-i\epsilon$};

        \draw[gray!60,thick,decoration={markings,mark=between positions 0.1 and 1 step 0.28 with \arrow{>}},postaction={decorate}] (-1.6,0.3) -- (0.7,0.3) arc (90:-90:0.3) (0.7,-0.3) -- (-1.6,-0.3);

        \draw[pyblue!60,thick,decoration={markings,mark=between positions 0.2 and 1 step 0.29 with \arrow{>}},postaction={decorate}] (1.,1.6) -- (1.,-1.6);

        \draw[gray, fill = gray] (-1.3, 0.3) circle (.03cm) node [above] {$\O_1^+$};
		\draw[gray, fill = gray] (-0.7, -0.3) circle (.03cm) node [below] {$\O_2^-$};
		\draw[gray, fill = gray] (-0.1, 0.3) circle (.03cm) node [above] {$\O_3^+$};
		\draw[gray, fill = gray] (0.5, -0.3) circle (.03cm) node [below] {$\O_4^-$};

        \draw[pyred, fill = pyred] (1., 1.3) circle (.03cm) node [right] {$\O_1^+$};
		\draw[pyred, fill = pyred] (1., -1.) circle (.03cm) node [right] {$\O_2^-$};
		\draw[pyred, fill = pyred] (1., 0.5) circle (.03cm) node [right] {$\O_3^+$};
		\draw[pyred, fill = pyred] (1., -0.3) circle (.03cm) node [right] {$\O_4^-$};
	\end{tikzpicture}
\end{subfigure}
\caption{Integration contour for the Schwinger-Keldysh path integral in Lorentzian (\textbf{left} panel) and Euclidean (\textbf{right} panel) regimes.}
\label{fig: SK Contour and Wick Rotation}
\end{figure}
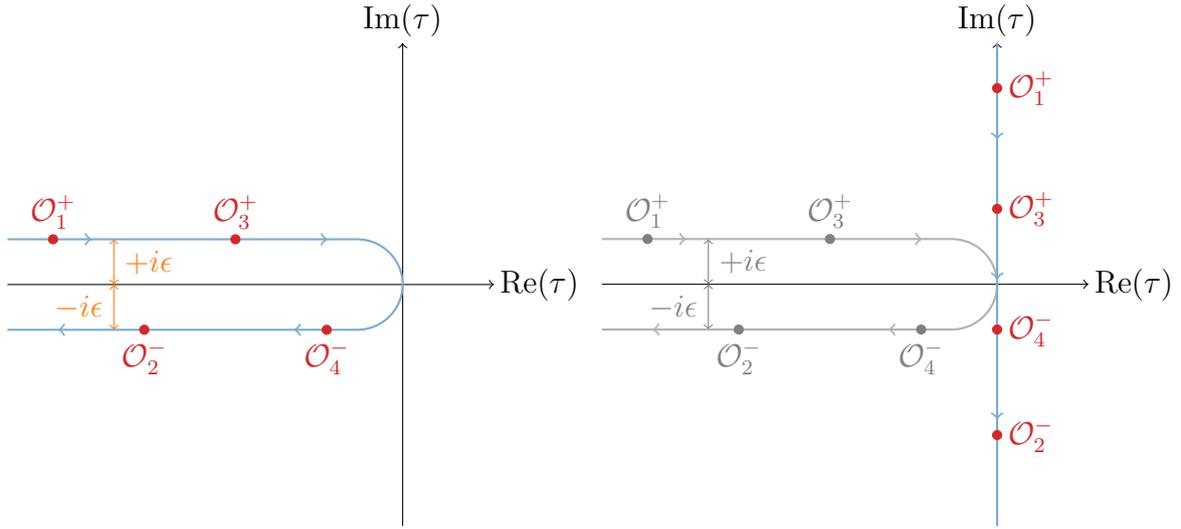

In inflationary quantum field theory, the observable quantities are given by equal-time correlation functions in the Bunch-Davies vacuum state $\ket{\Omega}$, evaluated on the conformal future boundary $\tau\to 0$:
\begin{equation}\label{eq: boundary correlators}
    \C_{\cal{B}}(\x_1,\ldots,\x_n)= \bra{\Omega}\hat{\mathcal{O}_1}(\x_1)\ldots\hat{\O}_n(\x_n)\ket{\Omega}\;,
\end{equation}
where the operators $\hat{\O}_i(x_i)$ are defined by the late-time leading behaviour of bulk local operators $\hat{\O}_i(\tau_i,\x_i)$.\footnote{See e.g.~\cite{SalehiVaziri:2024joi} for a detailed discussion of the subtleties raised by this definition.} These correlation functions are naturally expressed via the so-called Schwinger-Keldysh (SK) formalism~\cite{Maldacena:2002vr,Weinberg:2005vy}. In the latter, we define the bulk correlation function as an inner product between two states resulting from two path integrals with different local operator insertions: 
\begin{equation}\label{eq: bulk correlator}
    \C_{\a_1\ldots\a_n}(\tau_1,\x_1;\ldots;\tau_n,\x_n) = \int_{\textrm{B.D.}}^{\varphi_+=\varphi_-}\mathcal{D}\varphi_+\mathcal{D}\varphi_- \mathcal{O}_1^{\a_1}\ldots\mathcal{O}_n^{\a_n}e^{i S_+[\varphi_+]-iS_-[\varphi_-]}\;,
\end{equation}
where $\O^{\a_j}_j \equiv \O^{\a_j}_j(\tau_j,\x_j)$ are functions of the physical field degrees of freedom $\varphi_{\a_j}(\tau_j,\x_j)$ standing for the insertion of a local operator $\hat{\O}^{\a_j}_j(\tau_j,\x_j)$ on the branch $\a_j=\pm$ of the path integral.\footnote{We consider that the Bunch-Davies vacuum state is normalised, hence there is no contribution from disconnected bubble diagrams.} In a unitary theory, these two copies of the evolution are perfectly independent and we only impose the two fields to coincide at final time in order to define the inner product.\footnote{Strictly speaking, one can impose the future boundary condition at any time $\tau_f$ in the future of the latest time involved in the correlation function \cite{Weinberg:2005vy}.} Therefore, the ordering subtleties only matter for operators inserted on the same branch. As a result, the operators living on the $(-)+$ branches are, respectively, (anti-)time ordered. For instance, as shown in the left panel of Fig.~\ref{fig: SK Contour and Wick Rotation}, in the case of a mixed four-point function, one has: 
\begin{equation}
    \C_{+-+-}(\tau_1,\x_1;\ldots;\tau_4,\x_4)= \bra{\Omega}\bar{\T}\left[\hat{\O}_{2}^-\hat{\O}^-_{4}\right]\T\left[\hat{\O}_{1}^+\hat{\O}^+_{3}\right]\ket{\Omega}\;.
\end{equation}
Since the two branches are sewed at the final stage of the evolution, one can define the boundary correlators \eqref{eq: boundary correlators} as the late time limit of a bulk correlator \eqref{eq: bulk correlator} with arbitrary values of the indices $\a_{j}$:
\begin{equation}\label{eq: push to boundary}
    \C_{\mathcal{B}}(\x_1,\ldots,\x_n)=\lim_{\tau\to 0}\C_{\a_1\ldots\a_n}(\tau,\x_1;\ldots;\tau,\x_n)\;.
\end{equation}
In practice, we are interested in the momentum space version of the boundary correlators defined as:
\begin{equation}
    \C_\mathcal{B}(\k_1,\ldots,\k_n) =\int\prod^n_{j=1}\d^d\x_j e^{i\sum_{j=1}^n\x_j\cdot\k_j}\C_\mathcal{B}(\x_1,\ldots,\x_n)\;.
\end{equation}
Most of the correlators we will consider actually vanish like power laws in this limit. Therefore, we will introduce a late-time regulator $-1\ll \tau_0<0$ to factorise this leading behaviour and the physical result of interest. 
\vskip 4pt
Both path integrals encode the time evolution from the Bunch-Davies vacuum state via the action $S[\varphi]$ of the theory. In order to enforce this initial condition, the time integration contour should be slightly tilted in the $\tau$ complex plane. For the two branches of \eqref{eq: bulk correlator}, this is achieved with the standard $i\epsilon$ prescription:
\begin{equation}
    S_\pm[\varphi_\pm] = \int^0_{-\infty_\pm}\frac{\d\tau\d^d\x}{(-H\tau)^{d+1}}\mathcal{L}(\varphi_\pm,\varphi'_\pm, \partial_j \varphi_\pm;\tau)\;.
\end{equation}
Performing the Wick rotation to EAdS \eqref{eq: Wick rotation to EAdS}, one obtains:
\begin{equation}
    \pm iS_\pm[\varphi_\pm]= e^{\pm\frac{i\pi(d+1)}{2}}S_E[\varphi^E_\pm]\;,
    \label{Spm-Se}
\end{equation}
where $S_E[\varphi^E_\pm]$ is the Euclidean action:
\begin{equation}
    S_E[\varphi^E_\pm]=\int^\infty_0\frac{\d z\d^d\x}{(H z)^{d+1}}\L(\varphi_\pm^E,\partial_z\varphi_\pm^E, i \partial_j\varphi^E_\pm;z)\;,
\end{equation}
and the $\pm$ versions of the fields now live, respectively, on the positive and negative real axis in the complex $\tau$-plane. One can always perform these Wick rotations if the Lagrangian is an analytic function of conformal time. If not, the presence of poles would induce additional terms coming from the residues of the Lagrangian. However, such peculiar features are usually motivated by dS breaking setups, such that emergent non-locality (see, e.g. \cite{Jazayeri:2023kji}), which are beyond the scope of the present work.

The resulting Euclidean path integral allows us to compute Euclidean correlation functions $\C^E_{\a_1\ldots\a_n}\left(^{z_1\ldots z_n}_{\x_1\ldots\x _n}\right)$ related to their Lorentzian counterparts by making use of the Wick rotation~\eqref{eq: z Lorentzian}:
\begin{equation}\label{eq: Wick rotated n point function}
    \C_{\a_1\ldots\a_n}\left(^{\tau_1\ldots \tau_n}_{\x_1\ldots\x _n}\right) = \C_{\a_1\ldots\a_n}^E\left(-\tau_1e^{i\a_1\left(\frac{\pi}{2}-\epsilon\right)},\x_1;\ldots;-\tau_n e^{i\a_n\left(\frac{\pi}{2}-\epsilon\right)},\x_n\right)\;.
\end{equation}
Notice that it does not make sense to add correlators with different $\a_j$ indices, since they do not live on the same copies of EAdS.\footnote{For instance, in the right panel of Fig.~\ref{fig: SK Contour and Wick Rotation}, $\O_1,\O_3$ and $\O_2,\O_4$ are inserted on two distinct copies of EAdS.} However, it makes sense at the final Lorentzian time slice due to the boundary condition $\varphi_+(\tau_0)=\varphi_-(\tau_0)$. 

\vskip 4pt
The path integral, being unaffected by the functional basis we use in order to express the fields, can be expressed in KLF space. This motivates the definition of the Euclidean correlators directly in terms of frequency and momentum variables:
\begin{equation}\label{eq: def KLF correlators}
    \G_{\a_1\ldots\a_n}\left(^{\mu_1\ldots\mu_n}_{\k_1\ldots\k_n}\right) = \left.\frac{(2\pi)^d}{\N_{\mu_1}}\frac{\delta}{\delta J_{\a_1,\k_1}^{\mu_1}}\ldots\frac{(2\pi)^d}{\N_{\mu_n}}\frac{\delta}{\delta J_{\a_n,\k_n}^{\mu_n}}Z[J_+,J_-]\right|_{J_\pm =0}\;,
\end{equation}
where the functional generator is defined in the usual way:
\begin{equation}
    Z[J_+,J_-]=\int^{\varphi_+=\varphi_-}_{\textrm{B.D.}}\D\varphi_+\D\varphi_- e^{i S_+[\varphi_+]-i S_-[\varphi_-] + \int_{\KLF}(J_+\varphi_++J_-\varphi_-)}\;.
\end{equation}
These correlators are simply related to their spacetime version $\C^E_{\a_1\ldots\a_n}\left(^{z_1\ldots z_n}_{\x_1\ldots\x _n}\right)$ by an inverse KLF transform in each variable:
\begin{equation}\label{eq: from KLF to real space correlators}
    \C^E_{\a_1\ldots\a_n}\left(^{z_1\ldots z_n}_{\x_1\ldots\x _n}\right)=\int_{\KLF_n}\prod^{n}_{j=1}\Phi^{\mu_j}_{\k_j}(z_j,\x_j)\G_{\a_1\ldots\a_n}\left(^{\mu_1\ldots\mu_n}_{\k_1\ldots\k_n}\right)\;.
\end{equation} 
Since the $\G_{\a_1\ldots\a_n}\left(^{\mu_1\ldots\mu_n}_{\k_1\ldots\k_n}\right)$ are defined by functional differentiation, they do not have to be square-integrable functions in their KLF variables. We have already seen an example of this in section~\ref{sec: Role of the other Series}: the two-point Whightman function is proportional to a KLF delta \begin{equation}
    \G^{\O}_{+-}\left(^{\mu_1\mu_2}_{\k_1\k_2}\right)=\delta\left(^{\mu_1\mu_2}_{\k_1-\k_2}\right)\frac{\rho_\O(\mu_1)}{H^2 \N_{\mu_1}}\;.
\end{equation} 
Furthermore, for the CFT, the spectral density itself is a sum of smooth and a distributional part, see Eq.~\eqref{eq: CFT spectral density}.
In the next sections, we explicitly work out the example of the free theory where the functional generator can be explicitly evaluated, and that of simple interacting theories with polynomial interactions involving test fields in $\dS$.

\subsection{Free Theories}\label{sec: Gaussian PI}

Let us consider the free theory of a massive scalar field in dS:
\begin{equation}\label{eq: free action scalar}
    S_0 = \frac{1}{2}\int \frac{\d\tau\d^d\x}{(-H\tau)^{d-1}}\left[\left(\partial_\tau\varphi\right)^2-\left(\partial_i\varphi\right)^2 -\frac{m_{\varphi}^2}{H^2 \tau^2}\varphi^2\right]\;,
\end{equation}
where the mass is related to the Casimir eigenvalue~\eqref{eq: Casimir eigenvalues} as $m_{\varphi} = H M_{\mu_{\varphi}}$.\footnote{Since dS invariance forbids non diagonalisable quadratic mixings among fields, the construction also holds for multiple fields.} We first perform the Wick rotation~\eqref{Spm-Se} to EAdS, and then use the KLF decomposition~\eqref{eq: KLF decomposition} to obtain the action in KLF space:
\begin{equation}
    \pm i S_{0,\pm}[\varphi] = - \frac{H^2 e^{\pm\frac{i(d-1)\pi}{2}}}{2}\int_{\KLF}\varphi^{\mu}_{\k}(\mu^2-\mu_\varphi^2)\varphi^{\mu}_{-\k}\;.
\end{equation}
Since both actions are now expressed as integrals over the same set of variables, one can write their difference in a compact way using a vector notation:
\begin{equation}
    i S_+[\varphi_+]-iS_-[\varphi_-] = -\frac{H^2}{2}\int_{\KLF_2}\boldsymbol{\varphi}^T\left(^{\mu_1}_{\k_1}\right)\cdot \boldsymbol{\D}^0\left(^{\mu_1\mu_2}_{\k_1\k_2}\right)\cdot\boldsymbol{\varphi}\left(^{\mu_2}_{\k_2}\right)\;,
\end{equation}
where we defined the vectors $\boldsymbol{\varphi}_j$ and $\boldsymbol{\D}^0_{i j}$ as
\begin{equation}
    \varphi_{\a}\left(^{\mu}_{\k}\right) = \varphi^\mu_{\a,\k}\;,\quad \D^0_{\a_1\a_2}\left(^{\mu_1\mu_2}_{\k_1\k_2}\right) = \delta_{\a_1\a_2}\delta\left(^{\mu_1\mu_2}_{\k_1-\k_2}\right)(\mu_1^2-\mu_\varphi^2)e^{\frac{i\a_1(d-1)\pi}{2}}\;.
\end{equation}
The matrix $\boldsymbol{{\cal D}}^0_{i j}$ does not take the full quadratic sector into account, as there is a non trivial contribution from the future boundary condition that imposes that the two branches of the SK path integral are sewed at late time. This condition can be expressed as a field space $\delta-$function~\cite{Weinberg:2005vy}:
\begin{equation}
    \delta\left[\varphi_+(0,\x)-\varphi_-(0,\x)\right] \propto \lim_{\epsilon\to 0}\exp\left(-\frac{1}{\epsilon}\int_{\Sigma_0}\d^d\x\left(\varphi_+(0,\x)-\varphi_-(0,\x)\right)^2\right)\;,
\end{equation}
where $\Sigma_0$ denotes the final spatial slice of the Poincaré patch. Since the harmonic functions vanish in the asymptotic future, we use the late-time regulator $0<z_0 \ll 1$ described above, and evaluate the harmonic functions on this time slice:
\begin{equation}
    \int_{\Sigma_{z_0}}\d^d\x\left(\varphi_+(z_0,\x)-\varphi_-(z_0,\x)\right)^2= H^2\int_{\KLF_2}\boldsymbol{\varphi}^T\left(^{\mu_1}_{\k_1}\right)\cdot \boldsymbol{C}\left(^{\mu_1\mu_2}_{\k_1\k_2}\right)\cdot\boldsymbol{\varphi}\left(^{\mu_2}_{\k_2}\right)\;,
\end{equation}
where the matrix $\boldsymbol{C}$ is defined as:
\begin{equation}
    \boldsymbol{C}_{\a_1\a_2}\left(^{\mu_1\mu_2}_{\k_1\k_2}\right) = \frac{\a_1\a_2}{H^2}(2\pi)^d\delta^{(d)}(\k_1+\k_2)\Phi^{\mu_1}_{\k_1}(z_0,\boldsymbol{0})\Phi^{\mu_2}_{\k_2}(z_0,\boldsymbol{0})\;.
\end{equation}
The functional generator of the free theory can now be completely rewritten in KLF space:
\begin{equation}
\begin{aligned}
    Z_0[J_+,J_-] = \int_{\textrm{B.D.}}\D\varphi_+\D\varphi_-\exp\Bigg(-\frac{1}{2}\int_{\KLF_2}\boldsymbol{\varphi}^T\left(^{\mu_1}_{\k_2}\right)\cdot \boldsymbol{\D}\left(^{\mu_1\mu_2}_{\k_1\k_2}\right)\cdot\boldsymbol{\varphi}\left(^{\mu_2}_{\k_2}\right)\\
    + \int_{\KLF}\boldsymbol{J}^T\left(^\mu_{\k}\right)\cdot\boldsymbol{\varphi}\left(^\mu_{\k}\right)\Bigg)\;,
    \label{Z0-before-shift}
\end{aligned}
\end{equation}
where the symbol $\boldsymbol{J}$ uses the same vector notation as before and the matrix $\boldsymbol{\D}$ is defined as the sum:
\begin{equation}
    \boldsymbol{\D}\left(^{\mu_1\mu_2}_{\k_1\k_2}\right) = H^2\left(\boldsymbol{\D}^0\left(^{\mu_1\mu_2}_{\k_1\k_2}\right)+ \frac{2}{\epsilon} \boldsymbol{C}\left(^{\mu_1\mu_2}_{\k_1\k_2}\right)\right)\;.
\end{equation}
The usual trick to evaluate this integral is to complete the square by shifting the field with a current-dependent quantity:
\begin{equation}
  \boldsymbol{\varphi}\left(^{\mu_1}_{\k_1}\right) = \boldsymbol{\psi}\left(^{\mu_1}_{\k_1}\right)+\int_{\KLF}\boldsymbol{\G}_f\left(^{\mu_1\mu_2}_{\k_1\k_2}\right)\cdot\boldsymbol{J}\left(^{\mu_2}_{\k_2}\right)\;,
  \label{shift-path-integral}
\end{equation}
where the matrix $\boldsymbol{\G}^f$ satisfies 
\begin{equation}\label{eq: def matrix-G}
    \int_{\KLF}\boldsymbol{\D}\left(^{\mu_1\;\mu}_{\k_1\;\k}\right)\cdot\boldsymbol{\G}_f\left(^{\mu\;\mu_2}_{\k\;\k_2}\right) = \delta\left(^{\mu_1\mu_2}_{\k_1\k_2}\right)\left(\mathds{1}_2+(\mu_1^2-\mu_\varphi^2)\boldsymbol{f}(\mu_1)\hat{\delta}_{\mu_\varphi}(\mu_1)\right)\;,
\end{equation}
and $f_{\a\b}(\mu)=\delta_{ab} f_a(\mu)$ is an arbitrary diagonal matrix with $f_\a(\mu)$ shadow-symmetric functions. Such an additional term is omitted in Minkowski as it turns out to be degenerate with contour prescriptions when going back to position space. As we will see, this also holds in our construction for fields in the principal series. However, we need to keep it in general.
Using \eqref{shift-path-integral}, the path-integral in \eqref{Z0-before-shift} can be computed to yield the functional generator:
\begin{equation}\label{eq: free KLF generator}
    Z_0[J_+,J_-]=\underbrace{Z_0[0,0]}_{\braket{\Omega|\Omega}\equiv1}\exp\left(\frac{1}{2}\int_{\KLF_2}\boldsymbol{J}^T\cdot\boldsymbol{\G}_f\cdot\boldsymbol{J}\right)\;.
\end{equation}
The extra factor is the current-less functional integral over the field variables $\psi_\pm$ without any operator insertion. By definition, this is the norm of the Bunch-Davies vacuum state, which we took to be one. From Eq.~\eqref{eq: free KLF generator} and using the general definition~\eqref{eq: def KLF correlators}, $\G_f$ coincides with the 2-point KLF correlator:
\begin{equation}
    \mathcal{G}^{\mu_\varphi}_{f,\a_1\a_2}\left(^{\mu_1\mu_2}_{\k_1\k_2}\right) =\left.\frac{(2\pi)^d}{\N_{\mu_1}}\frac{\delta}{\delta J^{\mu_1}_{\a_1,k_1}}\frac{(2\pi)^d}{\N_{\mu_2}}\frac{\delta}{\delta J^{\mu_2}_{\a_2,k_2}}Z[J_+,J_-]\right|_{J_\pm = 0}\;.
\end{equation}
The components of the matrix $\boldsymbol{\G}_f$ can be found by solving equation~\eqref{eq: def matrix-G}. In the limit where $\epsilon$ goes to zero, this gives two sets of equations:
\begin{equation}\label{eq: def matrix-G-diagonal}
\int_{\KLF}\boldsymbol{\D}_0\left(^{\mu_1\mu}_{\k_1\k}\right)\cdot\boldsymbol{\G}_f\left(^{\mu\;\mu_2}_{\k\;\k_2}\right) = \frac{1}{H^2}\delta\left(^{\mu_1\mu_2}_{\k_1\k_2}\right)\left(\mathds{1}_2+(\mu_1^2-\mu_\varphi^2)\boldsymbol{f}(\mu_1)\hat{\delta}_{\mu_\varphi}(\mu_1)\right)
\end{equation}
and
\begin{equation}\label{eq: def matrix-G-off-diagonal}
    \int_{\KLF}\boldsymbol{C}\left(^{\mu_1\mu}_{\k_1\k}\right)\cdot\boldsymbol{\G}_f\left(^{\mu\;\mu_2}_{\k\;\k_2}\right) = 0\;.
\end{equation}

\paragraph{Diagonal propagators.} Let us start with the diagonal part of \eqref{eq: def matrix-G-diagonal}: 
\begin{equation}
    (\mu_1^2-\mu_\varphi^2)\G_{f,\a \a}^{\mu_\varphi}\left(^{\mu_1\mu_2}_{\k_1\k_2}\right)=\frac{e^{-\a\frac{i\pi(d-1)}{2}}}{H^2}\delta\left(^{\mu_1\;\mu_2}_{\k_1-\k_2}\right)\left(1+(\mu_1^2-\mu_\varphi^2)\hat{\delta}_{\mu_\varphi}(\mu_1)f_{\a}(\mu_1)\right)\;.
\end{equation}
The solution is given by:
\begin{equation}\label{eq: reduced propagators time ordered}
    \G^{\mu_\varphi}_{f,\a\a}\left(^{\mu_1\mu_2}_{\k_1\k_2}\right)=\frac{1}{H^2}\delta\left(^{\mu_1\mu_2}_{\k_1-\k_2}\right)\G^{\mu_\varphi}_{f,\a\a}(\mu_1)\;,
\end{equation}
where we defined the reduced propagator as:
\begin{equation}\label{eq: diagonal propagator KLF function f}
    \G^{\mu_\varphi}_{f,\a\a}(\mu)\equiv e^{-\frac{i\a\pi(d-1)}{2}}\left(\frac{1}{\mu^2-\mu_\varphi^2}+\hat{\delta}_{\mu_\varphi} (\mu)f_{\a}(\mu)\right)\;.
\end{equation}
The corresponding position-space objects $\C_{\a\a}^E(X_1^E,X_2^E)$ we find by applying the inverse transformation~\eqref{eq: from KLF to real space correlators} are nothing but the (anti-)time-ordered Green functions such that:
\begin{equation}
    \left(H^2\Box_{\EAdS}-m_{\varphi}^2\right)\C_{\a\a}^E(X_1^E,X_2^E) = (H z_1)^{d+1}\delta(z_1-z_2)\delta^{(d)}(\x_1-\x_2)\;.
\end{equation}
We stress that this holds for any choice of functions $f_\a(\mu)$, and that the latter only affect boundary conditions. 
\vskip 4pt
Like in flat spacetime, the reduced two-point function \eqref{eq: diagonal propagator KLF function f} has poles when the propagating quanta goes on-shell, i.e.~at $\mu=\pm \mu_\varphi$. Therefore, for principal series fields, the inverse KLF transform must be regulated with an appropriate $i\epsilon$ prescription. As explained in appendix~\ref{sec: propagators in real space}, the usual Feynman prescription breaks the shadow symmetry. Thus, the only way to proceed is to add up two pole prescriptions as:
\begin{equation}\label{eq: arbitrary i epsilon function g}
\begin{aligned}
     \frac{1}{\mu^2-\mu_\varphi^2}\to &\frac{1}{(\mu^2-\mu_\varphi^2)^{g}_{i\a\epsilon}}\equiv \frac{g(\mu_\varphi)}{\mu^2-\mu_\varphi^2+i\a\epsilon}+\frac{g(-\mu_\varphi)}{\mu^2-\mu_\varphi^2-i\a\epsilon}\;,
\end{aligned}
\end{equation}
in terms of a function $g(\mu)$. Like for the functions $f_\a(\mu)$ above, it should be determined by imposing appropriate boundary conditions. In order to produce the (anti-)time ordered propagators, it is shown in appendix \ref{sec: propagators in real space} that there is a large freedom in the choice of these functions. The simplest choice turns out to be:
\begin{equation}
    \left\{\begin{array}{crc}
        f_{\a}(\mu) & = & 0 \\
        g(\mu) & = & \frac{e^{\pi\mu}}{2\sinh(\pi\mu)}
    \end{array}\right.\;.
\end{equation}
This makes manifest the possibility of absorbing the effect of the functions $f_{\a}(\mu)$ in the contour prescription for heavy fields. As a result, we recover the same convention as~\cite{Melville:2023kgd,Werth:2024mjg}:
\begin{equation}\label{eq: i epsilon prescription}
    \frac{1}{(\mu^2-\mu_\varphi^2)_{i\a\epsilon}}\equiv\frac{i}{2\sinh(\pi\mu_\varphi)}\left(\frac{e^{+\pi\mu_\varphi}}{\mu^2-\mu_\varphi^2+i\a\epsilon}-\frac{e^{-\pi\mu_\varphi}}{\mu^2-\mu_\varphi^2-i\a\epsilon}\right)\;.
\end{equation}
\vskip 4pt
In the case of non-principal fields where $\mu_\varphi=i\nu_\varphi$ with real $\nu_\varphi$, the poles are located on the imaginary axis and there is no need for a particular contour prescription. Because of this, the boundary condition is fully encoded in the functions $f_\a$. It is shown in appendix \ref{sec: propagators in real space} that the (anti-)time ordered propagators are defined by the following choice:
\begin{equation}
    f_{\a}(i \nu_{\varphi})= - i\a\frac{e^{i\a\pi|\nu_\varphi|}}{\N_{i\nu_\varphi}}\;.
\end{equation}
Physically, we need this additional contribution compared to principal series fields because the (anti-)time ordered SK propagators do not define a Green function in $L^2$. As a result, the reduced propagator should include non-principal contributions, as we have seen in section~\ref{sec: Role of the other Series}.

Now that the functions $f_{\a}$ and $g$ are known, we will remove their explicit mentions in the propagators. Our conventions for the KLF (anti-)time ordered propagators are thus expressed as:
\begin{equation}
    \G^{\mu_\varphi}_{\a\a}\left(^{\mu_1\mu_2}_{\k_1\k_2}\right) = \frac{1}{H^2}\delta\left(^{\mu_1\mu_2}_{\k_1\k_2}\right)\G^{\mu_\varphi}_{\a\a}(\mu_1)\;,
    \label{general-to-reduced-propagators-link-same-a}
\end{equation}
where the reduced propagators $\G^{\mu_\varphi}_{\a\a}(\mu)$ are of the form:
\begin{FramedBox}
    \begin{equation}\label{eq: KLF (anti-)time ordered}
    \begin{aligned}
        \G_{\a\a}^{\mu_\varphi}(\mu)&=e^{-\frac{i\a\pi(d-1)}{2}}\frac{1}{(\mu^2-\mu_\varphi^2)_{i\a\epsilon}} \quad \textrm{Principal Series}\\
        \G_{\a\a}^{i\nu_\varphi}(\mu)&=e^{-\frac{i\a\pi(d-1)}{2}}\left(\frac{1}{\mu^2+\nu_\varphi^2} - i\a\frac{e^{i\a\pi|\nu_\varphi|}}{\N_{i\nu_\varphi}}\hat{\delta}_{i\nu_\varphi}(\mu)\right) \quad \textrm{Other UIRs}\;.
    \end{aligned}
    \end{equation}
\end{FramedBox}
We show in appendix~\ref{sec: propagators in real space} that this is consistent with the usual expression of the (anti-) time ordered propagators in Lorentzian position space (see e.g.~\cite{DiPietro:2021sjt}): 
\begin{equation}
\begin{aligned}
    \C_{\a\a}(\tau_1,\x_1;\tau_2,\x_2)&=\frac{1}{H^2}\int_{\KLF}\Phi^{\mu}_{\k}\left(- \tau_1 e^{i \a \left( \frac{\pi}{2}-\epsilon \right)} ,\x_1 \right)\Phi^{\mu}_{-\k}\left(- \tau_2 e^{i \a \left( \frac{\pi}{2}-\epsilon \right)},\x_2\right)\G_{\a\a}^{\mu_\varphi}(\mu)\\
    &= W_{\mu_\varphi}(\sigma^{\textrm{dS}}-i\a\epsilon)\;,
\end{aligned}
\end{equation}
where $W_{\mu_\varphi}(\sigma)$ is defined in equation~\eqref{eq:definition of W}, $\sigma^{\textrm{dS}}$ is defined in~\eqref{eq: def two-point invariant dS}, and the $i\epsilon$ prescription is here to regulate the propagator for time-like separations, where $W_{\mu_\varphi}(\sigma)$ has a branch-cut.
\vskip 4pt
\paragraph{Off-diagonal propagators.} The off-diagonal part of \eqref{eq: def matrix-G-diagonal} gives the following constraint
\begin{equation}\label{eq: off-diag propagators 1}
    (\mu_1^2-\mu_\varphi^2)\G^f_{-\a\a}\left(^{\mu_1\mu_2}_{\k_1\k_2}\right) = (\mu_2^2-\mu_\varphi^2)\G^f_{-\a\a}\left(^{\mu_1\mu_2}_{\k_1\k_2}\right) =0\;,
\end{equation}
which is the KLF-space counterpart of the Casimir equation. On the other hand, Eq.~\eqref{eq: def matrix-G-off-diagonal} gives
\begin{equation}\label{eq: off-diag propagators 2}
    \int\displaylimits^\infty_{-\infty}\d\mu\N_{\mu}\Phi^{\mu_1}_{\k_1}(z_0,\boldsymbol{0})\Phi^{\mu}_{-\k_1}(z_0,\boldsymbol{0})\left(\G_{-\a\a}\left(^{\;\;\mu\;\;\mu_2}_{-\k_1\k_2}\right)-\G_{\a\a}\left(^{\;\;\mu\;\;\mu_2}_{-\k_1\k_2}\right) \right)=0\;.
\end{equation}
We deduce from~\eqref{eq: off-diag propagators 1} that $\G_{-\a\a}$ is proportional to the distribution $\hat{\delta}_{\mu_\varphi}(\mu_1)$. Besides, Eq.~\eqref{eq: off-diag propagators 2} implies that the propagator has the same dependence as in~\eqref{eq: reduced propagators time ordered}. Therefore, the off-diagonal two-point function can be expressed as 
\begin{equation}
\G^{\mu_\varphi}_{-\a\a}\left(^{\mu_1\mu_2}_{\k_1\k_2}\right) = \frac{1}{H^{2}}\delta\left(^{\mu_1\mu_2}_{\k_1-\k_2}\right)\frac{\hat{\delta}_{\mu_\varphi}(\mu_1)}{\N_{\mu_1}}\F_{-\a\a}^{\mu_\varphi}.
\label{step-F}
\end{equation}
In the insert below, we show that boundary conditions impose $\F_{-\a\a}^{\mu_\varphi}=1$. To summarise, the KLF space Wightman function takes the same reduced form as the (anti-)time ordered propagators:
\begin{equation}
    \G^{\mu_\varphi}_{-\a\a}\left(^{\mu_1\mu_2}_{\k_1\k_2}\right) = \frac{1}{H^2}\delta\left(^{\mu_1\mu_2}_{\k_1\k_2}\right)\G^{\mu_\varphi}_{-\a\a}(\mu_1)\;,
\label{general-to-reduced-propagators-link-different-a}    
\end{equation}
where  $\G^{\mu_\varphi}_{-\a\a}(\mu)$ is identified as:
\begin{FramedBox}
    \begin{equation}\label{eq: KLF wightman function}
    \G_{\pm\mp}^{\mu_\varphi}(\mu) = \frac{\hat{\delta}_{\mu_\varphi}(\mu)}{\N_{\mu}}\;.
\end{equation}
\end{FramedBox}
The position-space version of this propagator can be found in appendix~\ref{sec: propagators in real space} to be:
\begin{equation}
    \C_{-\a\a}(\tau_1,\x_1;\tau_2,\x_2)=W_{\mu_\varphi}(\sigma^{\textrm{dS}}-i\a\epsilon\;\textrm{sgn}(\tau_1-\tau_2))\;.
\end{equation}

\begin{framed}
In this insert, we use boundary conditions to determine $\F_{-\a\a}^{\mu_\varphi}$ in \eqref{step-F}. For this, we plug this equation into \eqref{eq: off-diag propagators 2} and perform the $\mu_1$ integral, yielding:
\begin{equation}
    \Phi^{\mu_1}_{\k_1}(z_0,\boldsymbol{0})\Phi^{\mu_2}_{-\k_1}(z_0,\boldsymbol{0})\hat{\delta}_{\mu_\varphi}(\mu_2)\F_{-\a\a}^{\mu_\varphi} = \N_{\mu_2}\Phi^{\mu_1}_{\k_1}(z_0,\boldsymbol{0})\Phi^{\mu_2}_{-\k_1}(z_0,\boldsymbol{0})\G^{\mu_\varphi}_{\a\a}(\mu_2)\;.
\end{equation}
Since this equation is true for any values of the KLF variables, it can be evaluated at $\mu_1=\mu_2 \equiv\mu$. Then, we integrate over the frequency and get:
\begin{equation}\label{eq: off diagonal interemediary steps}
    \Phi^{\mu_\varphi}_{\k_1}(z_0,\boldsymbol{0})\Phi^{\mu_\varphi}_{-\k_1}(z_0,\boldsymbol{0})\F^{\mu_\varphi}_{-\a\a} = \int\displaylimits^\infty_{-\infty}\d\mu\N_\mu\Phi^{\mu}_{\k_1}(z_0,\boldsymbol{0})\Phi^{\mu}_{-\k_1}(z_0,\boldsymbol{0})\G_{\a\a}^{\mu_\varphi}(\mu)\;.
\end{equation}
With the time-ordered boundary conditions encoded in the choice of the functions $f_{\a}$ and $g$ discussed above, it is shown in appendix~\ref{sec: propagators in real space} that the right-hand side is:
\begin{equation}
    \int\displaylimits^\infty_{-\infty}\d\mu\N_\mu\Phi^{\mu}_{\k_1}(z_0,\boldsymbol{0})\Phi^{\mu}_{-\k_1}(z_0,\boldsymbol{0})\G_{\a\a}^{\mu_\varphi}(\mu) = \Phi^{\mu_\varphi}_{\k_1}(z_0,\boldsymbol{0})\Phi^{\mu_\varphi}_{-\k_1}(e^{-i\a\pi}z_0,\boldsymbol{0})\;.
\end{equation}
Therefore, equation~\eqref{eq: off diagonal interemediary steps} becomes:
\begin{equation}\label{eq: equation for Fmuaa Euclidean}
    \Phi^{\mu_\varphi}_{\k_1}(z_0,\boldsymbol{0})\Phi^{\mu_\varphi}_{-\k_1}(z_0,\boldsymbol{0})\F^{\mu_\varphi}_{-\a\a}=\Phi^{\mu_\varphi}_{\k_1}(z_0,\boldsymbol{0})\Phi^{\mu_\varphi}_{-\k_1}(e^{-i\a\pi}z_0,\boldsymbol{0})\;.
\end{equation}
At first sight, equation~\eqref{eq: equation for Fmuaa Euclidean} has no solution. This is because we try to compare two propagators living on different copies of EAdS. Since the late-time boundary condition is imposed at $\tau_0$, Eq.~\eqref{eq: equation for Fmuaa Euclidean} is valid either in the exact limit where $\tau_0,z_0\equiv 0$, or after performing the inverse rotation to Lorentzian signature. As the harmonic functions are singular in the late time limit, we pick the second option and apply the rotations~\eqref{eq: Lorentzian harmonic functions}, consistently with the signs of the indices in~\eqref{eq: Wick rotated n point function}. Since the left-hand side lives on the branch $-\a\a$, the rotation gives:
\begin{equation}
\begin{aligned}
    \Phi^{\mu_\varphi}_{\k_1}(z_0,\boldsymbol{0})\Phi^{\mu_\varphi}_{-\k_1}(z_0,\boldsymbol{0})&\equiv  \Phi^{\mu_\varphi}_{\k_1}\left(-\tau_0 e^{-\frac{i\a\pi}{2}},\boldsymbol{0}\right)\Phi^{\mu_\varphi}_{-\k_1}\left(-\tau_0 e^{\frac{i\a\pi}{2}},\boldsymbol{0}\right)\\
    &= H^{d+1}u^\a_{k_1}(\tau_0,\mu)u^{-\a}_{k_1}(\tau_0,\mu)\;.
\end{aligned}
\end{equation}
On the other hand, the right-hand side lives on the $\a\a$ branch, so the rotation reads:
\begin{equation}
\begin{aligned}
    \Phi^{\mu_\varphi}_{\k_1}(z_0,\boldsymbol{0})\Phi^{\mu_\varphi}_{-\k_1}\left(e^{-ia\pi}z_0,\boldsymbol{0}\right)&\equiv  \Phi^{\mu_\varphi}_{\k_1}\left(-\tau_0 e^{\frac{i\a\pi}{2}},\boldsymbol{0}\right)\Phi^{\mu_\varphi}_{-\k_1}\left(-\tau_0 e^{-\frac{i\a\pi}{2}},\boldsymbol{0}\right)\\
    &= H^{d+1}u^{-\a}_{k_1}(\tau_0,\mu)u^{\a}_{k_1}(\tau_0,\mu)\;.
\end{aligned}
\end{equation}
One thus deduces that the consistent normalisation is:
\begin{equation}
    \F_{-\a\a}^{\mu_\varphi}= 1\;.
\end{equation}
\end{framed}

\subsection{Polynomial Interactions}
Let us now turn to interactions. First, let us consider an $n$-point polynomial interaction in a single-field theory:
\begin{equation}
    S_I[\varphi] = -\lambda\int\frac{\d\tau\d^d\x}{(-H\tau)^{d+1}}\varphi^n\;.
\end{equation}
As in the free-theory case, we perform the Wick rotation to $\EAdS$ and we expand the fields in KLF modes. The actions on the two branches are therefore expressed as:
\begin{equation}\label{eq: N point contact term action}
\begin{aligned}
     \pm i S_I[\varphi]= -\frac{\lambda H^{\frac{(n-2)(d+1)}{2}}e^{\pm\frac{i(d+1)\pi}{2}}}{\pi^{\frac{n}{2}}}\int_{\KLF_n}(2\pi)^d\delta^{(d)}\left(\sum_{j=1}^n \k_j\right)\mathcal{I}^{\mu_1\ldots\mu_n}_{k_1\ldots k_n}\prod_{j=1}^n\varphi^{\mu_j}_{\k_j}\;,
\end{aligned}
\end{equation}
where we defined the $n$-point vertex function as an integral over $n$ functions $K_{i\mu}$:
\begin{equation}
\label{eq: def vertex function}
    \mathcal{I}^{\mu_1\ldots\mu_n}_{k_1\ldots k_n} = \int^\infty_0\d z\; z^{\frac{d(n-2)}{2}-1} K_{i\mu_1}(k_1 z)\ldots K_{i\mu_n}(k_n z)\;.
\end{equation}

\vskip 4pt
Similar-looking quantities arise in in-out computations in flat space, with harmonic functions replaced by $4$-momentum plane waves, yielding an energy-conserving delta function, in addition to the $3$-momentum conservation.
However, in flat space, the true conceptual analogues of the frequencies $\mu_j$, labelling UIRs, are the masses $m_j$ of the particles rather than their energies $\omega_j$. The nonconservation of mass at vertices in flat space has a counterpart here: the nonconservation of $\mu$'s at vertices, which is precisely encoded in the vertex functions \eqref{eq: def vertex function}. The latter are purely kinematic quantities, independent of the nature of the interacting fields, and are central in our formalism.

\vskip 4pt 
The integrals \eqref{eq: def vertex function} naturally 
arise when evaluating the late-time $n$-point function generated by a contact interaction between $n$ general massive scalars~\cite{Sleight:2019hfp}.
They are known in terms of Lauricella hypergeometric series~\cite{Lauricella1893SulleFI}. However, their evaluation usually requires analytic continuation, as the physical domain does not always fall within the convergence domain of the series. 
Besides, as we show in section~\ref{sec:KLF-rules} and as we will see with examples in section~\ref{sec:examples}, computing cosmological correlators also requires to analytically continue the vertex functions \eqref{eq: def vertex function} for complex $\mu$, i.e.~outside the principal series. This is not problematic: from their writing as Lauricella series, we know that $\mathcal{I}^{\mu_1\ldots\mu_n}_{k_1\ldots k_n}$ is meromorphic in the $\mu_j$ planes, which also allows us to deform the KLF contour integrations and compute correlators by collecting residues, see section \ref{subsection_single_exchange}.

\vskip 4pt
In the simplest case, $n=3$, the vertex function reads:
\begin{equation}
    \mathcal{I}^{\mu_1\mu_2\mu_3}_{k_1 k_2 k_3}= \int^\infty_0\d z \;z^{\frac{d}{2}-1}K_{i\mu_1}(k_1 z)K_{i\mu_2}(k_2 z)K_{i\mu_3}(k_3 z)\;.
\label{eq:triple-K}    
\end{equation}
These integrals, known as triple-$K$ integrals, are known to arise in the evaluation of the conformal three-point function in momentum space~\cite{Bzowski:2012ih,Bzowski:2013sza,Bzowski:2015pba,Bzowski:2015yxv,Bzowski:2017poo}. They can be expressed in terms of the function Appell $F_4$~\cite{Bzowski:2013sza}, but whose series representation only converges when the momenta violate the triangle inequality, see appendix \ref{appendix-triple-K}. Therefore, the evaluation of these integrals requires a suitable analytical continuation of the function $F_4$ to the physical domain. For generic complex $\mu$'s, the (late-time) convergence of \eqref{eq:triple-K} requires that $ |\text{Im}(\mu_1)| + |\text{Im}(\mu_2)| + |\text{Im}(\mu_3)|< \frac{d}{2}$.
In the opposite regime, the vertex function admits a unique analytical continuation.
Furthermore, by expanding the integrand around $z\sim 0$, one sees that analyticity is violated at the points such that:
\begin{equation}\label{eq: poles vertex function}
    \frac{d}{2}\pm i \mu_1  \pm i\mu_2 \pm i \mu_3 \in -2\;\mathbb{N}\;.
\end{equation}
This manifests as simple poles in the analytically continued vertex function~\cite{Bzowski:2015yxv}. 

\vskip 4pt 
Eventually, notice that
the above construction can be trivially extended to multifield theories. For instance, for a cubic interaction of the type $\L_I=- \frac12 g\varphi^2\chi$ the KLF-space version of the interaction action reads: 
\begin{equation}\label{eq: SI KLF two fields}
    \pm i S_I = -\frac{g H^{\frac{d+1}{2}}e^{\pm\frac{i(d+1)\pi}{2}}}{2\pi^{\frac{3}{2}}}\int_{\KLF_3}(2\pi)^ d\delta^{(d)}\left(\sum_{j=1}^ 3\k_j\right)\mathcal{I}^{\mu_1\mu_2\mu_3}_{\k_1,\k_2,\k_3}\,\varphi^{\mu_1}_{\k_1}\varphi^{\mu_2}_{\k_2}\chi^{\mu_3}_{\k_3}\;.
\end{equation}

\subsection{KLF Feynman Rules}
\label{sec:KLF-rules}
Now that we have simple expressions for the KLF space propagators and vertices, let us explain how to use this in order to write every possible perturbative Feynman diagram. In the path integral formalism, the standard way is to evaluate the functional generator as an asymptotic series in the interacting action. In this picture, the vacuum and the dynamical evolution are those of the free theory and the interactions are viewed as operator insertions in the Gaussian path integral~\cite{Chen:2017ryl}: 
\begin{equation}\label{eq: generator PT}
\begin{aligned}
     Z[J_+,J_-]&=\braket{0|e^{iS_I^+[\varphi_+]-iS_I^-[\varphi_-]+\int_{\KLF}J_+\varphi_++J_-\varphi_-}|0}\\
     &=\sum_{n=0}^\infty\frac{1}{n!}\left(iS_I^+\left[\frac{\delta}{\delta J_+}\right]-iS_I^-\left[\frac{\delta}{\delta J_-}\right]\right)^n\,Z_0[J_+,J_-]\;,
\end{aligned}
\end{equation}
where $\ket{0}$ is the Bunch-Davies vacuum associated with the free theory $S_0$, and $S_I=S-S_0$ is the interacting action. This is fully equivalent to the operator formalism where the fields variables are in interaction picture~\cite{Weinberg:2005vy,Chen:2017ryl}.
\vskip 4pt
The KLF correlators can then be obtained by applying the definition~\eqref{eq: def KLF correlators}, leading to the following set of Feynman rules:
\begin{itemize}
    \item A line in a diagram denotes the propagation of a free particle. Therefore, it corresponds to a free propagator: 
    \begin{equation}
    \begin{aligned}
        &\begin{tikzpicture}[baseline={(0,0)}]
            \draw[thick, black] (-1,0) to (1,0);
            \filldraw[color=black,fill=black] (-1,0) circle (3pt);
            \filldraw[color=black,fill=black] (1,0) circle (3pt);
            \node[black] at (-1,0.4) {$\a$};
            \node[black] at (-1,-0.4) {$\mu_1\;,\k_1$};
            \node[black] at (1,-0.4) {$\mu_2\;,\k_2$};
            \node[black] at (1,0.4) {$\b$};
            \node[black] at (0.,+0.4) {$\varphi,\;\mu_\varphi$};
            \end{tikzpicture}
            & \hspace*{-0.8cm}= \frac{1}{H^2}\delta\left(^{\mu_1\mu_2}_{\k_1-\k_2}\right)\G_{\a\b}^{\mu_\varphi}(\mu_1)\;.
    \end{aligned}
    \end{equation}
    \item An $n$-point vertex living on the branch $\a=\pm$ is expressed by a momentum-conserving delta function and a vertex function $\I^{\mu_1\ldots\mu_n}_{\k_1\ldots\k_n}$:
    \begin{equation}\label{eq: Feynman rule vertex}
\begin{aligned}
    &\begin{tikzpicture}[baseline={(0,0)}]
    \draw[thick, black] (0, 0) to (-1,1) node[above] {\scriptsize $\mu_1,\k_1$};
    \draw[thick, black] (0, 0) to (1,1) node[above] {\scriptsize $\mu_n,\k_n$};
    \draw[thick, black] (0,0) to (-0.33,1) node[above, xshift=1mm] {\scriptsize $\mu_2,\k_2$};
    \node[black] at (0.2,0.7) {$\ldots$};
    \draw[color=black] (0,0) circle (3pt) node[below] {\scriptsize$\aa=\pm$};
    \filldraw[color=black,fill=black] (0,0) circle (3pt);
    \end{tikzpicture}
    &\hspace*{-0.8cm}= - i\,\aa\,\C\,\lambda(2\pi)^d\delta^{(d)}\left(\sum_{j=1}^n\k_j\right)\frac{H^{\frac{(n-2)(d+1)}{2}}e^{\aa\frac{i \pi d}{2}}}{\pi^\frac{n}{2}}\mathcal{I}^{\mu_1 \ldots \mu_n}_{k_1 \ldots k_n} \,,
    \end{aligned}
\end{equation}
where $\C$ is a combinatorial factor fixed by the form of the interaction term. 

\item A KLF space correlator $\G_{\a_1\ldots\a_n}\left(^{\mu_1\ldots\mu_n}_{\k_1\ldots\k_n}\right)$ is obtained by setting the external point to the corresponding branch and summing over the possible branch $\a^I_j$ of the internal vertices.

\item All the frequency and momenta attached to the internal lines should be integrated over with the KLF measure. At tree level, the internal momentum integrals are straightforward because of momentum conservation at vertices. However, due to the lack of frequency conservation, even tree-level diagrams feature one frequency integral per internal line.
\end{itemize}
According to these rules, a diagram can be represented as:
\begin{equation}
    \G_{\a_1\ldots\a_n}\left(^{\mu_1\ldots\mu_n}_{\k_1\ldots\k_n}\right)=\underbrace{(2\pi)^{d}\delta^{(d)}\left(\sum_{m=1}^n\k_m\right)}_{\textrm{Momentum Conservation}}\sum_{\{a^V_E\}}\G_{\{a_E^V\}}\left(^{\mu_1\ldots\mu_n}_{\k_1\ldots\k_n}\right)\;,
    \label{KLF-correlator-split}
\end{equation}
where 
\begin{equation}\label{eq: G AVE}
    \G_{\{a_E^V\}}\left(^{\mu_1\ldots\mu_n}_{\k_1\ldots\k_n}\right)=\underbrace{\left(\prod_{j=1}^{n_E}\prod_{\ell=1}^{n_j}\frac{\G^{\mu_{\varphi_\ell}}_{\a_\ell\a_j}(\mu_\ell)}{H^2}\right)}_{\textrm{External Legs}}\underbrace{\G^A_{\{\a^V_E\}}\left(^{\mu_1\ldots\mu_n}_{\k_1\ldots\k_n}\right)}_{\textrm{Amputated Diagram}}\;.
\end{equation}
The amputated diagram $\G^A_{\{\a^V_E\}}\left(^{\mu_1\ldots\mu_n}_{\k_1\ldots\k_n}\right)$ is labeled by a configuration of $n_E$ external vertices $\a_j\in\{\a^V_E\}$ to which $n_j$ external legs are attached. Consistently, we have $\sum_{j=1}^{n_E}n_j=n$ and the $\a_\ell$ refer to the external legs of $\G_{\a_1\ldots\a_n}\left(^{\mu_1\ldots\mu_n}_{\k_1\ldots\k_n}\right)$.

This object represents the non trivial part of the KLF correlator. It implicitly contains a sum over all possible internal vertices $\{\a^V_I\}$ and integrals over every internal frequency and momentum $\{\mu_I,\k_I\}$. For instance, in the case of a four-point function with three external vertices $\{\a_E^V\}=\{\a_1,\a_2,\a_4\}$, the amputated diagram can be represented as: 
\begin{equation}
\begin{aligned}
&\begin{tikzpicture}[baseline={(0,0)}][rotate=90]
  \draw[color=black] (0.8,0) -- (1.3,-0.8) node[right] {$\mu_3,\k_3$};
  \draw[color=black] (0.8,0) -- (1.3,+0.8) node[right] {$\mu_2,\k_2$};

   \draw[color=black] (-0.45,-0.45) -- (-1.3,-1.3) node[left] {$\mu_4,\k_4$};

  \draw[color=black] (-0.45,0.45) -- (-1.3,1.3) node[left] {$\mu_1,\k_1$};
  
  \draw[fill=lightgray] (0,0) circle (0.8);

  \draw[color=black] (0.8,0) circle (3pt) node[right] {$\a_2$};
  \filldraw[color=black,fill=black] (0.8,0) circle (3pt);

  \draw[color=black] (-0.56,0.56) circle (3pt) node[left] {$\a_1$};
  \filldraw[color=black,fill=black] (-0.56,0.56) circle (3pt);

  \draw[color=black] (-0.56,-0.56) circle (3pt) node[left] {$\a_4$};
  \filldraw[color=black,fill=black] (-0.56,-0.56) circle (3pt);
\end{tikzpicture}
& = \SumInt_{\{\a_I^V\},\{\mu_I,\k_I\}}&
\begin{tikzpicture}[baseline={(0,0)}][rotate=90]
  \draw[color=black] (0.8,0) -- (1.3,-0.8) node[right] {$\mu_3,\k_3$};
  \draw[color=black] (0.8,0) -- (1.3,+0.8) node[right] {$\mu_2,\k_2$};

   \draw[color=black] (-0.45,-0.45) -- (-1.3,-1.3) node[left] {$\mu_4,\k_4$};

  \draw[color=black] (-0.45,0.45) -- (-1.3,1.3) node[left] {$\mu_1,\k_1$};
  
  \draw[fill=lightgray] (0,0) circle (0.8) node[above] {$\{\a^V_I\}$};

  \node at (0,-0.2) {$\{\mu_I,\k_I\}$};

  \draw[color=black] (0.8,0) circle (3pt);
  \filldraw[color=black,fill=black] (0.8,0) circle (3pt) node[right] {$\a_2$};

  \draw[color=black] (-0.56,0.56) circle (3pt);
  \filldraw[color=black,fill=black] (-0.56,0.56) circle (3pt) node[left] {$\a_1$};

  \draw[color=black] (-0.56,-0.56) circle (3pt) node[left] {$\a_4$};
  \filldraw[color=black,fill=black] (-0.56,-0.56) circle (3pt);
\end{tikzpicture}\;.
\end{aligned}
\end{equation}

\vskip 4pt
Finally, the corresponding bulk correlator is obtained by successively applying the KLF inverse transformation~\eqref{eq: from KLF to real space correlators} and the Wick rotation~\eqref{eq: Wick rotated n point function}. See appendix~\ref{sec: propagators in real space} and section~\ref{sec: loop two point} for applications of this procedure, respectively, to the free propagator and its one-loop correction.

\paragraph{Boundary Correlators.}
Let us now explain how to extract the boundary correlator $\C_\mathcal{B}\left(\k_1,\ldots,\k_n\right)$ from the KLF space correlator $\G_{\a_1\ldots\a_n}(^{\mu_1\ldots\mu_n}_{\k_1\ldots\k_n})$. As explained above, they are defined as the late time limit of bulk correlators with any value of the external SK indices. In order to simplify things, we make use of this freedom to turn all the external legs in~\eqref{eq: G AVE} into off-diagonal propagators. Following \eqref{KLF-correlator-split}, the contribution of this diagram thus reads:
\begin{equation}\label{eq: definition amputated correlator}
    \G_{\{a_E^V\}}\left(^{\mu_1\ldots\mu_n}_{\k_1\ldots\k_n}\right)\equiv \underbrace{\left(\prod_{j=1}^{n_E}\prod_{\ell=1}^{n_j}\frac{\hat{\delta}_{\mu_{\varphi_\ell}}(\mu_\ell)}{H^2\N_{\mu_\ell}}\right)}_{\textrm{External Legs}}\underbrace{\G^{A}_{\{\a_E^V\}}\left(^{\mu_1\ldots\mu_n}_{\k_1\ldots\k_n}\right)}_{\textrm{Amputated Correlator}}\;.
\end{equation}
Then, as stated before, we apply the inverse KLF transform~\eqref{eq: from KLF to real space correlators} and the Wick rotation~\eqref{eq: Wick rotated n point function} in the limit where $\tau_0\to0$. Since the external legs are fully expressed in terms of $\hat{\delta}_{\mu_{\varphi_j}}$, the frequency integrals are straightforward to perform. Their effect is to send the KLF frequency of the amputated diagram $\G^A_{\{\a^V_E\}}\left(^{\mu_1\ldots\mu_n}_{\k_1\ldots\k_n}\right)$ on-shell and to add an extra factor $\E_{\{\a_E^V\}}(\tau_0)$ coming from the late time limit of the harmonic function and the Wick rotation. Diagrammatically, we will denote the contribution coming from a single external line attached to an external vertex $\a\in\{\a^V_E\}$ as:
\begin{equation}
\begin{aligned}
&\begin{tikzpicture}[baseline={(0,0)}]
    
    \draw[color=black] (-0.8,0) -- (0.8,0);

    \draw[color=black] (0.8,0.) circle (3pt);
  \filldraw[color=black,fill=black] (0.8,0.) circle (3pt) node[below] {$\a$};
    \filldraw[color=black,fill=white] (-0.8,-2pt) rectangle ++(4pt,4pt);
\end{tikzpicture}
& = \frac{H^{\frac{d-3}{2}}}{\sqrt{\pi}}e^{-\frac{i\a\pi d}{4}}(-\tau_0)^{\frac{d}{2}}K_{i\mu_\varphi}(-k\tau_0 e^{-\frac{i\a\pi}{2}})\;,
\end{aligned}
\end{equation}
where, as customary, a $\begin{tikzpicture}
    \filldraw[color=black,fill=white] (-0.8,-2pt) rectangle ++(4pt,4pt);
\end{tikzpicture}$ denotes an evaluation at the boundary time $\tau_0$.
The full factor $\E_{\{\a_E^V\}}(\tau_0)$ is then given by the product of all the external line contributions:
\begin{equation}\label{eq: contribution external legs general}
\begin{aligned}
    \E_{\{\a_E^V\}}(\tau_0) &= \prod_{j=1}^{n_E}\prod_{\ell=1}^{n_j}\left(\begin{tikzpicture}[baseline={(0,0)}]
    
    \draw[color=black] (-0.8,0) -- (0.8,0);

    \node[color=black] at (0,0.3){$\mu_{\varphi_\ell},k_\ell$};

    \draw[color=black] (0.8,0.) circle (3pt);
  \filldraw[color=black,fill=black] (0.8,0.) circle (3pt) node[below] {$\a_j$};
    \filldraw[color=black,fill=white] (-0.8,-2pt) rectangle ++(4pt,4pt);
\end{tikzpicture}\right) \\
&=\frac{H^{\frac{n(d-3)}{2}}}{\pi^{\frac{n}{2}}}(-\tau_0)^{\frac{n d}{2}}\prod_{j=1}^{n_E} e^{-\frac{in_j\a_j\pi d}{4}}\prod_{\ell=1}^{n_j}K_{i\mu_{\varphi_\ell}}(-k_\ell\tau_0 e^{-\frac{i\a_j\pi}{2}})\;,
\end{aligned}
\end{equation}
where $\a_j\in\{\a^V_E\}$, $n_j$ is the number of external legs attached to it, and we have $\sum_{j=1}^{n_E}n_j=n$.
Thus, the full momentum-space contribution, labeled by the set $\{\a^V_E\}$, to a late-time dS correlator involving fields with masses $\{\mu_{\varphi_j}\}_{j=1}^n$, is:
\begin{FramedBox}
    \begin{equation}\label{eq: general late time subdiagrams}
     F^{\varphi_1\ldots\varphi_n}_{\{\a^V_E\}}\left(\k_1\ldots\k_n\right) = (2\pi)^d\delta^{(d)}\left(\sum_{j=1}^n\k_j\right)\E_{\{\a_E^V\}}(\tau_0)\G^A_{\{\a^V_E\}}\left(^{\mu_{\varphi_1}\ldots\mu_{\varphi_n}}_{\k_1\;\ldots\;\k_n}\right)\;.
\end{equation}
\end{FramedBox}
The complete boundary correlator is then given by the sum over all values of the set $\{\a_E^V\}$:
\begin{equation}
    \C^{\varphi_1\ldots\varphi_n}_\mathcal{B}(\k_1,\ldots,\k_n) \equiv \sum_{\{\a^V_E\}}F^{\varphi_1\ldots\varphi_n}_{\{\a_E^V\}}\left(\k_1,\ldots,\k_n\right)\;.
\end{equation}
In most applications, we will consider the correlators of $n$ identical conformally coupled fields with $\mu_\varphi=i/2$, for which the Bessel function reads:
\begin{equation}\label{eq: Bessel CC field}
    K_{\frac{1}{2}}(z)=\sqrt{\frac{\pi}{2 z}} e^{-z}\;.
\end{equation}
In this case, the contribution from the external legs can be written as: 
\begin{equation}\label{eq: contribution external legs CC}
    \lim_{\tau_0\to 0}\E_{\{\a_E^V\}}(\tau_0) = \frac{H^{\frac{n(d-3)}{2}}}{2^{\frac{n}{2}}}\frac{(-\tau_0)^{\frac{n(d-1)}{2}}}{(k_1\ldots k_n)^{\frac{1}{2}}}e^{-\frac{i\pi(d-1)}{4}\sum_{j=1}^{n_E} n_j\a_j}\;, \quad \textrm{c.c. external legs}
\end{equation}

\section{Applications with Cubic Interactions}
\label{sec:examples}
Let us see some applications of these Feynman rules in three situations.

\subsection{Boundary Three-Point Function}
Let us start with the boundary three-point function of conformally coupled scalars. In a $\frac{1}{3!}\lambda\varphi^3$ self-interacting theory, the set $\{a^V_E\}$ only contains a single vertex $\a$ and the different indices are $n=3$, $n_E=1$ and $n_{j=1}=3$. The only relevant diagrams are:
\begin{equation}
    \begin{aligned}
        F^{\varphi}_{\a}(\k_1,\k_2\,\k_3) =\begin{tikzpicture}[baseline={(0,0)}]
            \draw[thick,black] (0,1) to (0,0);
            \draw[thick,black] (-0.71,-0.71) to (0,0);
            \draw[thick,black] (0.71,-0.71) to (0,0);
            \filldraw[color=black,fill=white] (-0.07,1) rectangle ++(4pt,4pt);
            \filldraw[color=black,fill=white] (-0.83,-0.83) rectangle ++(4pt,4pt);
            \filldraw[color=black,fill=white] (0.71,-0.83) rectangle ++(4pt,4pt);
            \filldraw[color=black,fill=black] (0,0) circle (2pt) node[left] {$\a$};
            \node at (0,1.4) {$\mu_\varphi,\k_1$};
            \node at (-0.71,-1.1) {$\mu_\varphi,\k_2$};
            \node at (0.71,-1.1) {$\mu_\varphi,\k_3$};
        \end{tikzpicture}
        = (2\pi)^d\delta^{(d)}(\k_1+\k_2+\k_3)\E_\a(\tau_0)\G^A_{\a}\left(^{\mu_\varphi\mu_\varphi\mu_\varphi}_{k_1\;k_2\;k_3}\right)\;.
    \end{aligned}
\end{equation}
The contribution from the external legs is simply given by equation~\eqref{eq: contribution external legs CC} with $n=3$:
\begin{equation}
    \lim_{\tau_0\to 0}\E_\a(\tau_0) =\frac{H^{\frac{3}{2}(d-3)}}{2\sqrt{2}\sqrt{k_1k_2k_3}}(-\tau_0)^{\frac{3}{2}(d-1)}e^{-\frac{3i\a\pi(d-1)}{4}}\;.
\end{equation}
On the other hand, the on-shell amputated diagram is fully determined by the vertex:
\begin{equation}
    \G^{A}_\a\left(^{\mu_\varphi\mu_\varphi\mu_\varphi}_{k_1\;k_2\;k_3}\right)=\begin{tikzpicture}[baseline={(0,0)}]
        \draw[thick,black] (0,1) to (0,0);
            \draw[thick,black] (-0.71,-0.71) to (0,0);
            \draw[thick,black] (0.71,-0.71) to (0,0);
            \filldraw[color=black,fill=black] (0,0) circle (2pt) node[left] {$\a$};
            \node at (0,1.4) {$\mu_\varphi,\k_1$};
            \node at (-0.71,-1.1) {$\mu_\varphi,\k_2$};
            \node at (0.71,-1.1) {$\mu_\varphi,\k_3$};
    \end{tikzpicture}= -\frac{i\a H^{\frac{d+1}{2}}\lambda e^{\frac{i\a\pi d}{2}}}{\pi^{\frac{3}{2}}}\I^{\mu_\varphi\mu_\varphi\mu_\varphi}_{k_1k_2k_3}\;.
\end{equation}
One can now use the very simple form of the Bessel function~\eqref{eq: Bessel CC field} to evaluate the vertex function as:
\begin{equation}
     \mathcal{I}^{\mu_\varphi\mu_\varphi\mu_\varphi}_{k_1k_2k_3}=\frac{\pi^{\frac{3}{2}}}{2\sqrt{2}}\frac{k_t^{\frac{3-d}{2}}}{\sqrt{k_1 k_2 k_3}}\Gamma\left(\frac{d-3}{2}\right)\;,
\end{equation}
where we defined the total energy as $k_t=k_1+k_2+k_3$. Notice that the vertex function is singular at $d=3$, highlighting the need of using dimensional regularisation. As we now show, this divergence is canceled by destructive interferences among the two channels, so that the full correlator remains finite.
\vskip 4pt
Omitting the momentum-conserving delta function, the contribution from the channel $\a$ is:
\begin{equation}
    F^{\mu_\varphi}_\a(\k_1,\k_2,\k_3)=-\frac{i\a\lambda H^{2(d-2)}k_t^{\frac{3-d}{2}}}{8k_1k_2k_3}e^{\frac{i\a\pi(3-d)}{4}}(-\tau_0)^{\frac{3}{2}(d-1)}\Gamma\left(\frac{d-3}{2}\right)\;.
\end{equation}
The full boundary three-point function is now given by the sum of the two channels:
\begin{equation}
    \C_\mathcal{B}(\k_1,\k_2,\k_3)=-\frac{\lambda}{4}H^{2(d-2)}(-\tau_0)^{\frac{3}{2}(d-1)}\frac{k_t^{\frac{3-d}{2}}}{k_1k_2k_3}\sin\left(\frac{\pi(d-3)}{4}\right)\Gamma\left(\frac{d-3}{2}\right)\;.
\end{equation}
One can now take the limit $d\to 3$ to obtain a finite result in agreement with the standard in-in computation~\cite{Creminelli:2011mw}: 
\begin{equation}
    \C_\mathcal{B}(\k_1,\k_2,\k_3)=\frac{\lambda H^2\pi}{8}\frac{\tau_0^3}{k_1k_2k_3}\;.
\end{equation}

\subsection{Boundary Four-Point Function}\label{subsection_single_exchange}

Let us now consider a two-field theory with the cubic interaction:
\begin{equation}
    \L_I = -\frac{g}{2}\varphi^2\chi\;,
\end{equation}
where the $1/2$ factor implies that there are no combinatorial factors in the vertex. As before, we take external fields $\varphi$ to be conformally coupled. We consider the second field $\chi$ to lie in the principal series, with other UIRs simply following by analytical continuation. The tree-level diagrams contributing to the boundary four-point function are labeled by a set of two external vertices $\{\a^V_E\}=\{\a,\b\}$ and are of the form:
\begin{equation}
    \begin{aligned}
        F_{\a\b}^{\mu_\varphi}(\left\{\k_j\right\}) = &\begin{tikzpicture}[baseline={(0,0)}]
            \draw[thick,black] (-1.,0.6) to (-0.5,0);
            \draw[thick,black] (-1.,-0.6) to (-0.5,0);
            \draw[thick,black] (1.,0.6) to (0.5,0);
            \draw[thick,black] (1.,-0.6) to (0.5,0);
            \draw[thick,pyblue] (-0.5,0) to (0.5,0);
            \filldraw[color=black,fill=black] (-0.5,0) circle (2pt);

            \node at (-0.5,-0.4) {$\a$};

            \node at (0.5,-0.36) {$\b$};
            \filldraw[color=black,fill=black] (0.5,0) circle (2pt);
            \filldraw[color=black,fill=white] (1,-0.75) rectangle ++(4pt,4pt);
            \filldraw[color=black,fill=white] (1,0.6) rectangle ++(4pt,4pt);
            \filldraw[color=black,fill=white] (-1.14,-0.75) rectangle ++(4pt,4pt);
            \filldraw[color=black,fill=white] (-1.14,0.6) rectangle ++(4pt,4pt);
            \node[black] at  (-1.14,1) {$\mu_\varphi,\k_1$};
            \node[black] at  (-1.14,-1) {$\mu_\varphi,\k_2$};
            \node[black] at  (1.14,1) {$\mu_\varphi,\k_3$};
            \node[black] at  (1.14,-1) {$\mu_\varphi,\k_4$};
            \node[pyblue] at  (0,0.5) {$\mu_\chi,\k_I$};
        \end{tikzpicture}
        & = \E_{\a\b}(\tau_0)\G^A_{\a\b}\left(^{\mu_\varphi\mu_\varphi\mu_\varphi\mu_\varphi}_{\k_1\,\k_2\, \k_3\,\k_4}\right)\;,
    \end{aligned}
    \label{eq: single_exch_diag}
\end{equation}
where we omitted the momentum-conserving delta function for simplicity. The contribution from the external legs is simply given by applying~\eqref{eq: contribution external legs CC} with $n_E=2$, $n_1=n_2=2$ and $n=4$:
\begin{equation}
    \lim_{\tau_0\to 0}\E_{\a\b}(\tau_0) = \frac{H^{2(d-3)}}{4}\frac{(-\tau_0)^{2(d-1)}}{\sqrt{k_1k_2k_3k_4}}e^{-\frac{i\pi(d-1)}{2}(\a+\b)}\;.
    \label{eq: single_exch_prefactor}
\end{equation}
On the other hand, the amputated diagram is given by the spectral integration :
\begin{equation}
    \G^A_{\a\b}\left(^{\mu_\varphi\mu_\varphi\mu_\varphi\mu_\varphi}_{\k_1\,\k_2\, \k_3\,\k_4}\right)
    = -\frac{\a\b g^2 H^{d-1}}{\pi^3}e^{\frac{i(\a+\b)\pi d}{2}}\int\displaylimits^\infty_{-\infty}\d\mu\,\N_\mu\,\I^{\mu_\varphi\mu_\varphi\mu}_{k_1\;k_2\;k_I}\,\I^{\mu\;\mu_\varphi\mu_\varphi}_{k_I\;k_3\;k_4}\,\G^{\mu_\chi}_{\a\b}(\mu)\;,
    \label{eq_amputated_single_exchange}
\end{equation}
where $\G^{\mu_\chi}_{\a\b}$ is the KLF propagator of the field $\chi$ and $\k_I=(\k_1+\k_2)$ is the exchanged momentum. Like in the previous example, the simple form of the Bessel function~\eqref{eq: Bessel CC field} for a conformally coupled field allows us to evaluate the vertex function as:
\begin{equation}
    \mathcal{I}^{\mu_\varphi\mu_\varphi\mu}_{k_1\;k_2\;k_I} = \frac{\pi^2}{8}\frac{2^{\frac{d}{2}}\Gamma\left(\frac{d-2}{2}\right)}{\sqrt{k_1 k_2}k_I^{\frac{d-2}{2}}}\frac{C_{i\mu-\frac{d-2}{2}}^{\left(\frac{d-2}{2}\right)}(u^{-1})}{i\sinh\left(\pi\mu+\frac{i\pi d}{2}\right)}\;,
    \label{vertex_function_I}
\end{equation}
where $C_\nu^{(\alpha)}(z)$ is the Gegenbauer function (see \eqref{A_Legendre_to_Gegenbauer}) and we defined the two cross-ratios $u$ and $v$ as:
\begin{equation}
    u=\frac{k_I}{k_1+k_2}\;,\quad v=\frac{k_I}{k_1+k_2}\;.
\end{equation}
In the physical region, one has $0\leq u,v\leq 1$.

\paragraph{Factorised contributions.}
When the two SK indices $\a$ and $\b$ are opposite, the KLF propagator is given by~\eqref{eq: KLF wightman function}. Consequently, the corresponding amputated diagram is factorised: 
\begin{equation}
\begin{aligned}
    \G^A_{-\a\a}\left(^{\mu_\varphi\mu_\varphi\mu_\varphi\mu_\varphi}_{\k_1\;\k_2\;\k_3\;\k_4}\right) &= \frac{g^2H^{d-1}}{\pi^3}\I^{\mu_\varphi\mu_\varphi\mu_\chi}_{k_1\;k_2\;k_I}\I^{\mu_\chi\mu_\varphi\mu_\varphi}_{k_I\;k_3\;k_4}\\
    & = -\frac{g^2 H^{d-1} \pi}{2^{6-d}}\frac{\Gamma\left(\frac{d-2}{2}\right)^2}{\sqrt{k_1k_2k_3k_4}k_I^{d-2}}\frac{C_{i\mu_\chi-\frac{d-2}{2}}^{\left(\frac{d-2}{2}\right)}(u^{-1})C_{i\mu_\chi-\frac{d-2}{2}}^{\left(\frac{d-2}{2}\right)}(v^{-1})}{\sinh^2\left(\pi\mu_\chi+\frac{i\pi d}{2}\right)}\;.
\end{aligned}
\end{equation}
This expression is analytic in the dimension $d$. The spurious second-order pole at $d=2$ coming from the Gamma factor is cancelled by some zeros in the Gegenbauer function. Since this expression does not depend on $\a$, one can simply sum up the $+-$ and $-+$ contributions as:
\begin{equation}
    F_{+-}^{\mu_\varphi}+F_{-+}^{\mu_\varphi} = -\frac{g^2 H^{3d-7} \pi}{2^{7-d}}\frac{(-\tau_0)^{2(d-1)}\Gamma\left(\frac{d-2}{2}\right)^2}{k_1k_2k_3k_4 k_I^{d-2}}\frac{C_{i\mu_\chi-\frac{d-2}{2}}^{\left(\frac{d-2}{2}\right)}(u^{-1})C_{i\mu_\chi-\frac{d-2}{2}}^{\left(\frac{d-2}{2}\right)}(v^{-1})}{\sinh^2\left(\pi\mu_\chi+\frac{i\pi d}{2}\right)}\;.
\end{equation}
\vskip 4pt
\paragraph{Nested $++$ component: KLF integral.}
The case where $\a=\b$ is more complicated since one has to carry out a non-trivial spectral integration. However, because the function $\mu\mapsto \mathcal{I}^{\mu_\varphi\mu_\varphi\mu}_{k_1\;k_2\;k_I}$ given in \eqref{vertex_function_I} is meromorphic, the spectral integral can be computed by collecting its poles. Analytical properties of $\mathcal{I}^{\mu_\varphi\mu_\varphi\mu}_{k_1\;k_2\;k_I}$ and other useful relations are detailed in Appendices \ref{sec:special functions} and \ref{appendix_single_exch}. We focus on the $\G^A_{++}$ component, since $\G^A_{--}$ can be obtained by complex conjugation.

From the reduced propagator \eqref{eq: KLF (anti-)time ordered} and the vertex functions \eqref{vertex_function_I}, one finds:
\begin{equation}
\begin{aligned}
    \G^A_{++} \left(^{\mu_\varphi\mu_\varphi\mu_\varphi\mu_\varphi}_{\k_1\;\k_2\;\k_3\;\k_4}\right)=
    &\frac{g^2 H^{d-1} e^{\frac{i\pi(d+1)}{2}}}{2^{6-d}}
    \frac{\Gamma\left(\frac{d-2}{2}\right)^2}{\sqrt{k_1 k_2 k_3 k_4}k_I^{d-2}}\\
    &\times
    \int\displaylimits^\infty_{-\infty}\d\mu\,\frac{\mu\,\sinh\left(\pi\mu\right)}{\sinh\left(\pi\mu+\frac{i\pi d}{2}\right)^2}
    \frac{C_{i\mu-\frac{d-2}{2}}^{\left(\frac{d-2}{2}\right)}(u^{-1})C_{i\mu-\frac{d-2}{2}}^{\left(\frac{d-2}{2}\right)}(v^{-1})}{(\mu^2-\mu_\chi^2)_{i\epsilon}}
    \,.
    \end{aligned}
    \label{eq: G++_explicit}
\end{equation}
As Gegenbauer functions have an exponentially growing large $\mu$ asymptotic behaviour outside the real axis (see Appendix \ref{appendix_single_exch}), the writing \eqref{eq: G++_explicit} is not convenient if one wants to perform a contour integration to use the residue theorem. Hence, we rewrite the integrand in terms of Legendre functions of the second kind (defined in \eqref{eq_def_legendre_Q}, with the asymptotic behaviour given in \eqref{large_mu_Q}), using the connection formula \eqref{connection_formula_C_Q}. The correlator \eqref{eq: G++_explicit} can thus be expressed in terms of the functions $Q_{\pm i\mu-\frac{1}{2}}^{-\frac{d-3}{2}}$: 
\begin{equation}
\begin{aligned}
    \G^A_{++} \left(^{\mu_\varphi\mu_\varphi\mu_\varphi\mu_\varphi}_{\k_1\;\k_2\;\k_3\;\k_4}\right)&=
    \frac{g^2 H^{d-1} e^{-\frac{i\pi(d+1)}{2}}}{4\sqrt{k_1 k_2 k_3 k_4}k_I^{d-2}}
    \left(u^{-2}-1\right)^{\frac{3-d}{4}}
    \left(v^{-2}-1\right)^{\frac{3-d}{4}}\\
    &\times\left(\hat{F}_{++}^0+\hat{F}_{++}^{1} \right)
    \,,
    \end{aligned}
    \label{eq: decomposition_F++}
\end{equation}
where
\begin{equation}
    \hat{F}_{++}^0\equiv\frac{1}{\pi}\int\displaylimits^\infty_{-\infty}\d\mu\,\frac{\mu}{\sinh\left(\pi\mu\right)}
    \frac{Q_{i\mu-\frac{1}{2}}^{\frac{d-3}{2}}\left(u^{-1}\right)
    Q_{i\mu-\frac{1}{2}}^{\frac{d-3}{2}}\left(v^{-1}\right)}{(\mu^2-\mu_\chi^2)_{i\epsilon}}\,,
    \label{F++0}
\end{equation}
\begin{equation}
    \hat{F}_{++}^{1}\equiv-\frac{1}{\pi}\int\displaylimits^\infty_{-\infty}\d\mu\,\frac{\mu}{\sinh\left(\pi\mu\right)}
    \frac{Q_{i\mu-\frac{1}{2}}^{\frac{d-3}{2}}\left(u^{-1}\right)
    Q_{-i\mu-\frac{1}{2}}^{\frac{d-3}{2}}\left(v^{-1}\right)}{(\mu^2-\mu_\chi^2)_{i\epsilon}}\,.
    \label{F++1}
\end{equation}
and we used that $\hat{F}_{++}^0$ and $\hat{F}_{++}^{1}$ are symmetric under $u\leftrightarrow v$ (for the latter, this can be checked by changing variables $\mu \leftrightarrow - \mu$ in the integrand).

We now focus on the computation of $\hat{F}_{++}^0$ and $\hat{F}_{++}^{1}$. Different terms in the above integrands have poles that contribute to $\G^A_{++}$. The analytic structures of the integrands are shown in Fig.~\ref{fig: KLF modes analytical structure} and are detailed in Appendix \ref{appendix_single_exch}. In the following, we call signal residues the ones that come from $\mu\mapsto\left(\mu^2-\mu_{\chi}^2\right)_{i\epsilon}^{-1}$ in $\hat{F}_{++}^{0}$ and $\hat{F}_{++}^{1}$, and we will denote $\hat{F}_{++}^{S}$ their two contributions.
All other residues in $\hat{F}_{++}^{0}$ and $\hat{F}_{++}^{1}$ are called background residues, and we gather them in the term $\hat{F}_{++}^{B}$ such that
\begin{equation}
    \hat{F}_{++}\equiv \hat{F}_{++}^{0}+\hat{F}_{++}^{1}=\hat{F}_{++}^{B}+\hat{F}_{++}^{S}\,.
    \label{eq: background+signal}
\end{equation}
\begin{figure}[h!]
\hspace{-1cm}
\begin{subfigure}[h!]{0.4\textwidth}
        \hspace{-0.25cm}
    	\begin{tikzpicture}[scale = 2]

        \draw[black, ->] (-1.8,0) -- (1.8,0) coordinate (xaxis);
		\draw[black, ->] (0,-1.8) -- (0,1.8) coordinate (yaxis);
		\node at (2.1, 0) {$\text{Re}(\mu)$};
		\node at (0, 1.95) {$\text{Im}(\mu)$};
        
        \draw[pyred, fill = pyred] (0, 0.25) circle (.03cm);
		\draw[pyred, fill = pyred] (0, 0.75) circle (.03cm);
		\draw[pyred, fill = pyred] (0, 1.25) circle (.03cm);
        \node at (-.7, 0.8) {\textcolor{pyred}{\footnotesize$Q_{i\mu-\frac{1}{2}}^{\frac{d-3}{2}}\left(u^{-1}\right)$}};
        \node at (0.7, 0.8) {\textcolor{pyred}{\footnotesize$Q_{i\mu-\frac{1}{2}}^{\frac{d-3}{2}}\left(v^{-1}\right)$}};

        \draw[pyblue, fill = pyblue] (0, -0.5) circle (.03cm);
		\draw[pyblue, fill = pyblue] (0, -1) circle (.03cm);
		\draw[pyblue, fill = pyblue] (0, -1.5) circle (.03cm);
        \draw[pyblue, fill = pyblue] (0, 0.5) circle (.03cm);
		\draw[pyblue, fill = pyblue] (0, 1) circle (.03cm);
		\draw[pyblue, fill = pyblue] (0, 1.5) circle (.03cm);
        \node at (0.5, -1.05) {\textcolor{pyblue}{$\frac{\mu}{\sinh\left(\pi\mu\right)}$}};

        \draw[pyorange, fill = pyorange] (-1.2, -0.2) circle (.03cm);
		\draw[pyorange, fill = pyorange] (-1.2, .2) circle (.03cm);
		\draw[pyorange, fill = pyorange] (1.2, -.2) circle (.03cm);
        \draw[pyorange, fill = pyorange] (1.2, 0.2) circle (.03cm);
        \node at (-0.9, -.5) {\footnotesize\textcolor{pyorange}{$\left(\mu^2-\mu_\chi^2\right)_{i\varepsilon}^{-1}$}};
        
        \draw[xshift=0,pyblue!80!black,decoration={markings,mark=between positions 0.1 and 1 step 0.2 with \arrow{>}},postaction={decorate}] (-1.70,0) -- (1.70,0) arc (0:-180:1.70);
    \end{tikzpicture}
\end{subfigure}
\hfill
\begin{subfigure}[h!]{0.4\textwidth}
    \hspace{-1.5cm}
    	\begin{tikzpicture}[scale = 2]

        \draw[black, ->] (-1.8,0) -- (1.8,0) coordinate (xaxis);
		\draw[black, ->] (0,-1.8) -- (0,1.8) coordinate (yaxis);
		\node at (2.1, 0) {$\text{Re}(\mu)$};
		\node at (0, 1.95) {$\text{Im}(\mu)$};
        
        \draw[pyred, fill = pyred] (0, 0.25) circle (.03cm);
		\draw[pyred, fill = pyred] (0, 0.75) circle (.03cm);
		\draw[pyred, fill = pyred] (0, 1.25) circle (.03cm);
        \node at (.7, 0.8) {\textcolor{pyred}{\footnotesize$Q_{i\mu-\frac{1}{2}}^{\frac{d-3}{2}}\left(u^{-1}\right)$}};

        \draw[pyred, fill = pyred] (0, -0.25) circle (.03cm);
		\draw[pyred, fill = pyred] (0, -0.75) circle (.03cm);
		\draw[pyred, fill = pyred] (0, -1.25) circle (.03cm);
        \node at (0.7, -0.75) {\textcolor{pyred}{\footnotesize$Q_{-i\mu-\frac{1}{2}}^{\frac{d-3}{2}}\left(v^{-1}\right)$}};

        \draw[pyblue, fill = pyblue] (0, -0.5) circle (.03cm);
		\draw[pyblue, fill = pyblue] (0, -1) circle (.03cm);
		\draw[pyblue, fill = pyblue] (0, -1.5) circle (.03cm);
        \draw[pyblue, fill = pyblue] (0, 0.5) circle (.03cm);
		\draw[pyblue, fill = pyblue] (0, 1) circle (.03cm);
		\draw[pyblue, fill = pyblue] (0, 1.5) circle (.03cm);
        \node at (-0.5, -1.05) {\textcolor{pyblue}{$\frac{\mu}{\sinh\left(\pi\mu\right)}$}};

        \draw[pyorange, fill = pyorange] (-1.2, -0.2) circle (.03cm);
		\draw[pyorange, fill = pyorange] (-1.2, .2) circle (.03cm);
		\draw[pyorange, fill = pyorange] (1.2, -.2) circle (.03cm);
        \draw[pyorange, fill = pyorange] (1.2, 0.2) circle (.03cm);
        \node at (-0.9, -.5) {\footnotesize\textcolor{pyorange}{$\left(\mu^2-\mu_\chi^2\right)_{i\varepsilon}^{-1}$}};
        
        \draw[xshift=0,pyblue!80!black,decoration={markings,mark=between positions 0.1 and 1 step 0.2 with \arrow{>}},postaction={decorate}] (-1.70,0) -- (1.70,0) arc (0:-180:1.70);
    \end{tikzpicture}
\end{subfigure}
\caption{Analytic structure in the $\mu$ complex plane of the integrands of $\hat{F}_{++}^0$, Eq.~\eqref{F++0} (left) and $\hat{F}_{++}^{1}$, Eq.~\eqref{F++1} (right). To perform the integrals, we always close the contour with a half-circle in the lower half-plane for $\hat{F}_{++}^0$. For $\hat{F}_{++}^1$, the contour is closed in the lower (resp. upper) half-plane for $u < v$ (resp. $u > v$).}
\label{fig: KLF modes analytical structure}
\end{figure}
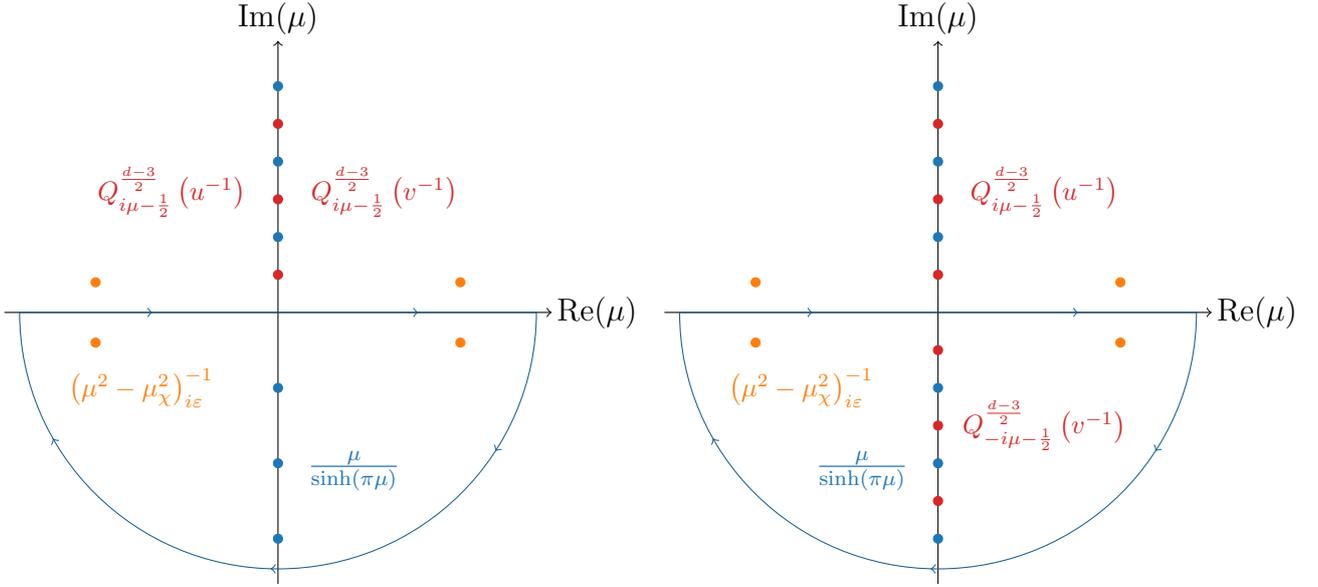 

\paragraph{Integration contour.} In the $\mu$-complex plane, we complete the integration contour with a half-circle, with a radius that is sent to infinity. To choose whether we close the contour in the upper or lower half-plane, let us examine the asymptotic behaviours of the integrands at $\mu\rightarrow+\infty$.
From the knowledge of the asymptotic expansion of $ Q_{\pm i\mu-\frac{d-2}{2}}^{\frac{d-3}{2}}(x)$ \eqref{large_mu_Q}, we have in the integrand of $\hat{F}_{++}^0$ \eqref{F++0}:
\begin{equation}
    Q_{i\mu-\frac{1}{2}}^{\frac{d-3}{2}}(u^{-1})\,
    Q_{i\mu-\frac{1}{2}}^{\frac{d-3}{2}}(v^{-1}) 
    \underset{\mu\rightarrow+\infty}{\propto}
    e^{- i\mu(\xi_u+\xi_v)}\,,
\end{equation}
where we defined $\xi_u\,,\,\xi_v > 0$ such that $u^{-1}\equiv\cosh(\xi_u)$ and $v^{-1}\equiv\cosh(\xi_v)$. As for the factor $\mu/\sinh(\pi\mu)$, it decays exponentially fast outside the imaginary axis, and simply oscillates there. As a result, for $\hat{F}_{++}^0$, we close the contour in the lower half-plane. 
For the Legendre functions entering $\hat{F}_{++}^{1}$ \eqref{F++1}, we have
\begin{equation}
    Q_{i\mu-\frac{1}{2}}^{\frac{d-3}{2}}(u^{-1})\,
    Q_{-i\mu-\frac{1}{2}}^{\frac{d-3}{2}}(v^{-1}) 
    \underset{\mu\rightarrow+\infty}{\propto}
    e^{- i\mu(\xi_u-\xi_v)}\,.
\end{equation}
Here, the contour prescription thus depends on the ordering of $u$ and $v$. In the following we choose $u<v$, and then also close the contour in the lower half-plane of $\mu$. 

Equipped with this contour prescription for $\hat{F}_{++}$, which is summarised in Fig.~\ref{fig: KLF modes analytical structure}, we can collect the residues coming from $\hat{F}_{++}^0$ and $\hat{F}_{++}^{1}$.
Let us start by the contribution coming from background residues, i.e.~$\hat{F}_{++}^{B}$.
Here, it appears that we have to make a distinction according to $d$.

\paragraph{Background residues for $d \in \mathbb{C} \setminus (2 \mathbb{Z})$.} 
When $d$ is a generic complex quantity that is not an even integer, the background residues in the integrals \eqref{F++0} and \eqref{F++1} come from poles of $\textcolor{pyblue}{\mu/\sinh(\pi\mu)}$ and $\textcolor{pyred}{Q_{\pm i\mu-\frac{1}{2}}^{\frac{d-3}{2}}\left(x\right)}$
(with $x=u^{-1},v^{-1}$) along the imaginary axis (see Fig.~\ref{fig: KLF modes analytical structure}). The different contributions are the following:
\begin{itemize}
\item \textbf{\textcolor{pyblue}{Poles from $\frac{\mu}{\sinh(\pi\mu)}$ in $\hat{F}_{++}^0$.}}
Given the contour prescription, the only background poles that contribute to $\hat{F}_{++}^0$ are the poles from $\mu\mapsto\frac{\mu}{\sinh(\pi\mu)}$ (see \eqref{residues_sinh_-1}) that are located in the lower half-plane:
\begin{equation}
    \hat{F}_{++}^0 \supset
    -\frac{2}{\pi} \sum_{n=0}^\infty (-1)^n\frac{n+1}{(n+1)^2+\mu_\chi^2} Q_{n+\frac{1}{2}}^{\frac{d-3}{2}}(u^{-1})Q_{n+\frac{1}{2}}^{\frac{d-3}{2}}(v^{-1})\,.
    \label{eq: poles_++0_sinh_d_odd}
\end{equation}
\item \textbf{\textcolor{pyblue}{Poles from $\frac{\mu}{\sinh(\pi\mu)}$ in $\hat{F}_{++}^{1}$.}}
Similarly, given the contour for $\hat{F}_{++}^{1}$, poles from $\frac{\mu}{\sinh(\pi\mu)}$ in the lower half-plane contribute as 
\begin{equation}
    \hat{F}_{++}^{1} \supset
    +\frac{2}{\pi} \sum_{n=0}^\infty (-1)^n\frac{n+1}{(n+1)^2+\mu_\chi^2} Q_{n+\frac{1}{2}}^{\frac{d-3}{2}}(u^{-1}) Q_{-n-\frac{3}{2}}^{\frac{d-3}{2}}(v^{-1})\,.
    \label{eq: poles_++1_sinh_d_odd}
\end{equation}
Now, we observe that the sum of these first two contributions \eqref{eq: poles_++0_sinh_d_odd}, \eqref{eq: poles_++1_sinh_d_odd} vanishes. Indeed, the expression obtained by the sum is a series that involves the difference $Q_{n+\frac{1}{2}}^{\frac{d-3}{2}}(v^{-1})-Q_{-n-\frac{3}{2}}^{\frac{d-3}{2}}(v^{-1})$, which vanishes according to the connection formula between associated Legendre functions for $d$ not an even integer \eqref{connection_formula_P_Q}-\eqref{connection_formula_P_Q_n_d_odd}. Therefore, the background term $\hat{F}_{++}^B$ will only be given by its last contribution.
\item\textbf{\textcolor{pyred}{Poles from $Q_{-i\mu-\frac{1}{2}}^{\frac{d-3}{2}}(v^{-1})$ in $\hat{F}_{++}^{1}$.}}
The last and only contribution to background poles is then from the residues \eqref{residues_Q_d_odd} of $\mu\mapsto Q_{-i\mu-\frac{1}{2}}^{\frac{d-3}{2}}(v^{-1})$ in $\hat{F}_{++}^{1}$ ($\mu\mapsto Q_{i\mu-\frac{1}{2}}^{\frac{d-3}{2}}(u^{-1})$ does not contribute since its poles are located in the upper half-plane). 
\end{itemize}
Finally, the background term is
\begin{equation}
\begin{aligned}
    \hat{F}_{++}^{B}&=e^{i\pi(d+1)}u^{d-2} \left[\left(u^{-2}-1\right)\left(v^{-2}-1\right)\right]^{\frac{d-3}{4}}\\
        &\times\sum_{n=0}^{+\infty}\frac{(-1)^n}{\left(n+\frac{d-2}{2}\right)^2+\mu_\chi^2}\left(\frac{u}{v}\right)^{n}\frac{\Gamma(n+d-2)}{\Gamma(n+1)}\\
        &\times\,_2F_1\left(
    \begin{matrix}
        \frac{1-n}{2}\,,\,-\frac{n}{2} \\ -n-\frac{d-4}{2}
    \end{matrix}\,,v^2\right)
    \,_2F_1\left(
    \begin{matrix}
        \frac{n+d-1}{2}\,,\,\frac{n+d-2}{2} \\ n+\frac{d}{2}
    \end{matrix}\,,u^2\right)
    \,.
    \end{aligned}
    \label{eq: F++B_d_odd}
\end{equation}

\paragraph{Background residues for $d$ even integer.}
When the the spatial dimension $d$ is even, the function $\mu\mapsto Q_{\pm i\mu-\frac{1}{2}}^{\frac{d-3}{2}}(x)$, $x>1$, is analytic in the complex $\mu$-plane, see Appendix \eqref{appendix_single_exch}. Consequently, the background term $\hat{F}_{++}^B$ is given by the sum of the two contributions \eqref{eq: poles_++0_sinh_d_odd} and \eqref{eq: poles_++1_sinh_d_odd}, which is now non-vanishing (except for a finite number of terms), due to the Legendre functions connection formula \eqref{eq_connection_P_Q_n_d_even} that holds when $d$ is an even integer. We then obtain\footnote{For numerical evaluation, it is convenient to use the form \eqref{eq_def_legendre_Q} for the function $Q_{n+\frac{1}{2}}^{\frac{d-3}{2}}$.}
\begin{equation}
    \hat{F}_{++}^B=+2 i \sum_{n=K-1}^{+\infty} (-1)^n \frac{\Gamma(n+1+K)}{\Gamma(n+2-K)}\frac{n+1}{(n+1)^2+\mu_\chi^2} P_{n+\frac{1}{2}}^{-\frac{d-3}{2}}\left(u^{-1}\right)\,Q_{n+\frac{1}{2}}^{\frac{d-3}{2}}\left(v^{-1}\right)
    \,,
    \label{eq: F++B_d_even}
\end{equation}
where $K\equiv\frac{d-2}{2}$, $K\in\mathbb{N}^*$ when $d$ is even. Remarkably, it turns out that Eq.~\eqref{eq: F++B_d_odd} has a smooth limit when $d$ approaches an even integer, and that this limit coincides with \eqref{eq: F++B_d_even}, which was not guaranteed a priori.

\paragraph{Signal residues.}
We now collect the signal residues, a reasoning that does not depend on $d$.
Using the $i\epsilon$-prescription \eqref{eq: i epsilon prescription}, which has poles and residues given in \eqref{eq_i_eps_residues}, we obtain $\hat{F}_{++}^{S}$ from the following two contributions:
\begin{itemize}
\item \textbf{\textcolor{pyorange}{Poles from $\frac{1}{\left(\mu^2-\mu_{\chi}^2\right)_{i\epsilon}}$ in $\hat{F}_{++}^0$.}}
From the contour prescription, the two signal poles that are located in the lower half-plane contribute to $\hat{F}_{++}^0$:
\begin{equation}
\begin{aligned}
    \hat{F}_{++}^0 \supset -\frac{i}{2\sinh(\pi\mu_{\chi})^2}
    &\biggl(e^{\pi\mu_{\chi}}\,Q_{i\mu_{\chi}-\frac{1}{2}}^{\frac{d-3}{2}}(u^{-1})Q_{i\mu_{\chi}-\frac{1}{2}}^{\frac{d-3}{2}}(v^{-1})\\
    &+e^{-\pi\mu_{\chi}}\,Q_{-i\mu_{\chi}-\frac{1}{2}}^{\frac{d-3}{2}}(u^{-1})
    Q_{-i\mu_{\chi}-\frac{1}{2}}^{\frac{d-3}{2}}(v^{-1})\biggr)\,.
    \label{eq: signal_poles_++0}
    \end{aligned}
\end{equation}
\item \textbf{\textcolor{pyorange}{Poles from $\frac{1}{\left(\mu^2-\mu_{\chi}^2\right)_{i\epsilon}}$ in $\hat{F}_{++}^{1}$.}}
The signal poles coming from $\hat{F}_{++}^{1}$ are those located in the lower half-plane: 
\begin{equation}
\begin{aligned}
    \hat{F}_{++}^{1} \supset \frac{i}{2\sinh(\pi\mu_{\chi})^2}
    &\biggl(e^{\pi\mu_{\chi}}\,Q_{-i\mu_{\chi}-\frac{1}{2}}^{\frac{d-3}{2}}(v^{-1})Q_{i\mu_{\chi}-\frac{1}{2}}^{\frac{d-3}{2}}(u^{-1})\\
    &+e^{-\pi\mu_{\chi}}\,Q_{i\mu_{\chi}-\frac{1}{2}}^{\frac{d-3}{2}}(v^{-1})
    Q_{-i\mu_{\chi}-\frac{1}{2}}^{\frac{d-3}{2}}(u^{-1})\biggr)\,.
    \end{aligned}
    \label{eq: signal_poles_++1}
\end{equation} 
\end{itemize}
The sum of these two terms can be simplified using the Legendre functions connection formula \eqref{connection_formula_P_Q} and the analytic continuation formula for the Legendre function of the first kind \eqref{eq_analytic_cont_P}. Finally, the signal term $\hat{F}_{++}^S$ is 
\begin{equation}
    \hat{F}_{++}^{S}=-\frac{e^{i\pi(d-3)}}{2}\left(\Gamma\left(\frac{d-2}{2}\pm i\mu_\chi\right)\right)^2 P_{i\mu_\chi-\frac{1}{2}}^{-\frac{d-3}{2}}(e^{+i\pi}u^{-1})\,P_{i\mu_\chi-\frac{1}{2}}^{-\frac{d-3}{2}}(v^{-1})\,.
    \label{eq: F++S}
\end{equation}

\paragraph{Final result.}
Gathering the expressions of the prefactor $\E_{++}(\tau_0)$ \eqref{eq: single_exch_prefactor} and of the amputated diagram $\G^A_{++}$ \eqref{eq: decomposition_F++}, the final result for $F_{++}$ \eqref{eq: single_exch_diag} is given by 
\begin{FramedBox}
    \begin{equation}\begin{aligned}
    F_{++}^{\mu_\varphi}(\left\{\k_j\right\}) &= \frac{\bar{g}^2 H^{2d-2} (-\tau_0)^{2(d-1)} e^{-\frac{3i\pi}{2}(d-1)}}{16 k_1k_2k_3k_4 k_I^{d-2}}
    \left(u^{-2}-1\right)^{\frac{3-d}{4}}
    \left(v^{-2}-1\right)^{\frac{3-d}{4}}\\
    &\times\left(\hat{F}_{++}^B+\hat{F}_{++}^{S} \right)
    \;,
    \end{aligned}
    \label{eq: final_result_F++}
\end{equation}
\end{FramedBox}
where $\hat{F}_{++}^B$ is given by \eqref{eq: F++B_d_odd} (or \eqref{eq: F++B_d_even} for $d$ an even integer), $\bar{g} = H^{-\frac{5-d}{2}}g$ is the dimensionless coupling, and $\hat{F}_{++}^S$ is given in \eqref{eq: F++S}.
The result found in \cite{Werth:2024mjg} is recovered for $d=3$. 

\subsection{Bulk Two-Point Function at One Loop}\label{sec: loop two point}
In this last section, we consider a bulk correlation function with loop corrections to the propagator. For definiteness, we look at an interaction of the kind $g\varphi\chi_1\chi_2$ and define the dimensionless coupling $\bar{g} = H^{-\frac{5-d}{2}}g$. The one-loop correction to the $\{\a_1,\a_2\}$ propagator of $\varphi$ is given by the sum over vertices $\{\a^V_E\}=\{\b_1,\b_2\}$ of the bubble diagram:
\begin{equation}\label{eq: self energy diagram}
    \begin{aligned}
        \G^{\mu_\varphi,1\;\textrm{loop}}_{\a_1\a_2}\left(^{\mu_1\mu_2}_{\k_1\k_2} \right) &= \sum_{\b_1,\b_2}\begin{tikzpicture}[baseline={(0,0)}]

        \draw[thick,black] (-1.5,0) to (-0.5,0);

        \draw[thick,black] (1.5,0) to (0.5,0);
        
        \draw [pyblue, thick] (-0.5,0) arc[start angle=180, end angle=0,radius=0.5];
        \draw [pyred, thick] (-0.5,0) arc[start angle=-180, end angle=0,radius=0.5];
        \filldraw[color=black,fill=black] (1.5,0) circle (2pt);
        \filldraw[color=black,fill=black] (-1.5,0) circle (2pt);
        \filldraw[color=black,fill=black] (0.5,0) circle (2pt);
        \filldraw[color=black,fill=black] (-0.5,0) circle (2pt);

        \node[black] at (-0.7,-0.2) {$\b_1$};

        \node[black] at (0.7,-0.2) {$\b_2$};

        \node[pyred] at (0,-0.7) {$\chi_1$};
        
        \node[pyblue] at (0,0.7) {$\chi_2$};
        
        \node[black] at (-1.6,0.3) {$\mu_1,\k_1$};

        \node[black] at (-1.6,-0.3) {$\a_1$};
        
        \node[black] at (1.6,0.3) {$\mu_2,\k_2$};

        \node[black] at (1.6,-0.3) {$\a_2$};
        
        \end{tikzpicture}\\
        & = (2\pi)^d\delta^{(d)}(\k_1+\k_2)\sum_{\b_1,\b_2}\frac{\G^{\mu_\varphi}_{\a_1\b_1}(\mu_1)}{H^2}\G^A_{\b_1\b_2}\left(^{\mu_1\mu_2}_{k_1k_2}\right)\frac{\G^{\mu_\varphi}_{\b_2\a_2}(\mu_2)}{H^2}\;.
    \end{aligned}
\end{equation}
First, let us look at the amputated diagram: 
\begin{equation}
\label{amputated-diagram-loop}    \G^A_{\a\b}\left(^{\mu_1\mu_2}_{k_1k_2}\right) = -\frac{\a\b g^2 H^{d-3}}{\pi^3}e^{\frac{i\pi(\a+\b)d}{2}}\int\displaylimits^\infty_{-\infty}\d\mu\N_{\mu}\d\mu'\N_{\mu'}\mathcal{Q}_{\mu_1\mu_2}(\mu,\mu')\G_{\a\b}^{\chi_1}(\mu)\G_{\a\b}^{\chi_2}(\mu')\;,
\end{equation}
where the loop momentum integral is:
\begin{equation}\label{eq: loop integral}
    \mathcal{Q}_{\mu_1\mu_2}(\mu,\mu') = \int\frac{\d^d\p}{(2\pi)^d}\I^{\mu_1\mu\mu'}_{k_1\; p|\k_1-\p|}\I^{\mu_2\mu\,\mu'}_{k_1\, p|\k_1-\p|}\;.
\end{equation}
Since the only scale of the integral is $k_1=k_2$ and the vertex functions scale as $\I \sim k_1^{-\frac{d}{2}}$, dimensional analysis implies that the integral~\eqref{eq: loop integral} is scale-invariant. The amputated diagram is therefore independent of $k_1$.  As we show in the following insert, the integral \eqref{eq: loop integral} can be evaluated thanks to the orthogonality relation among the Clebsch-Gordan coefficients of $\SO(1,d+1)$.
\begin{framed}
\paragraph{Loop Momentum Integral and tensor product of UIRs.} Despite being a $d$-dimensional integral over a product of Appell $F_4$ functions, the loop-integral~\eqref{eq: loop integral} can be fixed by symmetry. The argument is to see the vertex function $\I^{\mu\mu_1\mu_2}_{p \;k_1\;k_2}$ with three frequencies within the principal series as related to the decomposition of the tensor product space $\V_{\mu_1}\otimes \V_{\mu_2}$ along UIRs of $\SO(1,d+1)$ using that~\cite{Dobrev:1975ru,Dobrev:1977qv}:
\begin{equation}
    \P_{\mu_1}\otimes\P_{\mu_2}\sim \int_\mu\P_\mu\;.
\end{equation}
For this, one defines the $3\mu$ symbol as:
\begin{equation}\label{eq: 3mu symbol}
    \left(\begin{array}{crc}
        \mu_1 & \mu_2 & \mu \\
        \k_1 & \k_2 & \k
    \end{array}\right) \equiv \braket{\mu_1,\k_1;\mu_2,\k_2|\mu,\k}\;.
\end{equation}
By definition, the tensor product states are orthogonal to each other:
\begin{equation}
    \braket{\mu_1,\mu_2,\k_1,\k_2|\mu_1,\mu_2,\k_1',\k_2'} = \frac{\delta(0)^2}{\N_{\mu_1}\N_{\mu_2}}(2\pi)^d\delta^{(d)}(\k_1-\k_1')(2\pi)^d\delta^{(d)}(\k_2-\k_2')\;.
\end{equation}
The completeness relation in the tensor product state is therefore given by:
\begin{equation}
    \mathds{1}_{\mu_1\otimes\mu_2} \equiv\int\frac{\d^d\k_1}{(2\pi)^d}\frac{\d^d\k_2}{(2\pi)^d}\frac{\N_{\mu_1}\N_{\mu_2}}{\delta(0)^2}\ket{\mu_1,\k_1;\mu_2,\k_2}\bra{\mu_1,\k_1;\mu_2,\k_2}\;.
\end{equation}
Since the tensor product state is isomorphic to the full KLF space, one can sandwich this expression with two single particle states, yielding an orthogonality relation for the $3\mu$ symbols:
\begin{equation}\label{eq: orthogonality relations C}
    \delta\left(^{\mu\, \mu'}_{\k\,\k'}\right)=\frac{\N_{\mu_1}\N_{\mu_2}}{\left[\delta(0)\right]^2}\int\frac{\d^d\k_1}{(2\pi)^d}\frac{\d^d\k_2}{(2\pi)^d}\left(\begin{array}{crc}
        \mu_1 & \mu_2 & \mu \\
        \k_1 & \k_2 & \k
    \end{array}\right)^*\left(\begin{array}{crc}
        \mu_1 & \mu_2 & \mu' \\
        \k_1 & \k_2 & \k'
    \end{array}\right)\;.
\end{equation}
\vskip 4pt
Let us use this to evaluate the momentum integral~\eqref{eq: loop integral}. To this aim, we will relate the $3\mu$ symbol~\eqref{eq: 3mu symbol} to the cubic vertex function. The overlap between the state produced by a product of two free principal series fields $\chi_1$ and $\chi_2$ and the vacuum can be written:
\begin{equation}\label{eq:overlap tensor prod basis}
    \bra{\Omega}\chi_1(z,\x)\chi_2(z,\x)\ket{\mu_1,\k_1,\mu_2,\k_2} = c_{\chi_1}(\mu_1)c_{\chi_2}(\mu_2)\Phi^{\mu_1}_{\k_1}(z,\x)\Phi^{\mu_2}_{\k_2}(z,\x)\;,
\end{equation}
where the coefficients $c_{\chi_j}(\mu)$ have been defined at the beginning of section~\ref{sec: Harmonic functions and Position Space} in terms of the spectral density $\rho_{\chi_j}(\mu)$. In the case of free fields, we have:
\begin{equation}
    |c_\chi(\mu)|^2 =\frac{\hat{\delta}_{\mu_\chi}(\mu)}{H^{2}\N_\mu}\;.
\end{equation}
On the other hand, one can see the product of two fields as a composite operator $\chi_1\chi_2(z,\x)$ with wavefunction:
\begin{equation}
    \bra{\Omega}\chi_1\chi_2(z,x)\ket{\mu,\k} = c_{\chi_1\chi_2}(\mu)\Phi^{\mu}_{\k}(z,\x)\;.
\end{equation}
The modulus square of the coefficient $c_{\chi_1\chi_2}(\mu)$ is given by:
\begin{equation}
    |c_{\chi_1\chi_2}(\mu)|^2 = \frac{\rho_{\chi_1\chi_2}(\mu)}{H^2\N_\mu}\;,
\end{equation}
where the special density of a product of two fields is such that:
\begin{equation}\label{eq: definition rho chi1 chi2}
    W_{\mu_{\chi_1}}(\sigma^E)W_{\mu_{\chi_2}}(\sigma^E) = \int\displaylimits^\infty_{-\infty}\d\mu\rho_{\chi_1\chi_2}(\mu)W_\mu(\sigma^E)\;,
\end{equation}
and the function $W_\mu(\sigma^E)$ is defined in~\eqref{eq:definition of W}. Inverting equation~\eqref{eq: definition rho chi1 chi2} gives an explicit expression for the spectral density~\cite{Bros:2009bz,Loparco:2023rug}: 
\begin{equation}\label{eq: specral density of the product}
     \rho_{\chi_1\chi_2}(\mu)=\frac{H^{d-1}\N_\mu}{32\pi^{\frac{d}{2}+2}\Gamma\left(\frac{d}{2}\right)\Gamma\left(\frac{d}{2}\pm i\mu\right)}\prod_{\pm,\pm,\pm}\Gamma\left(\frac{\frac{d}{2}\pm i\mu\pm i\mu_{\chi_1}\pm i\mu_{\chi_2}}{2}\right)\;.
\end{equation}
Equivalent expressions can be found in terms of a $_7 F_6$ hypergeometric function, see e.g.~\cite{Marolf:2010zp,DiPietro:2021sjt,Xianyu:2022jwk}.
As explained in~\cite{Loparco:2023rug}, other UIRs in the density~\eqref{eq: specral density of the product} arise if there exists some integer $n$ such that $\frac{d}{2}+2n<i\mu_{\chi_1}+i\mu_{\chi_2} < d$, and should be taken into account exactly like in the CFT example in section~\ref{sec: Role of the other Series} by taking the residues of~\eqref{eq: specral density of the product}. We ignore this as we assumed $\chi_1$ and $\chi_2$ to be in the principal series.
\vskip 4pt
Inserting the resolution of identity~\eqref{eq: KLF resolution of indentity} into~\eqref{eq:overlap tensor prod basis} and using the orthogonality relation among the KLF harmonic functions yields:
\begin{equation}
   c_{\chi_1\chi_2}(\mu)\, \left(\begin{array}{crc}
        \mu_1 & \mu_2 & \mu \\
        \k_1 & \k_2 & \k
    \end{array}\right)^* = \frac{H^{\frac{d+1}{2}}}{\pi^{3/2}}c_{\chi_1}(\mu_1)c_{\chi_2}(\mu_2) 
    (2\pi)^d\delta^{(d)}(\k_1+\k_2-\k)\I^{\mu_1\mu_2\mu}_{k_1k_2\;k}\;.
\end{equation}
Plugging this into the orthogonality relation~\eqref{eq: orthogonality relations C} gives:
\begin{equation}
   \frac{\rho_{\chi_1\chi_2}(\mu)}{\N_{\mu}^2}\hat{\delta}_{\mu'}(\mu) = \frac{H^{d-1}}{\pi^3}\frac{\hat{\delta}_{\mu_{\chi_1}}(\mu_1)\hat{\delta}_{\mu_{\chi_2}}(\mu_2)}{\delta(0)^2}\int\frac{\d^d\p}{(2\pi)^d}\I^{\mu\mu_1\mu_2}_{k p|\k-\p|}\I^{\mu'\mu_1\mu_2}_{k\, p|\k-\p|}\;.
\end{equation}
We get rid of the delta functions on the right-hand side by evaluating it at $\mu_j=\mu_{\chi_j}$. After relabelling, this gives the following result for the loop momentum-integral \eqref{eq: loop integral}:
\begin{equation}
    \mathcal{Q}_{\mu_1\mu_2}(\mu,\mu')=\frac{\pi^3}{H^{d-1}}\frac{\rho_{\mu\mu'}(\mu_1)}{\N_{\mu_1}} \frac{\hat{\delta}_{\mu_1}(\mu_2)}{\N_{\mu_2}}\;,
\label{result-Q}    
\end{equation}
where $\rho_{\mu \mu'}$ denotes the spectral density of the product of two fields with frequencies $\mu$ and $\mu'$.
\end{framed}
Using \eqref{result-Q}, the amputated diagram \eqref{amputated-diagram-loop} can therefore be rewritten as a double layer spectral integral: 
\begin{equation}\label{eq: double spectral integral GA}
    \G^A_{\a\b}\left(^{\mu_1\mu_2}_{k_1k_2}\right)=-\frac{\a\b g^2 e^{\frac{i\pi(\a+\b) d}{2}}}{H^2}\frac{\hat{\delta}_{\mu_2}(\mu_1)}{\N_{\mu_1}^2}\int\displaylimits^\infty_{-\infty}\d\mu\,\N_\mu\,\d\mu'\,\N_{\mu'}\,\rho_{\mu\mu'}(\mu_1)\G_{\a\b}^{\chi_1}(\mu)\G_{\a\b}^{\chi_2}(\mu')\;.
\end{equation}
Instead of evaluating this integral by direct computation, as shown in appendix~\ref{sec:loop-from-2-1-spectral-integral}, one can use the property~\eqref{eq: definition rho chi1 chi2} to write~\eqref{eq: double spectral integral GA} as a single spectral integral: 
\begin{equation}
   \G_{\a\b}^A\left(^{\mu_1\mu_2}_{k_1k_2}\right) \equiv -\frac{\a\b g^2 e^{\frac{i\pi(\a+\b)d}{2}}}{H^2}\frac{\hat{\delta}_{\mu_2}(\mu_1)}{\N_{\mu_1}}\int\displaylimits^\infty_{-\infty}\d\mu\rho_{\chi_1\chi_2}(\mu)\G_{\a\b}^{\mu}(\mu_1)\;.
\label{eq: single spectral integral GA}   
\end{equation}
Then, the full KLF two-point function~\eqref{eq: self energy diagram} can be written as: 
\begin{equation}\label{eq: loop as mixing equation}
    \G^{\mu_\varphi,\textrm{1 loop}}_{\a_1\a_2}\left(^{\mu_1\mu_2}_{\k_1\k_2}\right)= \int\displaylimits^\infty_{-\infty}\d\mu\rho_{\chi_1\chi_2}(\mu)\G^{\mu,\textrm{mix}}_{\a_1\a_2}\left(^{\mu_1\mu_2}_{\k_1\k_2}\right)\;,
\end{equation}
where $\G^{\mu,\textrm{mix}}_{\a_1\a_2}\left(^{\mu_1\mu_2}_{\k_1\k_2}\right)$ is: 
\begin{equation}\label{eq: mixing equation}
    \G^{\mu,\textrm{mix}}_{\a_1\a_2}\left(^{\mu_1\mu_2}_{\k_1\k_2}\right) = -g^2\delta\left(^{\mu_1\;\mu_2}_{\k_1-\k_2}\right)\sum_{\b_1,\b_2}\b_1\b_2 e^{\frac{i\pi(\b_1+\b_2)d}{2}}\frac{\G^{\mu_\varphi}_{\a_1\b_1}(\mu_1)}{H^2}\frac{\G^\mu_{\b_1\b_2}(\mu_1)}{H^2}\frac{\G^{\mu_\varphi}_{\b_2\a_2}(\mu_1)}{H^2}\;.
\end{equation}
One can check that this corresponds to the $\O(g^2)$ contribution to the $\varphi$ two-point function in an alternative theory featuring a quadratic mixing $\L_\textrm{mix} = -g\varphi\chi$ where the mass of $\chi$ is given by $m_\chi = H M_\mu$. Diagrammatically, for fixed values of $\b_1$ and $\b_2$, equation~\eqref{eq: loop as mixing equation} becomes:
\begin{equation}
    \begin{tikzpicture}[baseline={(0,0)}]

        \draw[thick,black] (-1.5,0) to (-0.5,0);

        \draw[thick,black] (1.5,0) to (0.5,0);
        
        \draw [pyblue, thick] (-0.5,0) arc[start angle=180, end angle=0,radius=0.5];
        \draw [pyred, thick] (-0.5,0) arc[start angle=-180, end angle=0,radius=0.5];
        \filldraw[color=black,fill=black] (1.5,0) circle (2pt);
        \filldraw[color=black,fill=black] (-1.5,0) circle (2pt);
        \filldraw[color=black,fill=black] (0.5,0) circle (2pt);
        \filldraw[color=black,fill=black] (-0.5,0) circle (2pt);

        \node[black] at (-0.7,-0.27) {$\b_1$};

        \node[black] at (0.7,-0.27) {$\b_2$};

        \node[pyred] at (0,-0.7) {$\chi_1$};
        
        \node[pyblue] at (0,0.7) {$\chi_2$};
        
        \node[black] at (-1.6,0.3) {$\mu_1,\k_1$};

        \node[black] at (-1.6,-0.3) {$\a_1$};
        
        \node[black] at (1.6,0.3) {$\mu_2,\k_2$};

        \node[black] at (1.6,-0.3) {$\a_2$};
        
        \end{tikzpicture}
        = \int\displaylimits^{+\infty}_{-\infty}\d\mu\rho_{\chi_1\chi_2}(\mu)\left(
        \begin{tikzpicture}[baseline={(0,0)}]

        \draw[thick,black] (-1.5,0) to (-0.5,0);

        \draw[thick,black] (1.5,0) to (0.5,0);

        \draw[thick,pyblue] (-0.5,0) to (0.5,0);
        
        \filldraw[color=black,fill=black] (1.5,0) circle (2pt);
        \filldraw[color=black,fill=black] (-1.5,0) circle (2pt);
        \filldraw[color=black,fill=black] (0.5,0) circle (2pt);
        \filldraw[color=black,fill=black] (-0.5,0) circle (2pt);

        \node[pyblue] at (0,0.3) {$\mu$};
        \node[black] at (-0.7,-0.27) {$\b_1$};

        \node[black] at (0.7,-0.27) {$\b_2$};
        
        \node[black] at (-1.6,0.3) {$\mu_1,\k_1$};

        \node[black] at (-1.6,-0.3) {$\a_1$};
        
        \node[black] at (1.6,0.3) {$\mu_2,\k_2$};

        \node[black] at (1.6,-0.3) {$\a_2$};
        
        \end{tikzpicture}\right)\;,
\end{equation}
This is consistent with the standard spectral representation techniques in the evaluation of loop diagrams in dS~\cite{Marolf:2010nz,Higuchi:2010xt,Xianyu:2022jwk,Zhang:2025nzd}. 
\vskip 4pt
As noted in~\cite{Higuchi:2010xt}, one can always get rid of such a quadratic mixing by diagonalising the mass matrix. As we show in appendix~\ref{sec: details PT}, the final sum over the indices $\b_1,\b_2$ can thus be simplified (see Eq.~\eqref{simplification-double-sum-loop}), and one finds that the one-loop reduced propagator takes the form:
\begin{FramedBox}
\begin{equation}
    \G^{\mu_\varphi,1\;\textrm{loop}}_{\a\b}\left(\mu_1\right)= \bar{g}^2 \int\displaylimits^\infty_{-\infty}\d\mu \frac{ \rho_{\chi_1\chi_2}(\mu)}{H^{d-1}}\frac{\G_{\a\b}^\mu(\mu_1)}{(\mu_\varphi^2-\mu^2)^2}\;.
\end{equation}
\end{FramedBox}

The UV divergences of this expression are contained in the Gamma factors in the spectral density $\rho_{\chi_1\chi_2}(\mu)$~\cite{Higuchi:2010xt}, and they can be renormalised by adding the proper set of local counterterms.

\section{Conclusions \& Future Directions}
\label{sec:conclusion}
In this paper, we constructed a new frequency-momentum space for dS correlators by simultaneously diagonalising the Casimir and the spatial translation operators. Being spanned by the states $\ket{\mu,\k}$, the resulting Kontorovich-Lebedev-Fourier (KLF) space is analogous to the familiar Hilbert space for Lorentz invariant QFT in flat spacetime. After showing that square-integrable functions decompose into the principal series UIRs of $\SO(1,d+1)$, we found how the corresponding KLF expansion can be extended to a wide class of functions by including non-principal contributions. In the example of a bulk CFT, we showed how the KLF decomposition of the Wightman two-point function recovers the K\"all\'en-Lehmann representation.
\vskip 4pt
At the level of perturbative computations, we reformulated the double-branch path integral formulation directly in KLF space and derived the appropriate set of Feynman diagrammatic rules. The parallel between our construction and the energy-momentum space in Minkowski makes manifest the similarities with flat space in the structure of perturbative computations. In particular, the rational form taken by the (anti-)time ordered propagators relates the propagation of a virtual internal particle to the non-analyticities of the two-point function. However, it also unveils an intrinsic difference between the two spacetimes: the non-conservation of the dS frequency at vertices implies vertex contributions made of generalised hypergeometric functions, and forces one to keep a spectral integration per internal line.
Unlike real-time computations where nested integrals are required at each interaction vertex, spectral integrations can be performed simply by closing the contours and collecting the residues of their meromorphic integrands, as we showed in the example of the four-point exchange diagram. At the loop level, the group theoretical nature of our new language allowed us to recast the momentum integral in the self-energy correction to the scalar propagator as fundamental relations among the Clebsch-Gordan coefficients of $\SO(1,d+1)$.
\vskip 4pt
As our work introduces a novel way to think about cosmological correlators, it naturally opens several avenues of research.
\begin{itemize}
    \item It would be interesting to apply the methods we developed in this paper to more complicated diagrams.
    \item For simplicity, we restricted our construction to scalar fields. The formulation of realistic QFTs in dS would require an extension of the formalism to spinning cosmological correlators.
    \item In flat spacetime, momentum-space methods provide the clearest path to understand low-energy effective field theories from a Wilsonian perspective. It would be interesting to seek for a notion of renormalisation group flow in KLF space.
    \item Here, we made a maximal use of $\SO(1,d+1)$ invariance. However, the space of consistent QFTs in dS is also strongly constrained by unitarity and causality. Characterising these constraints in the KLF formulation would be an interesting starting point for a non-perturbative bootstrap in dS.
    \item Many phenomenologically relevant inflationary models involving dS-breaking interactions can be deduced from dS invariant correlators, to which our formalism applies. Besides, it would be interesting to study scenarios where the background itself deviates from pure dS.
\end{itemize}

\paragraph{Acknowledgements.} We thank Cliff Burgess, Mariana Carrillo Gonz\'alez, Thomas Colas, Guillaume Faye, Sadra Jazayeri, Denis Karateev, Elias Kiritsis, Francesco Nitti, Nadine Nussbaumer, Piotr Tourkine, Ugo Moschella, Guilherme Pimentel, Luca Santoni, Pierre Vanhove and Richard Woodard for useful discussions. The research of DW is funded by the European Union (ERC, \raisebox{-2pt}{\includegraphics[height=0.9\baselineskip]{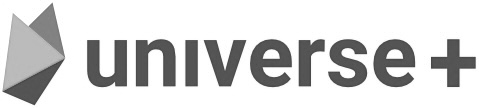}}, 101118787). Views and opinions expressed are however those of the author(s) only and do not necessarily reflect those of the European Union or the European Research Council Executive Agency. Neither the European Union nor the granting authority can be held responsible for them.

\appendix

\section{Kontorovich-Lebedev Transform}\label{sec: Rigged Hilbert Space}
In this appendix, we show how the Kontorovich-Lebedev transform can be derived using spectral theory. 

\vskip 4pt
Let us first recall the simpler situation of Sturm-Liouville theory. The latter applies to bounded linear operators $\O$ acting on $L^2([a,b])$.\footnote{An operator $O$ is bounded if there exists a constant $K$ such that $\lVert\O f\rVert \leq K \lVert f \rVert$ for all $f$.} When such an operator is self-adjoint, the Sturm-Liouville theorem states that its eigenvectors constitute an orthogonal basis of $L^2([a,b])$. Here or below in the unbounded case, $\O$ being self-adjoint implies that the eigenvalues are real and that the eigenfunctions are orthogonal to one another, the point here is to build a \textit{basis} made of such orthogonal functions.

\vskip 4pt
The situation where $\O$ is unbounded is more subtle. The Hilbert space $L^2$ is not the proper mathematical structure to formulate that problem, but one should instead work with the {\it{rigged Hilbert space}} \cite{Gelfand-Vilenkin,Maurin}, see for instance \cite{delaMadrid:2005qdg} for a pedagogical discussion. The latter is constituted of a Gelfand triple forming the following structure:
\begin{equation}
    \Phi \subset L^2\subset\Phi^\times\cong\Phi'\;,
\end{equation}
where $\Phi$ is the maximal dense subset of $L^2$ which is closed under the action on $\O$, and $\Phi^\times$ and $\Phi'$ are, respectively, the set of (anti)linear functionals acting
on $\Phi$. Alternatively, one can think of these two sets as the distributions defined on the space of test functions $\Phi$. The elements of $\Phi^\times(\Phi')$ are identified with a ket $\ket{f}$ (a bra $\bra{f}$) such that their actions on functions are:
\begin{equation}\label{eq: def element dual space}
    \F_f[\varphi]:=\braket{f|\varphi} = \int_0^\infty\frac{\d z}{z}f^*(z)\varphi(z)\quad\text{and}\quad \tilde{\F}_f[\varphi]:=\braket{\varphi|f} = \int_0^\infty\frac{\d z}{z}f(z)\varphi^*(z)\;.
\end{equation}
Therefore, these spaces can be identified with the set of functions for which the integral \eqref{eq: def element dual space} can be defined:
\begin{equation}\label{eq: def function in Phi star}
    \Phi^\times \cong \left\{f(z)\left|\int^\infty_0\frac{\d z}{z}f^*(z)\varphi(z)<\infty\;,\quad\forall\; \varphi \in \Phi\right.\right\}\;.
\end{equation}
When $\O$ is unbounded, its  eigenfunctions are not all in $L^2$, nor all in $\Phi^\times$. However, the generalised Sturm-Liouville theorem of interest to us states that the eigenfunctions that belong to $\Phi^\times$ form an orthogonal basis of $\Phi^\times$.  

\vskip 4pt
The relevant operator $\O$ for the Kontorovitch-Lebedev transform is the modified Bessel differential operator defined as:
\begin{equation}\label{eq: operator O Bessel}
    \O f(z)\equiv z\frac{\d}{\d z}\left(z\frac{\d f}{\d z}\right)-z^2 f(z)\;.
\end{equation}
This operator is self-adjoint in the Hilbert space $L^2\left(\mathbb{R}_+,\frac{\d z}{z}\right)$ with the inner product
\begin{equation}
    \braket{f|g} = \int\displaylimits^\infty_{0}\frac{\d z}{z}f^*(z)g(z)\;,
\end{equation}
for functions that satisfy the appropriate boundary conditions $\left[ z (g' f^* -g f'^{*}) \right]^\infty_0$. The test functions in that case can be thought of as the functions that decay faster than any polynomial at infinity, and behave around $0$ like $z^\alpha$ with $\Re(\alpha)>0$ around zero. As for the eigenfunctions of $\O$ with eigenvalue $\lambda$, they are given by the modified Bessel functions $K_{\sqrt{\lambda}}(z)$ and $I_{\pm\sqrt{\lambda}}(z)$. However, the $I_{\pm\sqrt{\lambda}}$ do not belong to $\Phi^\times$. For the $K$'s, given the behaviour $K_{\mu}(z) \underset{0}{\sim} 2^{-\mu -1} \Gamma (-\mu ) x^{\mu }+2^{\mu -1} \Gamma (\mu ) x^{-\mu }$, only the $K_{i \mu}$ with real $\mu$ belong to $\Phi^\times$. As a result, the generalised Sturm-Liouville theorem thus tells us that the $K_{i \mu},\mu \in \mathbb{R}$, constitute the orthogonal basis of $\Phi^\times$ we were after.

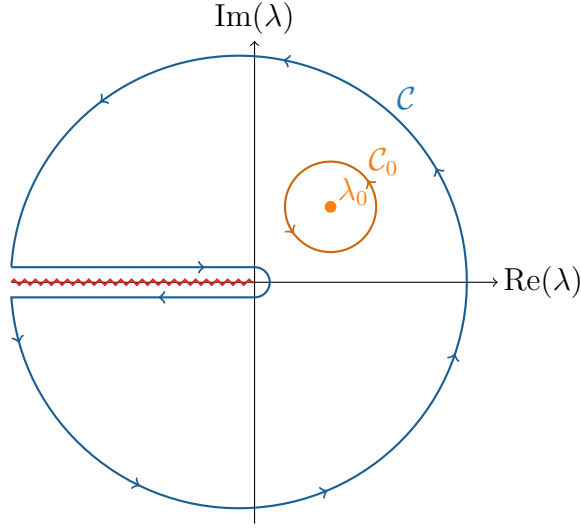
\begin{figure}
\centering
    \begin{tikzpicture}[scale = 2]

        \draw[black, ->] (-1.6,0) -- (1.6,0) coordinate (xaxis);
		\draw[black, ->] (0,-1.6) -- (0,1.6) coordinate (yaxis);
		\node at (1.9, 0) {$\text{Re}(\lambda)$};
		\node at (0, 1.75) {$\text{Im}(\lambda)$};

        \draw [decorate,decoration={zigzag,segment length=4,amplitude=1,post=lineto,post length=0},color=pyred,thick] (-1.6,0) -- (0,0);
        \node[pyblue] at (1.,1.22) {$\C$};

         \draw[xshift=0,pyorange!80!black,decoration={markings,mark=between positions 0.1 and 1 step 0.5 with \arrow{>},},postaction={decorate},thick] (0.8,0.5) arc[start angle=0, end angle=360,radius=0.3];
         \filldraw[color=pyorange,fill=pyorange] (0.5,0.5) circle (1pt);
         \node[pyorange] at (0.65,0.6) {$\lambda_0$};
         \node[pyorange] at (0.85,0.8) {$\C_0$};

        \draw[xshift=0,pyblue!80!black,decoration={markings,mark=between positions 0.1 and 1 step 0.1 with \arrow{>},},postaction={decorate},thick] (-1.6,0.1) -- (0,0.1) arc (90:-90:0.1) -- (-1.6,-0.1) arc (-176.1:176.1:1.5);
    \end{tikzpicture}
    \caption{Integration contour for the dispersive integral \eqref{Cauchy}.}
    \label{fig: contour Green function BC}
\end{figure}

\vskip 4pt
In the following, we go one step further and derive the completeness relation among these basis functions, and hence the associated spectral measure. For this, we define, for $\lambda \in \mathbb{C}$, the Green function $G^{(\lambda)}(z,z')$
such that
\begin{equation}\label{eq: definition Green funcion}
    \left(\O-\lambda\right)_z G^{(\lambda)}(z,z') = z' \;\delta(z-z')\;
\end{equation}
and with vanishing boundary conditions.
Following the standard method, one can find the explicit expression:
\begin{equation}
\label{eq: Green Function in terms of I K}
    G^{(\lambda)}(z,z') = -K_{\sqrt{\lambda}}(z)I_{\sqrt{\lambda}}(z')\Theta(z-z')-I_{\sqrt{\lambda}}(z)K_{\sqrt{\lambda}}(z')\Theta(z'-z)\;,
\end{equation}
where we picked the positive square root for $I$ in order to satisfy the boundary conditions at $z\to 0$. One can see that the location where the Green function fails to be analytic, along the branch cut on the negative real axis, coincides with the values we identified above according to the generalised Sturm-Liouville theorem. Now, using Cauchy theorem, one can write, for $\lambda_0\notin \mathbb{R}_-$:
\begin{equation}
    G^{(\lambda_0)}(z,z') = \oint_{\C_0}\frac{\d \lambda}{2\pi i}\frac{G^{(\lambda)}(z,z')}{\lambda-\lambda_0}\;,
    \label{Cauchy}
\end{equation}
where $\C_0$ is a contour that does not cross $\mathbb{R}_-$, as shown in figure~\ref{fig: contour Green function BC}.
Then, by blowing up the contour to $\C$ like in the figure, and as the contribution from the arc at infinity vanishes, we are left with the following dispersive integral:
\begin{equation}
    G^{(\lambda_0)}(z,z') = \int_{\mathbb{R}_-}\frac{\d\lambda}{2\pi i}\frac{\text{Disc}_{\lambda'}(G^{(\lambda)}(z,z'))}{\lambda-\lambda_0}\;.
\end{equation}
Writing $\lambda=-\mu^2$ and for $z>z'$ ($z< z'$ is similar), this discontinuity is given by:
\begin{equation}
    \text{Disc}_{\lambda}(G^{(\lambda')}(z,z')) = \lim_{\epsilon\to 0}\left(G^{(-\mu^2+i\epsilon)}(z,z')-G^{(-\mu^2-i\epsilon)}(z,z')\right)
\end{equation}
\begin{equation}
    = K_{-i\mu}(z)I_{-i\mu}(z')-K_{i\mu}(z)I_{i\mu}(z') = \frac{2 i}{\pi}\sinh(\pi\mu)K_{i\mu}(z)K_{i\mu}(z')\;.
\end{equation}
After the change of variable $\lambda=-\mu^2$, one finds that the Green function can be written as the spectral integral:
\begin{equation}
    G^{(\lambda_0)}(z,z') = \frac{2}{\pi}\int^\infty_0\d\mu \frac{\mu\sinh(\pi\mu)}{\pi}\frac{K_{i\mu}(z)K_{i\mu}(z')}{-\mu^2+\mu^2_0}\;.
\end{equation}
Now, one can apply the operator $\O-\lambda_0$ to the left in order to get rid of the denominator:
\begin{equation}
    \left(\O_z-\lambda_0\right)G^{(\lambda_0)}(z,z') = \frac{2}{\pi}\int^\infty_0\d\mu\frac{\mu\sinh(\pi\mu)}{\pi}K_{i\mu}(z)K_{i\mu}(z') = z'\delta(z-z')\;.
\end{equation}
Because of the shadow symmetry of the integral, one can rewrite it as:
\begin{equation}
\label{eq: proof completeness}
    \frac{1}{\pi}\int_{-\infty}^{+\infty}\d\mu \, \N_\mu K_{i\mu}(z)K_{i\mu}(z') = z'\delta(z-z')\, \quad \textrm{with} \quad  \N_{\mu} = \frac{\mu}{\pi}\sinh(\pi\mu)\,.
\end{equation}
This is the completeness relation we were after. From it, one also deduces the coefficient entering into the orthogonality relation among these eigenfunctions:
\begin{equation}
\label{eq: orthogonality McDonald}
    \int^\infty_0\frac{\d z}{z}K_{i\mu}(z)K_{i\mu'}(z) = \frac{ \pi |\mu|}{\N_{\mu}}\delta(\mu^2-\mu'^2)\;, \quad \mu \in \mathbb{R}\,.
\end{equation}  
The two relations \eqref{eq: proof completeness} and \eqref{eq: orthogonality McDonald} thus enable us derive the Kontorovitch-Lebedev transform and its inverse:
\begin{equation}
     f(z) = \int_{-\infty}^{+\infty}\d\mu \;\N_{\mu}\Phi^\mu(z)f^{\mu}\;,\quad
f^{\mu} = \int_0^\infty\frac{\d z}{z}\Phi^\mu(z)f(z)   
\end{equation}
where the harmonic function is given by:
\begin{equation}
    \Phi^\mu(z)=\frac{1}{\sqrt{\pi}}K_{i\mu}(z)\;.
\end{equation}
This transformation was first introduced in~\cite{Kontorovich1938} and later studied in~\cite{Lebedev1946}.  The Parseval identity was proved in~\cite{Lebedev1949}, the KL transform for distributions was considered in~\cite{Yakubovich1994, Zemanian_1975}, and a table for KL transform can be found in~\cite{Erdelyi1954}.

\vskip 4 pt
Together with the Fourier transform, the relations \eqref{eq: proof completeness} and \eqref{eq: orthogonality McDonald} imply the following completeness and orthogonality relations among the harmonic functions \eqref{eq: harmonic function expression} in the principal series:
\begin{equation}
\begin{aligned}
\int_{\KLF} 
 \left[\Phi^{\mu}_{\k}(z',\x')\right]^* 
 \Phi^{\mu}_{\k}(z,\x)&=(H z)^{d+1}\delta(z-z')\delta^{(d)}(\x-\x')
\,, \\
  \int_{\EAdS}\left[\Phi^{\mu}_{\k}(z,\x)\right]^*\Phi^{\mu'}_{\k'}(z,\x)&=\frac{(2\pi)^d}{\N_\mu}\delta^{(d)}(\k-\k')\frac12(\delta_{\mu'}(\mu)+\delta_{\mu'}(-\mu))\;,
  \label{completeness-orthogonality-harmonic-principal}
\end{aligned}
\end{equation}
with the integration measures
\begin{equation}
    \begin{aligned}
        \int_{\KLF} &\equiv \int \displaylimits_{-\infty}^{+\infty}\d\mu \, \N_{\mu}\int_{\mathbb{R}^d}\frac{\d^d\k}{(2\pi)^d} \,, \\
        \int_{\EAdS} &\equiv \int \displaylimits_0^\infty \frac{\d z}{(Hz)^{d+1}}\int_{\mathbb{R}^d}\d^d\x \,.
    \end{aligned}
\end{equation}
The KLF transform and its inverse thus read:
\begin{equation}
\begin{aligned}
    f(z,\x) &= \int_{\KLF} \Phi^{\mu}_{\k}(z,\x)f^{\mu}_{\k}\,,\\
        f^{\mu}_{\k} &= \int_\EAdS \left[\Phi^{\mu}_{\k}(z,\x)\right]^*f(z,\x)\,.
\end{aligned}
\end{equation}

\section{From KLF to Position-Space Propagators}
\label{sec: propagators in real space}

In section \ref{sec: Gaussian PI}, we showed that the four possible KLF propagators are given by Eqs.~\eqref{eq: KLF (anti-)time ordered} and \eqref{eq: KLF wightman function}. Here, in order to make contact with results in the literature, we explicitly perform the KLF inverse transform and the Wick rotation in order to go back to configuration Lorentzian space.

For this, let us recall that KLF space is defined in Euclidean Anti de-Sitter space (EAdS). The latter is defined in embedding space as the set of points at fixed negative Lorentzian distance from the origin:
\begin{equation}
    \eta_{A B}X^{E,A}X^{E,B}=-R_{\EAdS}^2\;,
\end{equation}
where $R_{\EAdS}$ is called the EAdS radius. Like dS, this can described in Poincar\'e coordinates, such that
\begin{equation}
    X^{E,0} = R_\EAdS \, \frac{z^2+\x^2+1}{2 z} \,, \quad X^{E,i} = -R_\EAdS\,\frac{x^i}{z}\,, \quad X^{E,d+1} = R_\EAdS\,\frac{z^2+\x^2-1}{2z} \,,
\end{equation}
where $i=1, \ldots, d$ and $z>0$, in which the EAdS metric reads
\begin{equation}
    \d s^2=R_\EAdS^2\frac{\d z^2+\d\x^2}{z^2}\,.
\end{equation}
When dealing with two-point correlation functions, it is natural to introduce the EAdS two-point invariant such that:
\begin{equation}
    \sigma^E \equiv \frac{X_1^{E,A}X_2^{E,B}\eta_{A B}}{R_\EAdS^2}=-\frac{z_1^2+z_2^2+|\x_{12}|^2}{2 z_1 z_2}\in(-\infty,-1)\;.
    \label{eq:Euclidean-two-point-invariant}   
\end{equation}
In what follows, the main building block will be the function defined by the momentum integral of two harmonic functions~\eqref{eq: harmonic function expression}: 
\begin{equation}\label{eq:definition of W}
\begin{aligned}
    W_\mu(\sigma^E) &=H^{-2}\int\frac{\d^d\k}{(2\pi)^d}\Phi^{\mu}_{\k}(X^E_1)\Phi^{\mu}_{-\k}(X^E_2)\\
    &= \frac{2H^{d-1}\pi^{\frac{d-3}{2}}(z_1 z_2)^{\frac{d}{2}}}{(2\pi)^d\Gamma\left(\frac{d-1}{2}\right)}\int\displaylimits_0^\infty \d k\; k^{d-1}K_{i\mu}(k z_1)K_{i\mu}(k z_2)\\
    &\qquad\times\int\displaylimits^1_{-1}\d\chi(1-\chi^2)^{\frac{d-3}{2}}e^{-i k|\x_{12}|\chi}\\
    &= \frac{2^{\frac{d}{2}}H^{d-1}\pi^{\frac{d}{2}-1}(z_1 z_2)^{\frac{d}{2}}}{(2\pi)^d|\x_{12}|^{\frac{d}{2}-1}}\int\displaylimits_0^\infty \d k\; k^{\frac{d}{2}}K_{i\mu}(k z_1)K_{i\mu}(k z_2)J_{\frac{d}{2}-1}(k|\x_{12}|)\\
    &=\frac{H^{d-1}}{(4\pi)^{\frac{d+1}{2}}}\Gamma\left(\begin{array}{c}
         \frac{d}{2}\pm i\mu  \\
         \frac{d+1}{2} 
         \end{array}\right)\;_2F_1\left(\begin{array}{cc}
        \frac{d}{2}+i\mu &  \frac{d}{2}-i\mu\\
         \frac{d+1}{2}& 
    \end{array};\frac{1+\sigma^E}{2}\right)\;.
\end{aligned}
\end{equation}
The integral over the angles and the norm are respectively given by Eq.~$(3.387, 2)$ and Eq.~$(6.578$, $10$) of \cite{gradshteyn2007}. We also used Eq.~$(14.3.15)$ of~\cite{NIST:DLMF} to connect the associated Legendre to hypergeometric functions. This is closely related to what are called the harmonic function of EAdS in
\cite{Costa:2011dw} as:
\begin{equation}
    \Omega_\mu(\sigma^E)=H^2\N_\mu W_\mu(\sigma^E)\;,
\end{equation}
which are known to form a complete basis of square integrable bi-local functions in EAdS. This fact easily derives from the orthogonality and completeness of KLF modes, from which one indeed finds: 
\begin{equation}
   H^2 \int\displaylimits^\infty_{-\infty}\d\mu \N_{\mu }W^{\mu}(\sigma^E_{12})= (H z_1)^{d+1}\delta(z_1-z_2)\delta^{(d)}(\x_1-\x_2)\;,
\end{equation}
and
\begin{equation}
   H^2 \int_{\EAdS}\frac{\d^d\x \d z}{(H z)^{d+1}}W_\mu(\sigma^E_{1x})W_{\mu'}(\sigma^E_{x2})=\frac{\hat{\delta}_{\mu'}(\mu)}{\N_\mu}W_\mu(\sigma^E_{12})\;.
\end{equation}
The integral~\eqref{eq:definition of W} is the EAdS counterpart of the addition theorem for spherical harmonics on the Euclidean sphere:
\begin{equation}\label{eq: addition theorem spherical harm}
    \sum_{\m}Y^{L}_{\m}(X_1^S)\left[Y^{L}_{\m}(X_2^S)\right]^*=\frac{-2L-d}{(4\pi)^{\frac{d-1}{2}}}\frac{\Gamma\left(
        -L , L+d\right)}{\Gamma\left(\frac{d+1}{2}\right)}\;_2F_1\left(\begin{array}{cc}
        -L &  L+d\\
         \frac{d+1}{2}& 
    \end{array};\frac{1-\sigma^S}{2}\right)\;,
\end{equation}
where $Y^{L}_{\m}(X^S)$ are the $(d+1)$-dimensional spherical harmonics and $\sigma^S$ is the two-point invariant on the sphere between the two points $X_1^S$ and $X_2^S$. See e.g.~\cite{Marolf:2010zp} for more details about obtaining dS propagators from~\eqref{eq: addition theorem spherical harm}.

\vskip 4pt
Let us now turn to the Euclidean position-space version of the four propagators. Using the general KLF decomposition of correlators \eqref{eq: from KLF to real space correlators}, the links \eqref{general-to-reduced-propagators-link-same-a}, \eqref{general-to-reduced-propagators-link-different-a} between general and reduced propagators, and the definition \eqref{eq:definition of W} of $W_\mu$, one has
\begin{equation}
\C_{\a\b}^{E,\varphi}(\sigma^E)  = \int\displaylimits^\infty_{-\infty}\d\mu\,\N_\mu\,
W_\mu(\sigma^E)\,\G^{\varphi}_{\a\b}(\mu)\,.
\label{Euclidean-two-point-spectral-integral}
\end{equation}
We begin with the simplest case of the $\pm \mp$ two-point function. Using the simple form \eqref{eq: KLF wightman function} of the KLF Wightman function straightforwardly gives 
\begin{equation}
    \C^E_{\pm\mp}(X_1^E,X_2^E) = W_{\mu_\varphi}(\sigma^E)\;.
\end{equation}
We now move on with the more subtle case of $\pm \pm$ propagators. In section~\eqref{sec: Gaussian PI}, we showed that these diagonal propagators in KLF are given by Eq.~\eqref{eq: diagonal propagator KLF function f} in terms of the arbitrary functions $f_\a(\mu)$. In the following, we show that there are physical choices for these functions corresponding to the (anti-)time ordered boundary conditions. In Lorentzian time, the Fourier space time-ordered two-point function is:
\begin{equation}
    \C_{\a\a}(\tau_1,\tau_2;k)=H^{d-1}u^{\a}_k(\tau_1,\mu)u^{-\a}_k(\tau_2,\mu)\Theta(\tau_1-\tau_2) + (\tau_1\leftrightarrow\tau_2)\;,
\end{equation}
where the mode functions $u^{\a}_k(\tau,\mu)$ are those defined in~\eqref{eq: Lorentzian mode functions}. After the Wick rotation, one finds:
\begin{equation}\label{eq: Euclidean (anti-)time ordered BC}
    \C_{\a\a}^E(z_1,z_2;k)=\frac{1}{H^2}\Phi^{\mu}_{\k}(z_1,\boldsymbol{0})\Phi^{\mu}_{\k}(e^{-i\a\pi}z_2,\boldsymbol{0})\Theta(z_1-z_2)+(z_1\leftrightarrow z_2)\;.
\end{equation}
\paragraph{Principal series.} When the propagating field is in the principal series, one needs an $i\epsilon$ prescription~\eqref{eq: arbitrary i epsilon function g} that also introduces an arbitrary function $g(\mu)$. In order to see why, let us notice that the usual Feynman prescription breaks the shadow symmetry:
\begin{equation}
\begin{aligned}
\label{Feynman-prescription}    \int\displaylimits^\infty_{-\infty}\d\mu\N_\mu\frac{K_{i\mu}(z_1)K_{i\mu}(z_2)}{\mu^2-\mu_\varphi^2+ia\epsilon}&= \pi\Theta(z_1-z_2)K_{i\mu_\varphi}(z_1)I_{i a\mu_\varphi}(z_2) \\
    &+ \pi\Theta(z_2-z_1)K_{i\mu_\varphi}(z_2)I_{i a\mu_\varphi}(z_1)\;.
\end{aligned}
\end{equation}
In order to perform this integral, we split the function $K_{i\mu}(z)$ into a sum of the function $I_{i\mu}(z)$ and its shadow using 
\begin{equation}\label{eq: connectio formula I vs K}
    {\cal N}_\mu K_{i\mu}(z) = \frac{i\mu}{2}(I_{i\mu}(z)-I_{-i\mu}(z))\;.
\end{equation}
When $z_1>z_2$, we have to split the function in $z_2$ so that $K_{i\mu}(k z_1)I_{\pm i\mu}(k z_2)$ decays in the lower (upper) half-plane. The case where $z_1<z_2$ can be obtained by permuting the $z$ variables. The shadow symmetry breaking comes from the function $I_{i\mu_\varphi}$, which is not symmetric under reflection of the frequency $\mu_\varphi$. Therefore, one needs to sum two opposite  Feynman pole prescriptions in order to restore it. To this aim, we introduce:
\begin{equation}\label{eq: general shadow sym i epsilon}
    \frac{1}{(\mu^2-\mu_\varphi^2)^{g}_{i\epsilon}}=\frac{g(\mu_\varphi)}{\mu^2-\mu_\varphi^2+i\epsilon}+\frac{g(-\mu_\varphi)}{\mu^2-\mu_\varphi^2-i\epsilon}\;,
\end{equation}
where $g(\mu)$ is an arbitrary function such that:
\begin{equation}
    g(\mu)+g(-\mu)=1\;.
\end{equation}
We impose the last equality so that the small $\epsilon$ limit gives back the propagator:
\begin{equation}
    \lim_{\epsilon\to0}\frac{1}{(\mu^2-\mu_\varphi^2)^g_{i\a\epsilon}}=\frac{1}{\mu^2-\mu_\varphi^2}\;.
\end{equation}

Let us see how to determine the functions $f$ and $g$ with physical boundary conditions. For concreteness, we look at the $++$ propagator, as the one $--$ can be simply obtained by complex conjugation. Let us first perform the spectral integration, i.e.~$\C^E_{++}(X^E_1,X^E_2)=\int \frac{\d^d\k}{(2\pi)^d} e^{-i \k \cdot(\x_1-\x_2)}\C_{++}^E(z_1,z_2;k)$, where 
\begin{equation}
\begin{aligned}
    \C_{++}^E(z_1,z_2;k)=\frac{H^{d-1}}{\pi}e^{-\frac{i\pi(d-1)}{2}}(z_1z_2)^\frac{d}{2}\int\displaylimits^\infty_{-\infty}\d\mu\N_\mu K_{i\mu}(k z_1)K_{i\mu}(k z_2)\\\times\Bigg(\frac{1}{(\mu^2-\mu_\varphi^2)^g_{i\epsilon}}+\hat{\delta}_{\mu_\varphi}(\mu)f_+(\mu)\Bigg)\;.
    \label{C++-momentum}
\end{aligned}
\end{equation}
To evaluate this, we use \eqref{Feynman-prescription} and find, for $z_1 > z_2$ (the other case is similar):
\begin{equation}
\begin{aligned}
    \C_{++}^E(z_1,z_2;k) =H^{d-1} e^{-i\frac{\pi(d-1)}{2}}(z_1z_2)^{\frac{d}{2}}K_{i\mu_\varphi}(k z_1)\Bigg(g(\mu_\varphi)I_{i\mu_\varphi}(k z_2)\\+g(-\mu_\varphi)I_{-i\mu_\varphi}(k z_2)+\frac{1}{\pi} \N_{\mu_\varphi} f_+(\mu_\varphi)K_{i\mu_\varphi}(k z_2)\Bigg)\;.
    \label{propagators-to-match}
\end{aligned}
\end{equation}
Now, we have to find all the possible functions $f_\a$ and $g$ compatible with~\eqref{eq: Euclidean (anti-)time ordered BC}. To this aim, we use Eq.~\eqref{eq: connectio formula I vs K} and the analytic continuation formula for the Bessel functions:
\begin{equation}
    K_{i\mu}\left(e^{-i\pi}z\right)=\frac{i\pi}{2\sinh(\pi\mu)}\left(e^{\pi\mu}I_{i\mu}(z)-e^{-\pi\mu}I_{-i\mu}(z)\right)\;.
\end{equation}
This fixes one of the two functions in terms of the other via the following condition:
\begin{equation}
\begin{aligned}
2 g(\mu)+\frac{i}{\pi} \mu f_+(\mu)&=\frac{e^{\pi \mu}}{\sinh(\pi \mu)}\,.
\end{aligned} 
\end{equation}
There is still a large freedom in the choice of the function $g$, and this can be used to set the functions $f_\a$ to zero:
\begin{equation}\label{eq: choice f and g}
    \left\{\begin{array}{crc}
        f_\a(\mu) & = & 0 \\
        g(\mu) & = & \frac{e^{\pi\mu}}{2\sinh(\pi\mu)}
    \end{array}\right.\;.
\end{equation}
This is equivalent to the prescription used in~\cite{Melville:2024ove,Werth:2024mjg} and yields the result~\eqref{eq: i epsilon prescription} quoted in the main text. Since the contribution from the functions $f_\a(\mu)$ are within the principal series,~\eqref{eq: choice f and g} amounts to absorbing it in the integral over the principal series, which is the simplest choice.

\paragraph{Other UIRs.} Let us now consider a propagating field that is not in the principal series, i.e.~its frequency is purely imaginary and we write $\mu_\varphi=i\nu_\varphi$ with real $\nu_\varphi$. In that case, the poles of the reduced two-point function \eqref{eq: diagonal propagator KLF function f} are on the imaginary axis. Hence, there is no need for an $i\epsilon$ prescription, i.e.~we can set $g=0$, and the boundary conditions are fully encoded in the functions $f_a(\mu)$. We look at the $++$ propagator, the other case being similar. To evaluate \eqref{C++-momentum}, like above, we use the formula~\eqref{eq: connectio formula I vs K} and close the contour in the lower half-plane for $z_1 > z_2$. As there is no $i\epsilon$ prescription here, the contour integral picks a single pole here, yielding:
\begin{equation}
\begin{aligned}
    \C^E_{++}(z_1,z_2;k)=H^{d-1}e^{-\frac{i\pi(d-1)}{2}}(z_1z_2)^{\frac{d}{2}}K_{\nu_\varphi}(k z_1)\Bigg(I_{|\nu_\varphi|}(k z_2)    +\frac{\N_{i\nu_\varphi}f_+(i\nu_\varphi)}{\pi}K_{\nu_\varphi}(k z_2)\Bigg)\;.
\end{aligned}
\end{equation}
Again using Eq.~\eqref{eq: connectio formula I vs K}, the matching with the behaviour \eqref{eq: Euclidean (anti-)time ordered BC}  now gives a unique solution for $f_+$, and similarly for $f_-$:
\begin{equation}
f_\a(i \nu_\varphi)=- i\a\frac{e^{i\a\pi|\nu_\varphi|}}{\N_{i\nu_\varphi}} \,. 
\end{equation}
Physically, this contribution corresponds to the component of the propagator that cannot be absorbed in the principal series integral.

\paragraph{Momentum integration.} Then, one has to carry out the momentum integral in order to obtain the Euclidean position-space propagator:
\begin{equation}
\begin{aligned}
    \C^E_{\a \a}(X_1^E,X_2^E)=\frac{1}{H^2}\int\frac{\d^d\k}{(2\pi)^d}\Bigg(\Phi^{\mu_\varphi}_{\k}(z_1,\x_1)\Phi^{\mu_\varphi}_{-\k}(e^{-i \a \pi}z_2,\x_2)\Theta(z_1-z_2)\\
    +\Phi^{\mu_\varphi}_{\k}(e^{-i\a \pi}z_1,\x_1)\Phi^{\mu_\varphi}_{-\k}(z_2,\x_2)\Theta(z_2-z_1)\Bigg)\;.
\end{aligned}
\end{equation}
Each integration can be carried out independently and are simply obtained by analytic continuation of the integral~\eqref{eq:definition of W}, yielding:
\begin{equation}
    \C^E_{++}(X_1^E,X_2^E)=W_{\mu_\varphi}(-\sigma^E-i \a \epsilon)\;.
\end{equation}
The ensemble of Euclidean two-point functions can thus be summarised as:
\begin{equation}\label{eq: Euclidean real space propagators}
    \boldsymbol{\C}^E(X_1^E,X_2^E)=\left(\begin{array}{cc}
        W_{\mu_\varphi}(-\sigma^E-i\epsilon) & W_{\mu_\varphi}(\sigma^E) \\
        W_{\mu_\varphi}(\sigma^E) & W_{\mu_\varphi}(-\sigma^E+i\epsilon)
    \end{array}\right)\;.
\end{equation}
\paragraph{Back to Lorentzian time.}
These propagators only make sense as Wick rotated versions of their Lorentzian dS counterparts. Let us explain how to carry the analytical continuation of the Euclidean two-point invariant~\eqref{eq:Euclidean-two-point-invariant} to Lorentzian signature. Since the building block $W_\mu(\sigma)$ has a branch cut for $\sigma\in(1,\infty)$, we must keep track of the $i\epsilon$ prescription and rotate to the tilted axis $\tau_c=\tau e^{i\a\epsilon}$ by applying~\eqref{eq: z Lorentzian} to the two Euclidean times $z_1$ and $z_2$. The four possible analytic continuations are given by:
\begin{equation}
\begin{aligned}
    \sigma_{\a\b} &= -\frac{\left(-\tau_1 e^{i\a\left(\frac{\pi}{2}-\epsilon\right)}\right)^2 + \left(-\tau_2 e^{i\b\left(\frac{\pi}{2}-\epsilon\right)}\right)^2+|\x_{12}|^2}{2\tau_1\tau_2 e^{i(\a+\b)\left(\frac{\pi}{2}-\epsilon\right)}}\\
    & = (-1)^{\a+\b}\left(\sigma^{\dS}-i\a\epsilon\tau_1^2-i\b\epsilon\tau_2^2\right)+i(\a+\b)\sigma^{\dS}\epsilon\;.
\end{aligned}
\end{equation}
On the one hand, the off-diagonal components give:
\begin{equation}\label{eq: Lorentzian two-point invariant off-diag}
    \sigma_{-\a\a}= \sigma^{\dS}-i\a\epsilon\;\textrm{sgn}(\tau_1-\tau_2)\;.
\end{equation}
On the other hand, the diagonal components are:
\begin{equation}
    \sigma_{\a\a}=-\sigma^{\dS}+i\a(1+\sigma^{\dS})\epsilon\;.
\end{equation}
From equation~\eqref{eq: Euclidean real space propagators}, it is clear that the regularisation will only be needed at time-like separations for $\sigma^{\dS}>1$. The final prescription is therefore:
\begin{equation}\label{eq: Lorentzian two-point invariant diag}
    \sigma_{\a\a} = -\sigma^{\dS}+i\a\epsilon\;.
\end{equation}
Inserting this in the Euclidean propagators yields the following result for the Lorentzian free two-point functions in dS:
\begin{equation}\label{eq:propagators real space}
    \boldsymbol{\C}(X_1,X_2)=\left(\begin{array}{cc}
        W_{\mu_\varphi}(\sigma^{\textrm{dS}}-i\epsilon) & W_{\mu_\varphi}(\sigma^{\textrm{dS}}-i\epsilon\;\textrm{sgn}(\tau_1-\tau_2)) \\
        W_{\mu_\varphi}(\sigma^{\textrm{dS}}+i\epsilon\;\textrm{sgn}(\tau_1-\tau_2)) & W_{\mu_\varphi}(\sigma^{\textrm{dS}}+i\epsilon)
    \end{array}\right)\;.
\end{equation}
This matches the usual results (see e.g. $3.3$ of \cite{DiPietro:2021sjt}). In particular, all these propagators agree at space-like separations as required by micro-causality. The different time orderings come from the regularisation prescription of the branch cut at time-like separations.
\section{Details of the KLF Decomposition of the CFT Two-Point Function}
\label{sec: CFT details}
In section \ref{sec: Role of the other Series}, we took the example of the CFT two-point function in order to illustrate the role of the other UIRs in the KLF decomposition of non square-integrable functions. Here is a detailed version of the computations.

\paragraph{KLF transform.} Let us first perform the KLF transform \eqref{eq: KLF tranform CFT 2points} explicitly, where we recall the expression \eqref{EAdS-CFT} of the Euclidean Wightman two-point function, and we write $X_2^E=(z,\boldsymbol{x})$:
\begin{equation}
\begin{aligned}
   & \int_{\EAdS}\frac{\d z_1\d^d\x_1}{(H z)^{d+1}}\left[\Phi^{\mu}_{\k}(z_1,\x_1)\right]^*\braket{\O(X_1^E)\O(X_2^E)}\\
    &=\int\frac{\d z'\d^d\x'}{(H z')^{d+1}}\frac{(z z')^{\Delta}}{((z+z')^2+|\x_{12}|^2)^\Delta}\left[\Phi^{\mu}_{\k}(z',\x')\right]^*\\
    &= \frac{H^{-\frac{d+1}{2}}z^{\Delta}e^{i\k\cdot\x}}{\pi^{\frac{1}{2}}}\int\d z'( z')^{\Delta-\frac{d}{2}-1}K_{i\mu}(p z')
        \int\d^d\x'\frac{e^{i\k\cdot\x'}}{{((z+z')^2+|\x'|^2)^\Delta}}\;.
\end{aligned}
\end{equation}
The spatial integral is:
\begin{equation}
    \begin{aligned}
    &\int\d^d\x'\frac{e^{i\k\cdot\x'}}{{((z+z')^2+|\x'|^2)^\Delta}} \\
    &= \frac{2\pi^{\frac{d-1}{2}}}{\Gamma\left(\frac{d-1}{2}\right)}\int_0^\infty\frac{\d r r^{d-1}}{((z+z')^2+r^2)^\Delta}\int^{1}_{-1}\d\chi(1-\chi^2)^{\frac{d-3}{2}}e^{i k r \chi}\\
    &=\frac{(2\pi)^{\frac{d}{2}}}{p^{\frac{d}{2}-1}}\int_0^\infty\frac{\d r\; r^{\frac{d}{2}}J_{\frac{d}{2}-1}(k r)}{((z+z')^2+r^2)^\Delta}\\
    &= \frac{(2\pi)^{\frac{d}{2}}}{2^{\Delta-1}p^{\frac{d}{2}-\Delta}\Gamma(\Delta)}(z+z')^{\frac{d}{2}-\Delta}K_{\Delta-\frac{d}{2}}(k(z+z'))\;.
    \end{aligned}
    \end{equation}
For the last step, we used Eq.~$(6.565$, $4$) of \cite{gradshteyn2007}. This integral converges if $d>0$, $\Delta>\frac{d-1}{4}$ and $z+z'>0$. The second condition gives a candidate bound for the square-integrability of the two-point function and the last one enforces the need to define this transformation in the Euclidean manifold.
\vskip 4pt
Now, let us perform the remaining time integral:
\begin{equation}
\begin{aligned}
    & \frac{2^{\frac{d}{2}-\Delta+1}\pi^{\frac{d-1}{2}}H^{-\frac{d+1}{2}}z^{\Delta}e^{i\k\cdot\x}}{k^{\frac{d}{2}-\Delta}\Gamma(\Delta)}\int_0^\infty\frac{\d z'}{z'}\left(\frac{ z'}{z+z'}\right)^{\Delta-\frac{d}{2}}K_{\Delta-\frac{d}{2}}(k (z+z'))K_{i\mu}(k z')\\
    &= H^{-(d+1)}\frac{2^{-2\Delta+d+1}\pi^{\frac{d-1}{2}}\Gamma\left(\Delta-\frac{d}{2}\pm i\mu\right)}{\Gamma(\Delta)\Gamma\left(\Delta-\frac{d}{2}+\frac{1}{2}\right)}\mu\sinh(\pi\mu)\frac{\left[\Phi_{\k}^{\mu}(z,\x)\right]^*}{\N_\mu}\;.
\end{aligned}
\end{equation}
We made use of $(6.583)$ from \cite{gradshteyn2007}. This converges when $\Delta-\frac{d}{2}>\text{Re}(i\mu)=0$ and $\text{arg}(z)\in]-\pi,\pi[$. The first condition gives a more stringent bound on the scaling dimension that the one we obtained from the spatial integral, and the second condition enforces the need of an $i\epsilon$ prescription when going back to dS. As a result, the inverse formula is defined for the following regime:
\begin{equation}
    \Delta>\frac{d}{2}\;,\quad \text{square integrable function}\;,
\end{equation}
and the spectral density along the principal series is given by: 
\begin{equation}
    \rho^{\P}_\O(\mu) =\frac{\N_\mu}{H^{d-1}} c_\Delta\Gamma\left(\Delta-\frac{d}{2}\pm i\mu\right)\;,
\end{equation}
where
\begin{equation}
    c_\Delta=\frac{2^{-2\Delta+d+1}\pi^{\frac{d+1}{2}}}{\Gamma(\Delta)\Gamma\left(\Delta-\frac{d}{2}+\frac{1}{2}\right)}\;.
\end{equation}
\paragraph{Evaluating the spectral integral.} Now, let us turn to the evaluation of the spectral integral \eqref{eq: KL CFT example integral}. To this aim, it is convenient to split the Wightman function into two EAdS bulk-to-bulk propagators~\cite{Sleight:2021plv,DiPietro:2021sjt}:
\begin{equation}\label{eq: splittin W in AdS}
    W_{\mu}(\sigma^E)=\frac{1}{2i\sinh(\pi\mu)}\left(\Pi^{\EAdS}_{-\mu}(\sigma^E)-\Pi^{\EAdS}_\mu(\sigma^E)\right)\;,
\end{equation}
where $\Pi^{\EAdS}_\mu$ is defined as: 
\begin{equation}
\label{PiEAdsS}
\begin{aligned}
    \Pi^{\EAdS}_\mu(\sigma^E) &\equiv H^{d-1}(z_1 z_2)^{\frac{d}{2}}\int\frac{\d^d\k}{(2\pi)^d}e^{-i\k\cdot(\x_1-\x_2)}I_{i\mu}(k z_1)K_{i\mu}(k z_2)\\
    &=\frac{H^{d-1}}{2\pi^{\frac{d}{2}}}\frac{\Gamma\left(\frac{d}{2}+i\mu\right)}{\Gamma(1+i\mu)}\frac{1}{(2|\sigma^E|-2)^{\frac{d}{2}+i\mu}}\;_2F_1\left(\begin{array}{cc}
        \frac{d}{2}+i\mu, & \frac{1}{2}+i\mu \\
         1+2i\mu
    \end{array};\frac{2}{1+\sigma^E}\right)\;.
\end{aligned} 
\end{equation}
This decays exponentially in the $\mu$ lower half-plane and it only has poles in the upper half-plane. It is equivalent to what is done in appendix~\ref{sec: propagators in real space} where we performed the spectral integral before the momentum one. After this step, the CFT two-point function becomes 
\begin{equation}\label{eq: CFT spectral int App}
\begin{aligned}
    \braket{\O(X_1^E)\O(X_2^E)}&=c_\Delta H^{-(d-1)}\int\d\mu\,\N_\mu\,\Gamma\left(\Delta-\frac{d}{2}\pm i\mu\right)W_\mu(\sigma^E)\\
    &=\frac{ic_\Delta}{\pi H^{d-1}}\int\d\mu\;\mu\;\Gamma\left(\Delta-\frac{d}{2}\pm i\mu\right)\Pi^{\EAdS}_\mu(\sigma^E)\;.
\end{aligned}
\end{equation}
Closing the contour in the lower half-plane and collecting the residues attached to the poles of the Gamma function yields the following expression for the two-point function:
\begin{equation}\label{eq: G-+ CFT Series App}
\begin{aligned}
     &\braket{\O(X_1^E)\O(X_2^E)}  =\frac{1}{2^\Delta(1-\sigma^E)^\Delta}
     \\ &= \frac{2 c_\Delta}{H^{d-1}}\sum_{n=0}^\infty\frac{(-1)^n}{n!}\left(\Delta-\frac{d}{2}+n\right)\Gamma\left(2\Delta-d+n\right)\Pi^{\EAdS}_{-i\left(\Delta-\frac{d}{2}+n\right)}(\sigma^E)\;.
\end{aligned}
\end{equation}
As explained in section \ref{sec: Role of the other Series}, this expansion is valid for any value of the dimension $\Delta$. However, for $\Delta<\frac{d}{2}$, the spectral integral \eqref{eq: CFT spectral int App} no longer coincides with the series of residues \eqref{eq: G-+ CFT Series App}. In that case, the integral is given by a mixture of the poles $\mu_{\pm}^n$: 
\begin{equation}
\begin{aligned}
    &\frac{c_{\Delta}}{H^{d-1}}\int\d\mu\,\N_\mu\Gamma\left(\Delta-\frac{d}{2}\pm i\mu\right)W_\mu(\sigma^E)\\
    &= \frac{2 c_\Delta}{H^{d-1}}\sum_{n=0}^{\lfloor\frac{d}{2}-\Delta\rfloor}\frac{(-1)^n}{n!}(\Delta-\frac{d}{2}+n)\Gamma\left(2\Delta-d+n\right)\Pi^{\EAdS}_{+i\left(\Delta-\frac{d}{2}+n\right)}(\sigma^E)\\
    &+\frac{2 c_\Delta}{H^{d-1}}\sum_{n=\lceil\frac{d}{2}-\Delta\rceil}^{\infty}\frac{(-1)^n}{n!}(\Delta-\frac{d}{2}+n)\Gamma\left(2\Delta-d+n\right)\Pi^{\EAdS}_{-i\left(\Delta-\frac{d}{2}+n\right)}(\sigma^E)\;.
\end{aligned}
\end{equation}
Using the connection formula~\eqref{eq: splittin W in AdS}, one finds:
\begin{equation}\label{eq: explicit expression full spectral decomp CFT}
\begin{aligned}
    &\braket{\O(X_1^E)\O(X_2^E)} = \frac{c_\Delta}{H^{d-1}}\int\d\mu\N_\mu\Gamma\left(\Delta-\frac{d}{2}\pm i\mu\right)W_\mu(\sigma^E)\\
    &+\frac{4 \pi c_\Delta}{H^{d-1}}\sum_{n=0}^{\lfloor\frac{d}{2}-\Delta\rfloor}\frac{(-1)^n}{n!}\N_{i\left(\Delta-\frac{d}{2}+n\right)}\Gamma\left(2\Delta-d+n\right)W_{i\left(\Delta-\frac{d}{2}+n\right)}(\sigma^E)\\
    & = \int\d\mu\,\rho_\O^\P(\mu)W_\mu(\sigma^E)+4 i \pi  \sum_{n=0}^{\lfloor\frac{d}{2}-\Delta\rfloor}\textrm{Res}\left(\rho_\O^\P(\mu)W_\mu(\sigma^E),i\left(\Delta-\frac{d}{2}+n\right)\right)\;,
\end{aligned}
\end{equation}
which is the result \eqref{eq: full spectral decomposition}.

\section{Special Functions}\label{sec:special functions}

Here we collect some definitions and mathematical identities involving special functions, in particular in view of the computation of the four-point diagram in subsection \ref{subsection_single_exchange}.

\paragraph{$\Gamma$-function.} The $\Gamma$-function, defined as
\begin{equation}
    \Gamma(z)=\int_0^\infty \d t\,e^{-t} t^{z-1}\quad\,,\quad \Re(z)>0\,,
\end{equation}
can be analytically continued to $\mathbb{C}\setminus \{-p,p\in\mathbb{N}\}$ using the functional relation
\begin{equation}
    \Gamma(z+1)=z\,\Gamma(z)\;.
\end{equation}
The Gamma function is meromorphic, with poles located at negative integers $-p\,,\,p\in\mathbb{N}$. The corresponding residue is
\begin{equation}
    \mathrm{Res}\left(\Gamma(z)\,,\,z=-p\right)=\frac{(-1)^p}{p!}\,.
    \label{eq_residue_gamma}
\end{equation}
Other useful functional relations are the reflection formula and the duplication formula:
\begin{equation}
    \Gamma(z)\Gamma(1-z)=\frac{\pi}{\sin(\pi z)}\,,
    \label{reflection_Gamma}
\end{equation}
\begin{equation}
    \Gamma(2z)=\frac{2^{2z-1}}{\sqrt{\pi}}\Gamma\left(z+\frac{1}{2}\right)\Gamma(z)\,,
    \label{eq_duplication_gamma}
\end{equation}
To avoid clutter, we define the following notation:
\begin{equation}\label{eq: definition Gamma notations}
    \Gamma(a\pm b) \equiv \Gamma(a+b)\Gamma(a-b)\;.
\end{equation}
\paragraph{Hypergeometric function.} The Gauss hypergeometric function is defined by the series
\begin{equation}
    _2F_1\left(
    \begin{matrix}
        a\,,\,b \\ c
    \end{matrix}\,,z\right)=\sum_{n=0}^\infty \frac{\left(a\right)_n\left(b\right)_n}{\left(c\right)_n}\frac{z^n}{n!}\;,
    \label{eq_2F1_def}
\end{equation}
where the Pochhammer symbol is
\begin{equation}
    \left(a\right)_n\equiv\frac{\Gamma(a+n)}{\Gamma(a)}\,.
\end{equation}
The series \eqref{eq_2F1_def} is convergent for $|z|<1$. $z\mapsto\, _2F_1\left(a,b;c;z\right)$ can be analytically continued in $\mathbb{C}\setminus\left[1;+\infty\right]$. Also, due to $\Gamma$-function factors, $a,b\mapsto\,_2F_1\left(a,b;c;z\right)$ is analytic and $c\mapsto\,_2F_1\left(a,b;c;z\right)$ is meromorphic with poles at $c=-p\,,\,p\in\mathbb{N}$.

\paragraph{Associated Legendre functions: definitions.} The associated Legendre function of the first kind is defined by
\begin{equation}
    P_{i\mu-\frac{1}{2}}^{-\frac{d-3}{2}}(x)=\frac{1}{\Gamma\left(\frac{d-1}{2}\right)}\left(\frac{x-1}{x+1} \right)^{\frac{d-3}{4}}\,
    _2F_1\left(
    \begin{matrix}
        \frac{1}{2}+i\mu\,,\,\frac{1}{2}-i\mu \\ \frac{d-1}{2}
    \end{matrix}\,;\frac{1-x}{2}\right)
    \quad,\quad x>1\,.
\end{equation}
It is related to the Gegenbauer function $C_{i\mu-\frac{d-2}{2}}^{\left(\frac{d-2}{2}\right)}$ as:
\begin{equation}
    P_{i\mu-\frac{1}{2}}^{-\frac{d-3}{2}}(x)=\frac{\Gamma(d-2)\Gamma\left(\frac{4-d}{2}+i\mu\right)}{2^{\frac{d-3}{2}}\Gamma\left(\frac{d-2}{2}+i\mu\right)\Gamma\left(\frac{d-1}{2}\right)}\left(x^2-1\right)^{\frac{d-3}{4}}C_{i\mu-\frac{d-2}{2}}^{\left(\frac{d-2}{2}\right)}(x)\,,\quad x>1.
    \label{A_Legendre_to_Gegenbauer}
\end{equation}
The associated Legendre of function of the second kind is defined by
\begin{equation}
    Q_{i\mu-\frac{1}{2}}^{\frac{d-3}{2}}(x)=\frac{e^{i\pi\frac{d+1}{2}}\sqrt{\pi}\left(x^2-1\right)^{\frac{d-3}{4}}}{x^{\frac{d-3}{2}}(2x)^{ i\mu+\frac{1}{2}}}\frac{\Gamma\left(\frac{d-2}{2}+ i\mu\right)}{\Gamma\left(1+ i\mu\right)}\,_2F_1\left(
    \begin{matrix}
        \frac{d}{4}+ \frac{i\mu}{2}\,,\,\frac{d-2}{4}+ \frac{i\mu}{2} \\ 1+i\mu
    \end{matrix}\,,\frac{1}{x^2}\right)\,,x>1\,.
    \label{eq_def_legendre_Q}
\end{equation}

\paragraph{Asymptotic behaviour.} The associated Legendre functions $P_{i\mu-\frac{1}{2}}^{-\frac{d-3}{2}}$ and $Q_{\pm i\mu-\frac{1}{2}}^{\frac{d-3}{2}}$ have the following large-$\mu$ asymptotic behaviour:
\begin{equation}
    P_{i\mu-\frac{1}{2}}^{-\frac{d-3}{2}}(\cosh(\xi)) \underset{\mu\rightarrow+\infty}{=}
    \sqrt{\frac{2}{\pi \sinh(\xi)}}\frac{1}{\mu^{\frac{d}{2}-1}}\sin\left(\mu\,\xi+\frac{\pi}{4}\left(4-d\right)\right)\left(1+O\left(\frac{1}{\mu}\right)\right)\,.
    \label{asymptotics_P}
\end{equation}
\begin{equation}
    Q_{\pm i\mu-\frac{1}{2}}^{\frac{d-3}{2}}(\cosh(\xi))\underset{\mu\rightarrow+\infty}{=}e^{i\frac{\pi}{2}(d+1)}\sqrt{\frac{\pi}{2 \sinh(\xi)}}\left(\pm i\mu\right)^{\frac{d}{2}-2}e^{\mp i\mu\xi}\left(1+O\left(\frac{1}{\mu}\right)\right)\;,\quad\xi>0\,.
    \label{large_mu_Q}
\end{equation}

\paragraph{Connection formulas.}
The connection formula between the two associated Legendre functions is given by
\begin{equation}
    P_{i\mu-\frac{1}{2}}^{-\frac{d-3}{2}}(x) =\frac{i e^{-i\pi\left(\frac{d-3}{2}\right)}}{\sinh(\pi\mu)\Gamma\left(\frac{d-2}{2}\pm i\mu\right)}\left(Q_{i\mu-\frac{1}{2}}^{\frac{d-3}{2}}(x)-Q_{-i\mu-\frac{1}{2}}^{\frac{d-3}{2}}(x)\right)\,,\quad x>1\,.
    \label{connection_formula_P_Q}
\end{equation}
From \eqref{A_Legendre_to_Gegenbauer}, this connection formula can be expressed in terms of the Gegenbauer function:
\begin{equation}
    C_{i\mu-\frac{d-2}{2}}^{\left(\frac{d-2}{2}\right)}(x) =\frac{e^{-i\pi\left(\frac{d-1}{2}\right)}\sinh\left(\pi\mu+i\pi\frac{d}{2}\right)}{\sqrt{\pi}\,2^{\frac{d-3}{2}}\Gamma\left(\frac{d-2}{2}\right)\sinh(\pi\mu)}\left(x^2-1\right)^{\frac{3-d}{4}}\left(Q_{i\mu-\frac{1}{2}}^{\frac{d-3}{2}}(x)-Q_{-i\mu-\frac{1}{2}}^{\frac{d-3}{2}}(x)\right)\,,\quad x>1\,.
    \label{connection_formula_C_Q}
\end{equation}

\paragraph{Analytic continuation formulas.}
The analytic continuation formula for the associated Legendre function of the first kind at $z=e^{+i\pi}x\,,\,x>1$, is
\begin{equation}
    P_{i\mu-\frac{1}{2}}^{-\frac{d-3}{2}}(e^{+i\pi}x)= \frac{e^{-i\pi\frac{d-3}{2}}}{\sinh(\pi\mu)\Gamma\left(\frac{d-2}{2}\pm i\mu\right)}\left(e^{\pi\mu}Q_{i\mu-\frac{1}{2}}^{\frac{d-3}{2}}(x)-e^{-\pi\mu}Q_{-i\mu-\frac{1}{2}}^{\frac{d-3}{2}}(x)\right)\,.
    \label{eq_analytic_cont_P}
\end{equation}

\section{Three-Point Vertex Function}
\label{appendix-triple-K}
The three-point vertex function is defined as the triple-$K$ integral~\eqref{eq:triple-K}. It can be evaluated by mean of the Appell $F_4$ function~\cite{gradshteyn2007}:
\begin{equation}\label{eq: explicit form vertex funcion}
\begin{aligned}
    \I^{\mu_1\mu_2\mu_3}_{\k_1\k_2\k_3}=\frac{2^{\frac{d}{2}-4}}{k_3^{\frac{d}{2}}}\Bigg(A^{\mu_1,\mu_2}_{\mu_3}\left(\frac{k_1}{k_3},\frac{k_2}{k_3}\right)+A^{\mu_1,-\mu_2}_{\mu_3}\left(\frac{k_1}{k_3},\frac{k_2}{k_3}\right)+A^{-\mu_1,\mu_2}_{\mu_3}\left(\frac{k_1}{k_3},\frac{k_2}{k_3}\right)\\
    +A^{-\mu_1,-\mu_2}_{\mu_3}\left(\frac{k_1}{k_3},\frac{k_2}{k_3}\right)\Bigg)\;,
\end{aligned}
\end{equation}
where we defined $A^{\mu_1,\mu_2}_{\mu_3}(u,v)$ in terms of the regularised Appell $F_4$ function:
\begin{equation}
\begin{aligned}
\label{A-F}
     A^{\mu_1,\mu_2}_{\mu_3}(u,v)= -\frac{\mu_1\mu_2u^{i\mu_1}v^{i\mu_2}}{\N_{\mu_1}\N_{\mu_2}}\Gamma\left(\frac{\frac{d}{2}+i\mu_1+i\mu_2\pm i\mu_3}{2}\right)\\
     \times \boldsymbol{F}_4\left(\begin{array}{crc}
         \frac{\frac{d}{2}+i\mu_1+i\mu_2+ i\mu_3}{2} &, &\frac{\frac{d}{2}+i\mu_1+i\mu_2- i\mu_3}{2} \\
          1+i\mu_1 & , &1+i\mu_2 
     \end{array};u^2,v^2\right)\;.
\end{aligned}
\end{equation}
The regularised Appell $F_4$ function admits the following series expansion:
\begin{equation}
    \boldsymbol{F}_4(\alpha,\beta;\gamma,\gamma';u^2,v^2) = \frac{1}{\Gamma(\gamma)\Gamma(\gamma')}\sum_{n,m=0}^\infty\frac{(\alpha)_{n+m}(\beta)_{m+n}}{(\gamma)_n(\gamma')_m n! m!}u^{2n} v^{2m}\;,\quad \textrm{for}\quad |u|+|v| <1\;.
   \label{def-F4-series} 
\end{equation}
This is analytic in all the parameters. Therefore, the only singularities in \eqref{A-F} come from the $\Gamma$ factors. The spurious poles coming from the $1/\N_\mu$ are cancelled by some zeros of the $F_4$ function.
\vskip 4pt
The physical domain of the vertex function is $\k_1+\k_2+\k_3=0$, for which the triangle inequality imposes $k_3 \leq k_1+k_2$. In Eq.~\eqref{eq: explicit form vertex funcion}, we thus need to evaluate $F_4$ for $u \equiv k_1/k_3$ and $v \equiv k_2/k_3$ such that $u+v  \geq 1$, i.e.~by analytically continuing $F_4$ outside the domain in \eqref{def-F4-series}.

\section{Details of the Perturbative Computations}\label{sec: details PT}

\subsection{Four-Point Function}\label{appendix_single_exch}

Here we give some details about the computation of the four-point diagram in subsection~\ref{subsection_single_exchange}.

\paragraph{Background residues.}

To simplify the sum of background residues \eqref{eq: poles_++0_sinh_d_odd} and \eqref{eq: poles_++1_sinh_d_odd}, we can use the connection formula \eqref{connection_formula_P_Q} with $\mu=-i (n+1)\,,\,n\in\mathbb{N}$:
\begin{equation}
\begin{aligned}
Q_{n+\frac{1}{2}}^{\frac{d-3}{2}}\left(x\right)-Q_{-n-\frac{3}{2}}^{\frac{d-3}{2}}\left(x\right)
&=+e^{-i\pi\frac{d-3}{2}}\Gamma\left(\frac{d-4}{2}-n\right)\Gamma\left(\frac{d}{2}+n\right)\\
&\times\cos\left(\pi\left(n+\frac{1}{2}\right)\right) P_{n+\frac{1}{2}}^{-\frac{d-3}{2}}\left(x\right)\,.
\end{aligned}
\label{eq_connection_P_Q_n}
\end{equation}
When $d$ is not an even integer, each term in the above right-hand side is finite, and since $\cos(\pi(n+\frac{1}{2}))=0$,
\begin{equation}
    Q_{n+\frac{1}{2}}^{\frac{d-3}{2}}\left(x\right)-Q_{-n-\frac{3}{2}}^{\frac{d-3}{2}}\left(x\right)=0\quad,\quad d\neq 2k\,k\in\mathbb{N}\,.
    \label{connection_formula_P_Q_n_d_odd}
\end{equation}
Then, the sum of \eqref{eq: poles_++0_sinh_d_odd} and \eqref{eq: poles_++1_sinh_d_odd} vanishes and the background residues are simply given by \eqref{eq: F++B_d_odd}.\\

When $d$ is an even integer, the term $\Gamma\left(\frac{d-4}{2}-n\right)$ in \eqref{eq_connection_P_Q_n} becomes infinite for $n\geq K-1$, with
\begin{equation}
    K\equiv\frac{d-2}{2}\,,\,K\in\mathbb{N}\,.
\end{equation}
However, we still have $\cos(\pi(n+\frac{1}{2}))=0$. Then, let us see if the connection formula \eqref{connection_formula_P_Q} can be analytically continued in $\mu=-i (n+1)\,,\,n\geq K-1$, when $d$ is even. Let write \eqref{connection_formula_P_Q} for $\mu=-i (n+1+\epsilon)$, with $\epsilon>0$:
\begin{equation}
\begin{aligned}
Q_{n+\epsilon+\frac{1}{2}}^{\frac{d-3}{2}}\left(x\right)-Q_{-n-\epsilon-\frac{3}{2}}^{\frac{d-3}{2}}\left(x\right)
&=e^{-i\pi\frac{d-3}{2}}\Gamma\left(\frac{d-4}{2}-n-\epsilon\right)\Gamma\left(\frac{d}{2}+n+\epsilon\right)\\
&\times\cos\left(\pi\left(n+\epsilon+\frac{1}{2}\right)\right) P_{n+\epsilon+\frac{1}{2}}^{-\frac{d-3}{2}}\left(x\right)\,.
\end{aligned}
\label{eq_connection_P_Q_n_eps}
\end{equation}
Using the asymptotic expansions for small $\varepsilon>0$ (in particular for the $\Gamma$-function from the residue expression \eqref{eq_residue_gamma}),
\begin{subequations}
\begin{equation}
    \cos\left(\pi\left(n+\epsilon+\frac{1}{2}\right)\right)\underset{\epsilon\rightarrow 0}{=}(-1)^{n+1}\pi\,\epsilon+O(\epsilon^2)\,,
\end{equation}
\begin{equation}
    \Gamma\left(\frac{d-4}{2}-n-\epsilon\right)=\Gamma\left(K-1-n-\epsilon\right)
    \underset{\epsilon\rightarrow 0}{=}\frac{(-1)^{n-K}}{\Gamma\left(n+2-K\right)}\frac{1}{\epsilon}+O(1)\,,
\end{equation}
\end{subequations}
and the continuity of $\mu\mapsto Q_\mu^\nu(x)$, the connection formula can be analytically continued by sending $\epsilon\rightarrow0$:
\begin{equation}
\begin{aligned}
Q_{n+\frac{1}{2}}^{\frac{d-3}{2}}\left(x\right)-Q_{-n-\frac{3}{2}}^{\frac{d-3}{2}}\left(x\right)
&=-i\pi\,\frac{\Gamma\left(n+1+K\right)}{\Gamma\left(n+2-K\right)}\,P_{n+\frac{1}{2}}^{-\frac{d-3}{2}}\left(x\right)\;,\;n\geq K-1\,.
\end{aligned}
\label{eq_connection_P_Q_n_d_even}
\end{equation}
This last formula can be used to simplify the sum of the background residues to find \eqref{eq: F++B_d_even}.

\paragraph{Analytical Properties of KLF integrands}

\paragraph{Analytic structure of $\mu\mapsto\frac{\mu}{\sinh(\pi\mu)}$.}
The background terms \eqref{eq: poles_++0_sinh_d_odd}, \eqref{eq: poles_++1_sinh_d_odd} in $\hat{F}_{++}$ come from the function $\mu\mapsto\frac{\mu}{\sinh(\pi\mu)}$, which has poles at values $\mu=i n\,,\,n\in\mathbb{Z}^*$, with residues:
\begin{equation}
    \mathrm{Res}\left(\mu\mapsto\frac{\mu}{\sinh(\pi\mu)}\,;\,\mu=i n\,,\,n\in\mathbb{Z}^*\right)=\frac{(-1)^n i n}{\pi}\,.
    \label{residues_sinh_-1}
\end{equation}

\paragraph{Analytic structure of $\mu\mapsto Q_{\pm i\mu-\frac{1}{2}}^{\frac{d-3}{2}}(x)$.}
The associated Legendre functions $Q_{\pm i\mu-\frac{1}{2}}^{\frac{d-3}{2}}$ \eqref{eq_def_legendre_Q} that appear in the integrands of $\hat{F}_{++}^0$ and $\hat{F}_{++}^{1}$ have a different analytic structure in the complex $\mu$ plane depending on $d$.

When $d$ is any complex number that is not an even integer, the singularities only come from the $\Gamma(\frac{d-2}{2}+i\mu)$ factor. Indeed, even if the Gauss hypergeometric function \eqref{eq_2F1_def} is singular for $c=-p\,,\,p\in\mathbb{N}$, the dressed hypergeometric function
\begin{equation}
    \,_2\mathcal{F}_1\left(
    \begin{matrix}
        a\,,\,b \\ c
    \end{matrix}\,,z\right)
    \equiv\frac{1}{\Gamma(c)}\,_2F_1\left(
    \begin{matrix}
        a\,,\,b \\ c
    \end{matrix}\,,z\right)\,
    \label{def_dressed_2F1}
\end{equation}
that enters the expression of $Q_{\pm i\mu-\frac{1}{2}}$ is analytic in $c$. 
Then, from the residues of the $\Gamma$-function, the residues of $\mu\mapsto Q_{\pm i\mu-\frac{1}{2}}^{\frac{d-3}{2}}$ are
\begin{equation}
\begin{aligned}
&\mathrm{Res}\left(\mu\mapsto Q_{\pm i\mu-\frac{1}{2}}^{\frac{d-3}{2}}(x)\,;\,\mu=\pm i\left(n+\frac{d-2}{2}\right)\,,\,n\in\mathbb{N}\right)\\
&=-i\frac{e^{i\pi\frac{d-3}{2}}\sqrt{\pi}\left(x^2-1\right)^{\frac{d-3}{4}}}{2^{-\frac{d-3}{2}}(2x)^{-n}}
\frac{(-1)^n}{n!}
\frac{1}{\Gamma\left(-n-\frac{d-4}{2}\right)}\,_2F_1\left(
    \begin{matrix}
        \frac{1-n}{2}\,,\,-\frac{n}{2} \\ -n-\frac{d-4}{2}
    \end{matrix}\,,\frac{1}{x^2}\right)\,,
\end{aligned}
\label{residues_Q_d_odd}
\end{equation}

When $d$ is even, $\mu\mapsto Q_{\pm i\mu-\frac{1}{2}}^{\frac{d-3}{2}}(x)$ is analytic in the whole complex plane. This is because each singularity coming from the $\Gamma(\frac{d-2}{2}+i\mu)$ factor is cancelled at the same point by the zero of the dressed hypergeometric function \eqref{def_dressed_2F1} that enters \eqref{eq_def_legendre_Q}. Let us see this explicitly for $\mu\mapsto Q_{i\mu-\frac{1}{2}}^{\frac{d-3}{2}}(x)$. First, it can be shown that for $m\in\mathbb{N}$,
\begin{equation}
    \lim_{c\rightarrow-n}\,_2\mathcal{F}_1\left(
    \begin{matrix}
        a\,,\,b \\ c
    \end{matrix}\,,z\right)=\frac{\left(a\right)_{n+1}\left(b\right)_{n+1}}{\left(n+1\right)!}\,x^{n+1}\,_2F_1\left(
    \begin{matrix}
        a+n+1\,,\,b+n+1 \\ n+2
    \end{matrix}\,,z\right)\,.
\end{equation}
Then, because the poles of $\mu\mapsto\Gamma\left(K+i\mu\right)$ are located at $\mu=i(n+K)$, we look at the following limit:
\begin{equation}
\begin{aligned}
    \frac{\Gamma\left(K+ i\mu\right)}{\Gamma\left(1+ i\mu\right)}&\,_2F_1\left(
    \begin{matrix}
        \frac{K+1+i\mu}{2}\,,\,\frac{K+i\mu}{2} \\ 1+i\mu
    \end{matrix}\,,\frac{1}{x^2}\right)\\
    \underset{\mu\rightarrow i(n+K)}{\sim}&
    \frac{\Gamma\left(K+ i\mu\right)}{\Gamma\left(\frac{K+1+i\mu}{2}\right)\Gamma\left(\frac{K+i\mu}{2}\right)}\frac{\Gamma\left(\frac{K+1+i\mu}{2}+n+K\right)\Gamma\left(\frac{K+i\mu}{2}+n+K\right)}{\left(n+K\right)!}\\
    &\times x^{-2(n+K)}\,_2F_1\left(
    \begin{matrix}
        \frac{K+1+i\mu}{2}+n+K\,,\,\frac{K+i\mu}{2}+n+K \\ n+K+1
    \end{matrix}\,,\frac{1}{x^2}\right)\\
    \underset{\mu\rightarrow i(n+K)}{\sim}
    &\frac{2^{-(n+1)}}{\sqrt{\pi}}\frac{\Gamma\left(\frac{K+1+i\mu}{2}+n+K\right)\Gamma\left(\frac{K+i\mu}{2}+n+K\right)}{\left(n+K\right)!}\\
    &\times x^{-2(n+K)}\,_2F_1\left(
    \begin{matrix}
        \frac{K+1+i\mu}{2}+n+K\,,\,\frac{K+i\mu}{2}+n+K \\ n+K+1
    \end{matrix}\,,\frac{1}{x^2}\right)\,,
\end{aligned}
\end{equation}
where we used the duplication formula \eqref{eq_duplication_gamma} in the last line. The last term is finite in the limit $\mu\rightarrow i(n+K)$, therefore $\mu\mapsto Q_{i\mu-\frac{1}{2}}^{\frac{d-3}{2}}$ is analytic when $d$ is even (a similar derivation can be done for $\mu\mapsto Q_{-i\mu-\frac{d-2}{2}}^{\frac{1}{2}}$).

\paragraph{Analytic structure of $\mu\mapsto\frac{1}{\left(\mu^2-\mu_\chi^2\right)_{i\varepsilon}}$.}
From the $i \epsilon$-prescription expression \eqref{eq: i epsilon prescription} and as shown in Fig.~\ref{fig: KLF modes analytical structure}, poles of $\mu\mapsto\frac{1}{\left(\mu^2-\mu_\chi^2\right)_{i\varepsilon}}$ are located at $\mu=\pm\mu_\chi\pm i \epsilon$. Their residues are the following:
\begin{equation}
\begin{aligned}
  &  \mathrm{Res}\left(\mu\mapsto\frac{1}{\left(\mu^2-\mu_\chi^2\right)_{i\epsilon}}\,;\,\mu=\pm \mu_\chi-i\epsilon\right)=
    \frac{e^{\pm \pi\mu_\chi}}{4\mu_\chi\sinh(\pi\mu_\chi)}\,, \\
   & \mathrm{Res}\left(\mu\mapsto\frac{1}{\left(\mu^2-\mu_\chi^2\right)_{i\epsilon}}\,;\,\mu=\pm \mu_\chi+i\epsilon\right)=
    -\frac{e^{\mp \pi\mu_\chi}}{4\mu_\chi\sinh(\pi\mu_\chi)}\,.
\label{eq_i_eps_residues}
\end{aligned}
\end{equation}

\subsection{Loop Diagram}
\label{sec:loop-from-2-1-spectral-integral}

In this section, we fill two gaps from section~\eqref{sec: loop two point} about the one-loop correction to the scalar propagator. 
\vskip 4pt
\paragraph{Reduction of the number of spectral integrations.}
First, let us show how the double spectral integral \eqref{eq: double spectral integral GA} can be turned into the single spectral integral \eqref{eq: single spectral integral GA}. Let us consider the product of two Euclidean position-space propagators. Using \eqref{Euclidean-two-point-spectral-integral}, this can be written as
\begin{equation}
     \C_{\a\b}^{E,\chi_1}(\sigma^E)\,\C_{\a\b}^{E,\chi_2}(\sigma^E)=\int\displaylimits^\infty_{-\infty}\d\mu\N_\mu\d\mu'\N_{\mu'}W_\mu(\sigma^E)W_{\mu'}(\sigma^E)\G^{\chi_1}_{\a\b}(\mu)\G^{\chi_1}_{\a\b}(\mu')\;.
\end{equation}
Using the property of the spectral density \eqref{eq: definition rho chi1 chi2}, this reads
\begin{equation}
     \C_{\a\b}^{E,\chi_1}(\sigma^E)\,\C_{\a\b}^{E,\chi_2}(\sigma^E)=\int\displaylimits^\infty_{-\infty}\d\nu W_\nu(\sigma^E)\int\displaylimits^\infty_{-\infty}\d\mu\N_\mu\d\mu'\N_{\mu'}\rho_{\mu\mu'}(\nu)\G^{\chi_1}_{\a\b}(\mu)\G^{\chi_1}_{\a\b}(\mu')\;.
\label{expression-1}     
\end{equation}

At the same time, the Euclidean position-space propagators are simply expressed in terms of analytic continuations of $W_\mu$, see Eq.~\eqref{eq: Euclidean real space propagators}. Considering the same continuations, the property \eqref{eq: definition rho chi1 chi2} thus implies that
\begin{equation}
 \C_{\a\b}^{E,\chi_1}(\sigma^E)\,\C_{\a\b}^{E,\chi_2}(\sigma^E) = \int\displaylimits^\infty_{-\infty}\d\mu\rho_{\chi_1\chi_2}(\mu)\C^{E,\mu}_{\a\b}(\sigma^E)  \,.  
\end{equation}
Using again \eqref{Euclidean-two-point-spectral-integral}, this gives
\begin{equation}
 \C_{\a\b}^{E,\chi_1}(\sigma^E)\,\C_{\a\b}^{E,\chi_2}(\sigma^E) = \int\displaylimits^\infty_{-\infty}\d\nu\N_\nu W_\nu(\sigma^E)\int\displaylimits^\infty_{-\infty}\d\mu\rho_{\chi_1\chi_2}(\mu)\G_{\a\b}^{\mu}(\nu)\;.
\label{expression-2}   
\end{equation}
We now use that $\N_\nu W_\nu(\sigma^E)$ form a basis of square integrable bi-local functions in EAdS and compare the expressions \eqref{expression-1} and \eqref{expression-2} to deduce that
\begin{equation}
    \int\displaylimits^\infty_{-\infty}\d\mu\N_\mu\d\mu'\N_{\mu'}\rho_{\mu\mu'}(\nu)
    \G^{\chi_1}_{\a\b}(\mu)\G^{\chi_1}_{\a\b}(\mu') = \N_{\nu} \int\displaylimits^\infty_{-\infty}\d\mu\rho_{\chi_1\chi_2}(\mu)\G_{\a\b}^{\mu}(\nu)\;,
\end{equation}
which is the desired expression.
\paragraph{Summation of the four $\b_1\b_2$ contributions.}
Let us explain how to make the final sum over the internal vertices in expression~\eqref{eq: mixing equation}. To this aim, we consider an alternative theory with a quadratic mixing between the external field $\varphi$ and a massive field $\chi$ of mass $m_\chi= H M_\mu$:
\begin{equation}\label{eq: quadratic mixing}
    \L_\textrm{mix}=-g\varphi\chi\;.
\end{equation}
According to the KLF space Feynman rules~\eqref{eq: Feynman rule vertex}, this generates a vertex of the form:
\begin{equation}
     \begin{tikzpicture}[baseline={(0,0)}]

        \draw[thick,black] (-1.,0) to (-0.,0);

        \draw[thick,black] (-0.,0) to (1,0);
        
        \filldraw[color=black,fill=black] (0,0) circle (2pt);
        
        \node[black] at (0,0.3) {$\a$};
        
        \node[black] at (-1.6,0.) {$\mu_1,\k_1$};
        
        \node[black] at (1.6,0.) {$\mu_2,\k_2$};
        
        \end{tikzpicture}
        = -i\a g e^{\frac{i\a\pi d}{2}}\delta\left(^{\mu_1\;\mu_2}_{\k_1-\k_2}\right)\;.
\end{equation}
Therefore, at first order in $g$, one can write the following KLF space mixed two-point function:
\begin{equation}
\begin{aligned}
    \braket{\Omega|\varphi^{\mu_1}_{\k_1}\chi^{\mu_2}_{\k_2}|\Omega}_{\a_1\b_2} &=\sum_{\b_1=\pm}\begin{tikzpicture}[baseline={(0,0)}]

        \draw[thick,black] (-1.,0) to (-0.,0);

        \draw[thick,pyblue] (-0.,0) to (1,0);
        
        \filldraw[color=black,fill=black] (1.,0) circle (2pt);
        \filldraw[color=black,fill=black] (-1.,0) circle (2pt);

        \filldraw[color=black,fill=black] (-0.,0) circle (2pt);

        \node[black] at (-1.7,0.) {$\mu_1,\k_2$};
        
        \node[black] at (1.7,0.) {$\mu_2,\k_2$};

        \node[black] at (0,0.3) {$\b_1$};

        \node[black] at (-1,0.3) {$\a_1$};

        \node[black] at (1,0.3) {$\b_2$};

        \node[black] at (-0.5,-0.3) {$\varphi$};

        \node[pyblue] at (0.5,-0.3) {$\chi$};
        
        \end{tikzpicture}\\
        &= -i g \delta\left(^{\mu_1\;\mu_2}_{\k_1-\k_2}\right)\sum_{\b_1=\pm}\b_1 e^{\frac{i\b_1 \pi d}{2}}\frac{\G^{\mu_\varphi}_{\a_1\b_1}(\mu_1)}{H^2}\frac{\G^{\mu}_{\b_1\b_2}(\mu_1)}{H^2}\;.
\end{aligned}
\label{mixed-two-point-form-1}
\end{equation}
As noted in~\cite{Higuchi:2010xt}, the presence of the mixing~\eqref{eq: quadratic mixing} in the interacting sector is an artefact of the choice of field variables. Indeed, if one consider a free theory with fields $\Psi_1$ and $\Psi_2$ of masses $m_1$ and $m_2$, it can be mapped to our $\varphi-\chi$ lagrangian by an $\SO(2)$ rotation in field space:
\begin{equation}\label{eq: SO(2) rotation field space}
    \left(\begin{array}{c}
         \varphi  \\
         \chi 
    \end{array}\right)=\left(\begin{array}{cc}
        \cos(\omega) & -\sin(\omega) \\
        \sin(\omega) & \cos(\omega)
    \end{array}\right)\cdot\left(\begin{array}{c}
         \Psi_1  \\
         \Psi_2 
    \end{array}\right)\;.
\end{equation}
The parameters in the initial Lagrangian are then related to $m_{1,2}$ and $\omega$ as:
\begin{equation}
\begin{aligned}
    m_\varphi^2 & = m_1^2\cos^2(\omega) + m_2^2\sin^2(\omega)\;,\\
    m_\chi^2 & = m_1^2\sin^2(\omega) + m_2^2\cos^2(\omega)\;,\\
    g &=  (m_1^2-m_2^2)\sin(\omega)\cos(\omega)\;.
\end{aligned}
\end{equation}
Using the rotation~\eqref{eq: SO(2) rotation field space}, the mixed two-point function can be written as a combination of free propagators:
\begin{equation}\label{eq: mixed two point rotation}
    \braket{\Omega|\varphi^{\mu_1}_{\k_1}\chi^{\mu_2}_{\k_2}|\Omega}_{\a_1\b_2} = \frac{g}{m_1^2-m_2^2}\left(\braket{\Omega|\Psi^{\mu_1}_{1,\k_1}\Psi^{\mu_2}_{1,\k_2}|\Omega}_{\a_1\b_2}-\braket{\Omega|\Psi^{\mu_1}_{2,\k_1}\Psi^{\mu_2}_{2,\k_2}|\Omega}_{\a_1\b_2}\right)\;,
\end{equation}
where we used the fact that the two-point functions between $\Psi_1$ and $\Psi_2$ vanishes. As the two terms in~\eqref{eq: mixed two point rotation} are simply given by free KLF propagators, one obtains:
\begin{equation}
    \braket{\Omega|\varphi^{\mu_1}_{\k_1}\chi^{\mu_2}_{\k_2}|\Omega}_{\a_1\b_2} = g\delta\left(^{\mu_1\;\mu_2}_{\k_1-\k_2}\right)\frac{\G^{\Psi_1}_{\a_1\b_2}(\mu_1)-\G^{\Psi_2}_{\a_1\b_2}(\mu_1)}{H^2(m_1^2-m_2^2)}\;.
\label{mixed-two-point-form-2}
\end{equation}
Now, comparing Eqs~\eqref{mixed-two-point-form-1} and \eqref{mixed-two-point-form-2} consistently at order $g$, one can identify that:
\begin{equation}
    -i \sum_{\b_1=\pm}\b_1 e^{\frac{i\b_1 \pi d}{2}}\G^{\mu_\varphi}_{\a_1\b_1}(\mu_1)\G^{\mu}_{\b_1\b_2}(\mu_1) = -\frac{\G^{\mu}_{\a_1\b_2}(\mu_1)}{\mu_\varphi^2-\mu^2}\;,
    \label{summation rule for the KLF propagators}
\end{equation}
where we discarded the second term as it does not contribute for principal series fields~\cite{Higuchi:2010xt}. Using twice Eq.~\eqref{summation rule for the KLF propagators}, the sum in \eqref{eq: mixing equation} therefore simplifies to:
\begin{equation}
    -g^2 \sum_{\b_1,\b_2}\b_1\b_2 e^{\frac{i\pi(\b_1+\b_2)d}{2}}\frac{\G^{\mu_\varphi}_{\a_1\b_1}(\mu_1)}{H^2}\frac{\G^\mu_{\b_1\b_2}(\mu_1)}{H^2}\frac{\G^{\mu_\varphi}_{\b_2\a_2}(\mu_1)}{H^2} = \frac{g^2}{H^6}\frac{\G^{\mu}_{\a_1\a_2}(\mu_1)}{(\mu_\varphi^2-\mu^2)^2}\;.
    \label{simplification-double-sum-loop}
\end{equation}
\bibliographystyle{JHEP}
\bibliography{Bibliography}

\providecommand{\href}[2]{#2}\begingroup\raggedright\begin{thebibliography}{100}

\bibitem{Froissart:1961ux}
M.~Froissart, \emph{{Asymptotic behavior and subtractions in the Mandelstam
  representation}}, \href{https://doi.org/10.1103/PhysRev.123.1053}{\emph{Phys.
  Rev.} {\bfseries 123} (1961) 1053}.

\bibitem{Martin:1962rt}
A.~Martin, \emph{{Unitarity and high-energy behavior of scattering
  amplitudes}}, \href{https://doi.org/10.1103/PhysRev.129.1432}{\emph{Phys.
  Rev.} {\bfseries 129} (1963) 1432}.

\bibitem{Bros:1965kbd}
J.~Bros, H.~Epstein and V.~Glaser, \emph{{A proof of the crossing property for
  two-particle amplitudes in general quantum field theory}},
  \href{https://doi.org/10.1007/BF01646307}{\emph{Commun. Math. Phys.}
  {\bfseries 1} (1965) 240}.

\bibitem{Martin:1965jj}
A.~Martin, \emph{{Extension of the axiomatic analyticity domain of scattering
  amplitudes by unitarity. 1.}},
  \href{https://doi.org/10.1007/BF02720568}{\emph{Nuovo Cim. A} {\bfseries 42}
  (1965) 930}.

\bibitem{Paulos:2017fhb}
M.~F. Paulos, J.~Penedones, J.~Toledo, B.~C. van Rees and P.~Vieira, \emph{{The
  S-matrix bootstrap. Part III: higher dimensional amplitudes}},
  \href{https://doi.org/10.1007/JHEP12(2019)040}{\emph{JHEP} {\bfseries 12}
  (2019) 040} [\href{https://arxiv.org/abs/1708.06765}{{\ttfamily
  1708.06765}}].

\bibitem{Bellazzini:2020cot}
B.~Bellazzini, J.~Elias~Mir{\'o}, R.~Rattazzi, M.~Riembau and F.~Riva,
  \emph{{Positive moments for scattering amplitudes}},
  \href{https://doi.org/10.1103/PhysRevD.104.036006}{\emph{Phys. Rev. D}
  {\bfseries 104} (2021) 036006}
  [\href{https://arxiv.org/abs/2011.00037}{{\ttfamily 2011.00037}}].

\bibitem{Correia:2020xtr}
M.~Correia, A.~Sever and A.~Zhiboedov, \emph{{An analytical toolkit for the
  S-matrix bootstrap}},
  \href{https://doi.org/10.1007/JHEP03(2021)013}{\emph{JHEP} {\bfseries 03}
  (2021) 013} [\href{https://arxiv.org/abs/2006.08221}{{\ttfamily
  2006.08221}}].

\bibitem{Kruczenski:2022lot}
M.~Kruczenski, J.~Penedones and B.~C. van Rees, \emph{{Snowmass White Paper:
  S-matrix Bootstrap}},  \href{https://arxiv.org/abs/2203.02421}{{\ttfamily
  2203.02421}}.

\bibitem{Rattazzi:2008pe}
R.~Rattazzi, V.~S. Rychkov, E.~Tonni and A.~Vichi, \emph{{Bounding scalar
  operator dimensions in 4D CFT}},
  \href{https://doi.org/10.1088/1126-6708/2008/12/031}{\emph{JHEP} {\bfseries
  12} (2008) 031} [\href{https://arxiv.org/abs/0807.0004}{{\ttfamily
  0807.0004}}].

\bibitem{Rychkov:2009ij}
V.~S. Rychkov and A.~Vichi, \emph{{Universal Constraints on Conformal Operator
  Dimensions}}, \href{https://doi.org/10.1103/PhysRevD.80.045006}{\emph{Phys.
  Rev. D} {\bfseries 80} (2009) 045006}
  [\href{https://arxiv.org/abs/0905.2211}{{\ttfamily 0905.2211}}].

\bibitem{Caracciolo:2009bx}
F.~Caracciolo and V.~S. Rychkov, \emph{{Rigorous Limits on the Interaction
  Strength in Quantum Field Theory}},
  \href{https://doi.org/10.1103/PhysRevD.81.085037}{\emph{Phys. Rev. D}
  {\bfseries 81} (2010) 085037}
  [\href{https://arxiv.org/abs/0912.2726}{{\ttfamily 0912.2726}}].

\bibitem{Rattazzi:2010gj}
R.~Rattazzi, S.~Rychkov and A.~Vichi, \emph{{Central Charge Bounds in 4D
  Conformal Field Theory}},
  \href{https://doi.org/10.1103/PhysRevD.83.046011}{\emph{Phys. Rev. D}
  {\bfseries 83} (2011) 046011}
  [\href{https://arxiv.org/abs/1009.2725}{{\ttfamily 1009.2725}}].

\bibitem{Dolan:2000ut}
F.~A. Dolan and H.~Osborn, \emph{{Conformal four point functions and the
  operator product expansion}},
  \href{https://doi.org/10.1016/S0550-3213(01)00013-X}{\emph{Nucl. Phys. B}
  {\bfseries 599} (2001) 459}
  [\href{https://arxiv.org/abs/hep-th/0011040}{{\ttfamily hep-th/0011040}}].

\bibitem{Dolan:2003hv}
F.~A. Dolan and H.~Osborn, \emph{{Conformal partial waves and the operator
  product expansion}},
  \href{https://doi.org/10.1016/j.nuclphysb.2003.11.016}{\emph{Nucl. Phys. B}
  {\bfseries 678} (2004) 491}
  [\href{https://arxiv.org/abs/hep-th/0309180}{{\ttfamily hep-th/0309180}}].

\bibitem{Costa:2011dw}
M.~S. Costa, J.~Penedones, D.~Poland and S.~Rychkov, \emph{{Spinning Conformal
  Blocks}}, \href{https://doi.org/10.1007/JHEP11(2011)154}{\emph{JHEP}
  {\bfseries 11} (2011) 154} [\href{https://arxiv.org/abs/1109.6321}{{\ttfamily
  1109.6321}}].

\bibitem{Callan:1989em}
C.~G. Callan, Jr. and F.~Wilczek, \emph{{INFRARED BEHAVIOR AT NEGATIVE
  CURVATURE}}, \href{https://doi.org/10.1016/0550-3213(90)90451-I}{\emph{Nucl.
  Phys. B} {\bfseries 340} (1990) 366}.

\bibitem{Paulos:2016fap}
M.~F. Paulos, J.~Penedones, J.~Toledo, B.~C. van Rees and P.~Vieira, \emph{{The
  S-matrix bootstrap. Part I: QFT in AdS}},
  \href{https://doi.org/10.1007/JHEP11(2017)133}{\emph{JHEP} {\bfseries 11}
  (2017) 133} [\href{https://arxiv.org/abs/1607.06109}{{\ttfamily
  1607.06109}}].

\bibitem{Carmi:2018qzm}
D.~Carmi, L.~Di~Pietro and S.~Komatsu, \emph{{A Study of Quantum Field Theories
  in AdS at Finite Coupling}},
  \href{https://doi.org/10.1007/JHEP01(2019)200}{\emph{JHEP} {\bfseries 01}
  (2019) 200} [\href{https://arxiv.org/abs/1810.04185}{{\ttfamily
  1810.04185}}].

\bibitem{Mack:1974jjo}
G.~Mack, \emph{{Group Theoretical Approach to Conformal Invariant Quantum Field
  Theory}}, \href{https://doi.org/10.1007/978-1-4615-8909-9_7}{\emph{NATO Sci.
  Ser. B} {\bfseries 5} (1974) 123}.

\bibitem{Mack:1974sa}
G.~Mack, \emph{{Osterwalder-Schrader Positivity in Conformal Invariant Quantum
  Field Theory}}, \href{https://doi.org/10.1007/3-540-07160-1_3}{\emph{Lect.
  Notes Phys.} {\bfseries 37} (1975) 66}.

\bibitem{Dobrev:1975ru}
V.~K. Dobrev, V.~B. Petkova, S.~G. Petrova and I.~T. Todorov, \emph{{Dynamical
  Derivation of Vacuum Operator Product Expansion in Euclidean Conformal
  Quantum Field Theory}},
  \href{https://doi.org/10.1103/PhysRevD.13.887}{\emph{Phys. Rev. D} {\bfseries
  13} (1976) 887}.

\bibitem{Dobrev:1977qv}
V.~K. Dobrev, G.~Mack, V.~B. Petkova, S.~G. Petrova and I.~T. Todorov,
  \emph{{Harmonic Analysis on the n-Dimensional Lorentz Group and Its
  Application to Conformal Quantum Field Theory}}, vol.~63. 1977,
  \href{https://doi.org/10.1007/BFb0009678}{10.1007/BFb0009678}.

\bibitem{Hogervorst:2017sfd}
M.~Hogervorst and B.~C. van Rees, \emph{{Crossing symmetry in alpha space}},
  \href{https://doi.org/10.1007/JHEP11(2017)193}{\emph{JHEP} {\bfseries 11}
  (2017) 193} [\href{https://arxiv.org/abs/1702.08471}{{\ttfamily
  1702.08471}}].

\bibitem{Simmons-Duffin:2017nub}
D.~Simmons-Duffin, D.~Stanford and E.~Witten, \emph{{A spacetime derivation of
  the Lorentzian OPE inversion formula}},
  \href{https://doi.org/10.1007/JHEP07(2018)085}{\emph{JHEP} {\bfseries 07}
  (2018) 085} [\href{https://arxiv.org/abs/1711.03816}{{\ttfamily
  1711.03816}}].

\bibitem{Karateev:2018oml}
D.~Karateev, P.~Kravchuk and D.~Simmons-Duffin, \emph{{Harmonic Analysis and
  Mean Field Theory}},
  \href{https://doi.org/10.1007/JHEP10(2019)217}{\emph{JHEP} {\bfseries 10}
  (2019) 217} [\href{https://arxiv.org/abs/1809.05111}{{\ttfamily
  1809.05111}}].

\bibitem{Hogervorst:2021uvp}
M.~Hogervorst, J.~Penedones and K.~S. Vaziri, \emph{{Towards the
  non-perturbative cosmological bootstrap}},
  \href{https://doi.org/10.1007/JHEP02(2023)162}{\emph{JHEP} {\bfseries 02}
  (2023) 162} [\href{https://arxiv.org/abs/2107.13871}{{\ttfamily
  2107.13871}}].

\bibitem{DiPietro:2021sjt}
L.~Di~Pietro, V.~Gorbenko and S.~Komatsu, \emph{{Analyticity and unitarity for
  cosmological correlators}},
  \href{https://doi.org/10.1007/JHEP03(2022)023}{\emph{JHEP} {\bfseries 03}
  (2022) 023} [\href{https://arxiv.org/abs/2108.01695}{{\ttfamily
  2108.01695}}].

\bibitem{SalehiVaziri:2024joi}
K.~Salehi~Vaziri, \emph{{A non-perturbative construction of the de Sitter
  late-time boundary}},  \href{https://arxiv.org/abs/2412.00183}{{\ttfamily
  2412.00183}}.

\bibitem{Maldacena:2011nz}
J.~M. Maldacena and G.~L. Pimentel, \emph{{On graviton non-Gaussianities during
  inflation}}, \href{https://doi.org/10.1007/JHEP09(2011)045}{\emph{JHEP}
  {\bfseries 09} (2011) 045} [\href{https://arxiv.org/abs/1104.2846}{{\ttfamily
  1104.2846}}].

\bibitem{Bzowski:2011ab}
A.~Bzowski, P.~McFadden and K.~Skenderis, \emph{{Holographic predictions for
  cosmological 3-point functions}},
  \href{https://doi.org/10.1007/JHEP03(2012)091}{\emph{JHEP} {\bfseries 03}
  (2012) 091} [\href{https://arxiv.org/abs/1112.1967}{{\ttfamily 1112.1967}}].

\bibitem{Mata:2012bx}
I.~Mata, S.~Raju and S.~Trivedi, \emph{{CMB from CFT}},
  \href{https://doi.org/10.1007/JHEP07(2013)015}{\emph{JHEP} {\bfseries 07}
  (2013) 015} [\href{https://arxiv.org/abs/1211.5482}{{\ttfamily 1211.5482}}].

\bibitem{Bzowski:2013sza}
A.~Bzowski, P.~McFadden and K.~Skenderis, \emph{{Implications of conformal
  invariance in momentum space}},
  \href{https://doi.org/10.1007/JHEP03(2014)111}{\emph{JHEP} {\bfseries 03}
  (2014) 111} [\href{https://arxiv.org/abs/1304.7760}{{\ttfamily 1304.7760}}].

\bibitem{Kundu:2014gxa}
N.~Kundu, A.~Shukla and S.~P. Trivedi, \emph{{Constraints from Conformal
  Symmetry on the Three Point Scalar Correlator in Inflation}},
  \href{https://doi.org/10.1007/JHEP04(2015)061}{\emph{JHEP} {\bfseries 04}
  (2015) 061} [\href{https://arxiv.org/abs/1410.2606}{{\ttfamily 1410.2606}}].

\bibitem{Kundu:2015xta}
N.~Kundu, A.~Shukla and S.~P. Trivedi, \emph{{Ward Identities for Scale and
  Special Conformal Transformations in Inflation}},
  \href{https://doi.org/10.1007/JHEP01(2016)046}{\emph{JHEP} {\bfseries 01}
  (2016) 046} [\href{https://arxiv.org/abs/1507.06017}{{\ttfamily
  1507.06017}}].

\bibitem{Ghosh:2014kba}
A.~Ghosh, N.~Kundu, S.~Raju and S.~P. Trivedi, \emph{{Conformal Invariance and
  the Four Point Scalar Correlator in Slow-Roll Inflation}},
  \href{https://doi.org/10.1007/JHEP07(2014)011}{\emph{JHEP} {\bfseries 07}
  (2014) 011} [\href{https://arxiv.org/abs/1401.1426}{{\ttfamily 1401.1426}}].

\bibitem{Shukla:2016bnu}
A.~Shukla, S.~P. Trivedi and V.~Vishal, \emph{{Symmetry constraints in
  inflation, $\alpha$-vacua, and the three point function}},
  \href{https://doi.org/10.1007/JHEP12(2016)102}{\emph{JHEP} {\bfseries 12}
  (2016) 102} [\href{https://arxiv.org/abs/1607.08636}{{\ttfamily
  1607.08636}}].

\bibitem{Arkani-Hamed:2018kmz}
N.~Arkani-Hamed, D.~Baumann, H.~Lee and G.~L. Pimentel, \emph{{The Cosmological
  Bootstrap: Inflationary Correlators from Symmetries and Singularities}},
  \href{https://doi.org/10.1007/JHEP04(2020)105}{\emph{JHEP} {\bfseries 04}
  (2020) 105} [\href{https://arxiv.org/abs/1811.00024}{{\ttfamily
  1811.00024}}].

\bibitem{Baumann:2019oyu}
D.~Baumann, C.~Duaso~Pueyo, A.~Joyce, H.~Lee and G.~L. Pimentel, \emph{{The
  cosmological bootstrap: weight-shifting operators and scalar seeds}},
  \href{https://doi.org/10.1007/JHEP12(2020)204}{\emph{JHEP} {\bfseries 12}
  (2020) 204} [\href{https://arxiv.org/abs/1910.14051}{{\ttfamily
  1910.14051}}].

\bibitem{Baumann:2020dch}
D.~Baumann, C.~Duaso~Pueyo, A.~Joyce, H.~Lee and G.~L. Pimentel, \emph{{The
  Cosmological Bootstrap: Spinning Correlators from Symmetries and
  Factorization}},
  \href{https://doi.org/10.21468/SciPostPhys.11.3.071}{\emph{SciPost Phys.}
  {\bfseries 11} (2021) 071}
  [\href{https://arxiv.org/abs/2005.04234}{{\ttfamily 2005.04234}}].

\bibitem{Pimentel:2022fsc}
G.~L. Pimentel and D.-G. Wang, \emph{{Boostless cosmological collider
  bootstrap}}, \href{https://doi.org/10.1007/JHEP10(2022)177}{\emph{JHEP}
  {\bfseries 10} (2022) 177}
  [\href{https://arxiv.org/abs/2205.00013}{{\ttfamily 2205.00013}}].

\bibitem{Jazayeri:2022kjy}
S.~Jazayeri and S.~Renaux-Petel, \emph{{Cosmological bootstrap in slow
  motion}}, \href{https://doi.org/10.1007/JHEP12(2022)137}{\emph{JHEP}
  {\bfseries 12} (2022) 137}
  [\href{https://arxiv.org/abs/2205.10340}{{\ttfamily 2205.10340}}].

\bibitem{Qin:2022fbv}
Z.~Qin and Z.-Z. Xianyu, \emph{{Helical inflation correlators: partial
  Mellin-Barnes and bootstrap equations}},
  \href{https://doi.org/10.1007/JHEP04(2023)059}{\emph{JHEP} {\bfseries 04}
  (2023) 059} [\href{https://arxiv.org/abs/2208.13790}{{\ttfamily
  2208.13790}}].

\bibitem{Qin:2023ejc}
Z.~Qin and Z.-Z. Xianyu, \emph{{Closed-form formulae for inflation
  correlators}}, \href{https://doi.org/10.1007/JHEP07(2023)001}{\emph{JHEP}
  {\bfseries 07} (2023) 001}
  [\href{https://arxiv.org/abs/2301.07047}{{\ttfamily 2301.07047}}].

\bibitem{Aoki:2024uyi}
S.~Aoki, L.~Pinol, F.~Sano, M.~Yamaguchi and Y.~Zhu, \emph{{Cosmological
  correlators with double massive exchanges: bootstrap equation and
  phenomenology}}, \href{https://doi.org/10.1007/JHEP09(2024)176}{\emph{JHEP}
  {\bfseries 09} (2024) 176}
  [\href{https://arxiv.org/abs/2404.09547}{{\ttfamily 2404.09547}}].

\bibitem{Qin:2025xct}
Z.~Qin, S.~Renaux-Petel, X.~Tong, D.~Werth and Y.~Zhu, \emph{{The exact and
  approximate tales of boost-breaking cosmological correlators}},
  \href{https://doi.org/10.1088/1475-7516/2025/09/058}{\emph{JCAP} {\bfseries
  09} (2025) 058} [\href{https://arxiv.org/abs/2506.01555}{{\ttfamily
  2506.01555}}].

\bibitem{Schwinger:1960qe}
J.~S. Schwinger, \emph{{Brownian motion of a quantum oscillator}},
  \href{https://doi.org/10.1063/1.1703727}{\emph{J. Math. Phys.} {\bfseries 2}
  (1961) 407}.

\bibitem{Feynman:1963fq}
R.~P. Feynman and F.~L. Vernon, Jr., \emph{{The Theory of a general quantum
  system interacting with a linear dissipative system}},
  \href{https://doi.org/10.1016/0003-4916(63)90068-X}{\emph{Annals Phys.}
  {\bfseries 24} (1963) 118}.

\bibitem{Keldysh:1964ud}
L.~V. Keldysh, \emph{{Diagram Technique for Nonequilibrium Processes}},
  \href{https://doi.org/10.1142/9789811279461_0007}{\emph{Sov. Phys. JETP}
  {\bfseries 20} (1965) 1018}.

\bibitem{Maldacena:2002vr}
J.~M. Maldacena, \emph{{Non-Gaussian features of primordial fluctuations in
  single field inflationary models}},
  \href{https://doi.org/10.1088/1126-6708/2003/05/013}{\emph{JHEP} {\bfseries
  05} (2003) 013} [\href{https://arxiv.org/abs/astro-ph/0210603}{{\ttfamily
  astro-ph/0210603}}].

\bibitem{Weinberg:2005vy}
S.~Weinberg, \emph{{Quantum contributions to cosmological correlations}},
  \href{https://doi.org/10.1103/PhysRevD.72.043514}{\emph{Phys. Rev. D}
  {\bfseries 72} (2005) 043514}
  [\href{https://arxiv.org/abs/hep-th/0506236}{{\ttfamily hep-th/0506236}}].

\bibitem{Qin:2023bjk}
Z.~Qin and Z.-Z. Xianyu, \emph{{Inflation correlators at the one-loop order:
  nonanalyticity, factorization, cutting rule, and OPE}},
  \href{https://doi.org/10.1007/JHEP09(2023)116}{\emph{JHEP} {\bfseries 09}
  (2023) 116} [\href{https://arxiv.org/abs/2304.13295}{{\ttfamily
  2304.13295}}].

\bibitem{Qin:2023nhv}
Z.~Qin and Z.-Z. Xianyu, \emph{{Nonanalyticity and on-shell factorization of
  inflation correlators at all loop orders}},
  \href{https://doi.org/10.1007/JHEP01(2024)168}{\emph{JHEP} {\bfseries 01}
  (2024) 168} [\href{https://arxiv.org/abs/2308.14802}{{\ttfamily
  2308.14802}}].

\bibitem{Xianyu:2023ytd}
Z.-Z. Xianyu and J.~Zang, \emph{{Inflation correlators with multiple massive
  exchanges}}, \href{https://doi.org/10.1007/JHEP03(2024)070}{\emph{JHEP}
  {\bfseries 03} (2024) 070}
  [\href{https://arxiv.org/abs/2309.10849}{{\ttfamily 2309.10849}}].

\bibitem{Liu:2024xyi}
H.~Liu, Z.~Qin and Z.-Z. Xianyu, \emph{{Dispersive bootstrap of massive
  inflation correlators}},
  \href{https://doi.org/10.1007/JHEP02(2025)101}{\emph{JHEP} {\bfseries 02}
  (2025) 101} [\href{https://arxiv.org/abs/2407.12299}{{\ttfamily
  2407.12299}}].

\bibitem{Fan:2025scu}
B.~Fan and Z.-Z. Xianyu, \emph{{Anatomy of family trees in cosmological
  correlators}}, \href{https://doi.org/10.1007/JHEP12(2025)179}{\emph{JHEP}
  {\bfseries 12} (2025) 179}
  [\href{https://arxiv.org/abs/2509.02684}{{\ttfamily 2509.02684}}].

\bibitem{Higuchi:1986wu}
A.~Higuchi, \emph{{Symmetric Tensor Spherical Harmonics on the $N$ Sphere and
  Their Application to the De Sitter Group SO($N$,1)}},
  \href{https://doi.org/10.1063/1.527513}{\emph{J. Math. Phys.} {\bfseries 28}
  (1987) 1553}.

\bibitem{Bros:1990cu}
J.~Bros, \emph{{Complexified de Sitter space: Analytic causal kernels and
  Kallen-Lehmann type representation}},
  \href{https://doi.org/10.1016/0920-5632(91)90119-Y}{\emph{Nucl. Phys. B Proc.
  Suppl.} {\bfseries 18} (1991) 22}.

\bibitem{Bros:1995js}
J.~Bros and U.~Moschella, \emph{{Two point functions and quantum fields in de
  Sitter universe}},
  \href{https://doi.org/10.1142/S0129055X96000123}{\emph{Rev. Math. Phys.}
  {\bfseries 8} (1996) 327}
  [\href{https://arxiv.org/abs/gr-qc/9511019}{{\ttfamily gr-qc/9511019}}].

\bibitem{Marolf:2010zp}
D.~Marolf and I.~A. Morrison, \emph{{The IR stability of de Sitter: Loop
  corrections to scalar propagators}},
  \href{https://doi.org/10.1103/PhysRevD.82.105032}{\emph{Phys. Rev. D}
  {\bfseries 82} (2010) 105032}
  [\href{https://arxiv.org/abs/1006.0035}{{\ttfamily 1006.0035}}].

\bibitem{Marolf:2010nz}
D.~Marolf and I.~A. Morrison, \emph{{The IR stability of de Sitter QFT: results
  at all orders}},
  \href{https://doi.org/10.1103/PhysRevD.84.044040}{\emph{Phys. Rev. D}
  {\bfseries 84} (2011) 044040}
  [\href{https://arxiv.org/abs/1010.5327}{{\ttfamily 1010.5327}}].

\bibitem{Higuchi:2010xt}
A.~Higuchi, D.~Marolf and I.~A. Morrison, \emph{{On the Equivalence between
  Euclidean and In-In Formalisms in de Sitter QFT}},
  \href{https://doi.org/10.1103/PhysRevD.83.084029}{\emph{Phys. Rev. D}
  {\bfseries 83} (2011) 084029}
  [\href{https://arxiv.org/abs/1012.3415}{{\ttfamily 1012.3415}}].

\bibitem{Chen:2016hrz}
X.~Chen, Y.~Wang and Z.-Z. Xianyu, \emph{{Standard Model Mass Spectrum in
  Inflationary Universe}},
  \href{https://doi.org/10.1007/JHEP04(2017)058}{\emph{JHEP} {\bfseries 04}
  (2017) 058} [\href{https://arxiv.org/abs/1612.08122}{{\ttfamily
  1612.08122}}].

\bibitem{Chakraborty:2023qbp}
P.~Chakraborty and J.~Stout, \emph{{Light scalars at the cosmological
  collider}}, \href{https://doi.org/10.1007/JHEP02(2024)021}{\emph{JHEP}
  {\bfseries 02} (2024) 021}
  [\href{https://arxiv.org/abs/2310.01494}{{\ttfamily 2310.01494}}].

\bibitem{Chakraborty:2023eoq}
P.~Chakraborty and J.~Stout, \emph{{Compact scalars at the cosmological
  collider}}, \href{https://doi.org/10.1007/JHEP03(2024)149}{\emph{JHEP}
  {\bfseries 03} (2024) 149}
  [\href{https://arxiv.org/abs/2311.09219}{{\ttfamily 2311.09219}}].

\bibitem{Chakraborty:2025myb}
P.~Chakraborty, \emph{{Primordial non-Gaussianity from light compact scalars}},
  \href{https://doi.org/10.1007/JHEP11(2025)023}{\emph{JHEP} {\bfseries 11}
  (2025) 023} [\href{https://arxiv.org/abs/2501.07672}{{\ttfamily
  2501.07672}}].

\bibitem{Loparco:2025azm}
M.~Loparco, J.~Penedones and Y.~Ulrich, \emph{{What is a photon in de Sitter
  spacetime?}},  \href{https://arxiv.org/abs/2505.00761}{{\ttfamily
  2505.00761}}.

\bibitem{Sleight:2019mgd}
C.~Sleight, \emph{{A Mellin Space Approach to Cosmological Correlators}},
  \href{https://doi.org/10.1007/JHEP01(2020)090}{\emph{JHEP} {\bfseries 01}
  (2020) 090} [\href{https://arxiv.org/abs/1906.12302}{{\ttfamily
  1906.12302}}].

\bibitem{Sleight:2019hfp}
C.~Sleight and M.~Taronna, \emph{{Bootstrapping Inflationary Correlators in
  Mellin Space}}, \href{https://doi.org/10.1007/JHEP02(2020)098}{\emph{JHEP}
  {\bfseries 02} (2020) 098}
  [\href{https://arxiv.org/abs/1907.01143}{{\ttfamily 1907.01143}}].

\bibitem{Sleight:2020obc}
C.~Sleight and M.~Taronna, \emph{{From AdS to dS exchanges: Spectral
  representation, Mellin amplitudes, and crossing}},
  \href{https://doi.org/10.1103/PhysRevD.104.L081902}{\emph{Phys. Rev. D}
  {\bfseries 104} (2021) L081902}
  [\href{https://arxiv.org/abs/2007.09993}{{\ttfamily 2007.09993}}].

\bibitem{Sleight:2021plv}
C.~Sleight and M.~Taronna, \emph{{From dS to AdS and back}},
  \href{https://doi.org/10.1007/JHEP12(2021)074}{\emph{JHEP} {\bfseries 12}
  (2021) 074} [\href{https://arxiv.org/abs/2109.02725}{{\ttfamily
  2109.02725}}].

\bibitem{Loparco:2023rug}
M.~Loparco, J.~Penedones, K.~Salehi~Vaziri and Z.~Sun, \emph{{The
  K{\"a}ll{\'e}n-Lehmann representation in de Sitter spacetime}},
  \href{https://doi.org/10.1007/JHEP12(2023)159}{\emph{JHEP} {\bfseries 12}
  (2023) 159} [\href{https://arxiv.org/abs/2306.00090}{{\ttfamily
  2306.00090}}].

\bibitem{Sun:2021thf}
Z.~Sun, \emph{{A note on the representations of SO(1,d + 1)}},
  \href{https://doi.org/10.1142/S0129055X24300073}{\emph{Rev. Math. Phys.}
  {\bfseries 37} (2025) 2430007}
  [\href{https://arxiv.org/abs/2111.04591}{{\ttfamily 2111.04591}}].

\bibitem{Dirac:1936fq}
P.~A.~M. Dirac, \emph{{Wave equations in conformal space}},
  \href{https://doi.org/10.2307/1968455}{\emph{Annals Math.} {\bfseries 37}
  (1936) 429}.

\bibitem{Bros:2001yw}
J.~Bros, H.~Epstein and U.~Moschella, \emph{{The Asymptotic symmetry of de
  Sitter space-time}},
  \href{https://doi.org/10.1103/PhysRevD.65.084012}{\emph{Phys. Rev. D}
  {\bfseries 65} (2002) 084012}
  [\href{https://arxiv.org/abs/hep-th/0107091}{{\ttfamily hep-th/0107091}}].

\bibitem{Cacciatori:2007in}
S.~Cacciatori, V.~Gorini, A.~Kamenshchik and U.~Moschella, \emph{{Conservation
  laws and scattering for de Sitter classical particles}},
  \href{https://doi.org/10.1088/0264-9381/25/7/075008}{\emph{Class. Quant.
  Grav.} {\bfseries 25} (2008) 075008}
  [\href{https://arxiv.org/abs/0710.0315}{{\ttfamily 0710.0315}}].

\bibitem{Rychkov:2016iqz}
S.~Rychkov, \emph{{EPFL Lectures on Conformal Field Theory in
  D{\ensuremath{>}}= 3 Dimensions}}, SpringerBriefs in Physics. 1, 2016,
  \href{https://doi.org/10.1007/978-3-319-43626-5}{10.1007/978-3-319-43626-5},
  [\href{https://arxiv.org/abs/1601.05000}{{\ttfamily 1601.05000}}].

\bibitem{Karateev:2017jgd}
D.~Karateev, P.~Kravchuk and D.~Simmons-Duffin, \emph{{Weight Shifting
  Operators and Conformal Blocks}},
  \href{https://doi.org/10.1007/JHEP02(2018)081}{\emph{JHEP} {\bfseries 02}
  (2018) 081} [\href{https://arxiv.org/abs/1706.07813}{{\ttfamily
  1706.07813}}].

\bibitem{Moschella:2024kvk}
U.~Moschella, \emph{{The Spectral Condition, Plane Waves, and Harmonic Analysis
  in de Sitter and Anti-de Sitter Quantum Field Theories}},
  \href{https://doi.org/10.3390/universe10050199}{\emph{Universe} {\bfseries
  10} (2024) 199} [\href{https://arxiv.org/abs/2403.15893}{{\ttfamily
  2403.15893}}].

\bibitem{Bargmann:1948ck}
V.~Bargmann and E.~P. Wigner, \emph{{Group Theoretical Discussion of
  Relativistic Wave Equations}},
  \href{https://doi.org/10.1073/pnas.34.5.211}{\emph{Proc. Nat. Acad. Sci.}
  {\bfseries 34} (1948) 211}.

\bibitem{Thomas1941}
L.~H. Thomas, \emph{On unitary representations of the group of de sitter
  space}, {\emph{Annals of Mathematics} {\bfseries 42} (1941) 113}.

\bibitem{Newton1950}
T.~D. Newton, \emph{A note on the representations of the de sitter group},
  {\emph{Annals of Mathematics} {\bfseries 51} (1950) 730}.

\bibitem{Dixmier1961}
J.~Dixmier, \emph{Représentations intégrables du groupe de de sitter},
  {\emph{Bulletin de la Société Mathématique de France} {\bfseries 89}
  (1961) 9}.

\bibitem{Hirai1962}
T.~Hirai, \emph{{On irreducible representations of the Lorentz group of $n$-th
  order}}, \href{https://doi.org/10.3792/pja/1195523378}{\emph{Proceedings of
  the Japan Academy} {\bfseries 38} (1962) 258 }.

\bibitem{Takahashi1963}
R.~Takahashi, \emph{Sur les représentations unitaires des groupes de lorentz
  généralisés}, {\emph{Bulletin de la Société Mathématique de France}
  {\bfseries 91} (1963) 289}.

\bibitem{Basile:2016aen}
T.~Basile, X.~Bekaert and N.~Boulanger, \emph{{Mixed-symmetry fields in de
  Sitter space: a group theoretical glance}},
  \href{https://doi.org/10.1007/JHEP05(2017)081}{\emph{JHEP} {\bfseries 05}
  (2017) 081} [\href{https://arxiv.org/abs/1612.08166}{{\ttfamily
  1612.08166}}].

\bibitem{Wigner1948}
V.~Bargmann and E.~P. Wigner, \emph{Group theoretical discussion of
  relativistic wave equations}, {\emph{Proceedings of the National Academy of
  Sciences of the United States of America} {\bfseries 34} (1948) 211}.

\bibitem{Tung:1985iqd}
W.-K. Tung, \emph{{Group Theory in Physics}}. 8, 1985,
  \href{https://doi.org/10.1142/0097}{10.1142/0097}.

\bibitem{Georgi:2000vve}
H.~Georgi, \emph{{Lie Algebras In Particle Physics : from Isospin To Unified
  Theories}}. Taylor {\&} Francis, Boca Raton, 2000,
  \href{https://doi.org/10.1201/9780429499210}{10.1201/9780429499210}.

\bibitem{Isaev:2018xcg}
A.~Isaev and V.~Rubakov, \emph{{Theory of Groups and Symmetries}}. WSP, 5,
  2018, \href{https://doi.org/10.1142/10898}{10.1142/10898}.

\bibitem{Bonifacio:2018zex}
J.~Bonifacio, K.~Hinterbichler, A.~Joyce and R.~A. Rosen, \emph{{Shift
  Symmetries in (Anti) de Sitter Space}},
  \href{https://doi.org/10.1007/JHEP02(2019)178}{\emph{JHEP} {\bfseries 02}
  (2019) 178} [\href{https://arxiv.org/abs/1812.08167}{{\ttfamily
  1812.08167}}].

\bibitem{Deser:1983tm}
S.~Deser and R.~I. Nepomechie, \emph{{Anomalous Propagation of Gauge Fields in
  Conformally Flat Spaces}},
  \href{https://doi.org/10.1016/0370-2693(83)90317-9}{\emph{Phys. Lett. B}
  {\bfseries 132} (1983) 321}.

\bibitem{Deser:1983mm}
S.~Deser and R.~I. Nepomechie, \emph{{Gauge Invariance Versus Masslessness in
  De Sitter Space}},
  \href{https://doi.org/10.1016/0003-4916(84)90156-8}{\emph{Annals Phys.}
  {\bfseries 154} (1984) 396}.

\bibitem{Higuchi:1986py}
A.~Higuchi, \emph{{Forbidden Mass Range for Spin-2 Field Theory in De Sitter
  Space-time}}, \href{https://doi.org/10.1016/0550-3213(87)90691-2}{\emph{Nucl.
  Phys. B} {\bfseries 282} (1987) 397}.

\bibitem{Brink:2000ag}
L.~Brink, R.~R. Metsaev and M.~A. Vasiliev, \emph{{How massless are massless
  fields in AdS(d)}},
  \href{https://doi.org/10.1016/S0550-3213(00)00402-8}{\emph{Nucl. Phys. B}
  {\bfseries 586} (2000) 183}
  [\href{https://arxiv.org/abs/hep-th/0005136}{{\ttfamily hep-th/0005136}}].

\bibitem{Zinoviev:2001dt}
Y.~M. Zinoviev, \emph{{On massive high spin particles in AdS}},
  \href{https://arxiv.org/abs/hep-th/0108192}{{\ttfamily hep-th/0108192}}.

\bibitem{Luty:2003vm}
M.~A. Luty, M.~Porrati and R.~Rattazzi, \emph{{Strong interactions and
  stability in the DGP model}},
  \href{https://doi.org/10.1088/1126-6708/2003/09/029}{\emph{JHEP} {\bfseries
  09} (2003) 029} [\href{https://arxiv.org/abs/hep-th/0303116}{{\ttfamily
  hep-th/0303116}}].

\bibitem{Nicolis:2008in}
A.~Nicolis, R.~Rattazzi and E.~Trincherini, \emph{{The Galileon as a local
  modification of gravity}},
  \href{https://doi.org/10.1103/PhysRevD.79.064036}{\emph{Phys. Rev. D}
  {\bfseries 79} (2009) 064036}
  [\href{https://arxiv.org/abs/0811.2197}{{\ttfamily 0811.2197}}].

\bibitem{Goodhew:2024eup}
H.~Goodhew, A.~Thavanesan and A.~C. Wall, \emph{{The Cosmological CPT
  Theorem}},  \href{https://arxiv.org/abs/2408.17406}{{\ttfamily 2408.17406}}.

\bibitem{Thavanesan:2025kyc}
A.~Thavanesan, \emph{{No-go Theorem for Cosmological Parity Violation}},
  \href{https://arxiv.org/abs/2501.06383}{{\ttfamily 2501.06383}}.

\bibitem{Thavanesan:2025ibm}
A.~Thavanesan and A.~C. Wall, \emph{{Kosmic Field Theories: Towards Holographic
  Duals for Unitary String Cosmologies}},
  \href{https://arxiv.org/abs/2510.21701}{{\ttfamily 2510.21701}}.

\bibitem{Grafe:2026qsm}
J.~Gr{\"a}fe and I.~Sachs, \emph{{Split Representations and Bubble Resummation
  for Massive de Sitter Correlators}},
  \href{https://arxiv.org/abs/2602.09977}{{\ttfamily 2602.09977}}.

\bibitem{Lee:2025kgs}
M.~H.~G. Lee and S.~Melville, \emph{{Propagator positivity bounds for
  cosmological correlators}},
  \href{https://arxiv.org/abs/2512.20706}{{\ttfamily 2512.20706}}.

\bibitem{Belrhali:2026ktb}
N.~Belrhali, A.~Poisson, S.~Renaux-Petel and D.~Werth, \emph{{De Sitter
  Momentum Space}},  \href{https://arxiv.org/abs/2601.15228}{{\ttfamily
  2601.15228}}.

\bibitem{gradshteyn2007}
I.~S. Gradshteyn and I.~M. Ryzhik, \emph{Table of integrals, series, and
  products}. Elsevier/Academic Press, Amsterdam, seventh~ed., 2007.

\bibitem{NIST:DLMF}
``{\it NIST Digital Library of Mathematical Functions}.''
  \url{https://dlmf.nist.gov/}, Release 1.2.5 of 2025-12-15.

\bibitem{Jazayeri:2023kji}
S.~Jazayeri, S.~Renaux-Petel, X.~Tong, D.~Werth and Y.~Zhu, \emph{{Parity
  violation from emergent nonlocality during inflation}},
  \href{https://doi.org/10.1103/PhysRevD.108.123523}{\emph{Phys. Rev. D}
  {\bfseries 108} (2023) 123523}
  [\href{https://arxiv.org/abs/2308.11315}{{\ttfamily 2308.11315}}].

\bibitem{Melville:2023kgd}
S.~Melville and G.~L. Pimentel, \emph{{de Sitter S matrix for the masses}},
  \href{https://doi.org/10.1103/PhysRevD.110.103530}{\emph{Phys. Rev. D}
  {\bfseries 110} (2024) 103530}
  [\href{https://arxiv.org/abs/2309.07092}{{\ttfamily 2309.07092}}].

\bibitem{Werth:2024mjg}
D.~Werth, \emph{{Spectral representation of cosmological correlators}},
  \href{https://doi.org/10.1007/JHEP12(2024)017}{\emph{JHEP} {\bfseries 12}
  (2024) 017} [\href{https://arxiv.org/abs/2409.02072}{{\ttfamily
  2409.02072}}].

\bibitem{Lauricella1893SulleFI}
G.~Lauricella, \emph{Sulle funzioni ipergeometriche a piu variabili},
  {\emph{Rendiconti del Circolo Matematico di Palermo} {\bfseries 7} (1893)
  111}.

\bibitem{Bzowski:2012ih}
A.~Bzowski, P.~McFadden and K.~Skenderis, \emph{{Holography for inflation using
  conformal perturbation theory}},
  \href{https://doi.org/10.1007/JHEP04(2013)047}{\emph{JHEP} {\bfseries 04}
  (2013) 047} [\href{https://arxiv.org/abs/1211.4550}{{\ttfamily 1211.4550}}].

\bibitem{Bzowski:2015pba}
A.~Bzowski, P.~McFadden and K.~Skenderis, \emph{{Scalar 3-point functions in
  CFT: renormalisation, beta functions and anomalies}},
  \href{https://doi.org/10.1007/JHEP03(2016)066}{\emph{JHEP} {\bfseries 03}
  (2016) 066} [\href{https://arxiv.org/abs/1510.08442}{{\ttfamily
  1510.08442}}].

\bibitem{Bzowski:2015yxv}
A.~Bzowski, P.~McFadden and K.~Skenderis, \emph{{Evaluation of conformal
  integrals}}, \href{https://doi.org/10.1007/JHEP02(2016)068}{\emph{JHEP}
  {\bfseries 02} (2016) 068}
  [\href{https://arxiv.org/abs/1511.02357}{{\ttfamily 1511.02357}}].

\bibitem{Bzowski:2017poo}
A.~Bzowski, P.~McFadden and K.~Skenderis, \emph{{Renormalised 3-point functions
  of stress tensors and conserved currents in CFT}},
  \href{https://doi.org/10.1007/JHEP11(2018)153}{\emph{JHEP} {\bfseries 11}
  (2018) 153} [\href{https://arxiv.org/abs/1711.09105}{{\ttfamily
  1711.09105}}].

\bibitem{Chen:2017ryl}
X.~Chen, Y.~Wang and Z.-Z. Xianyu, \emph{{Schwinger-Keldysh Diagrammatics for
  Primordial Perturbations}},
  \href{https://doi.org/10.1088/1475-7516/2017/12/006}{\emph{JCAP} {\bfseries
  12} (2017) 006} [\href{https://arxiv.org/abs/1703.10166}{{\ttfamily
  1703.10166}}].

\bibitem{Creminelli:2011mw}
P.~Creminelli, \emph{{Conformal invariance of scalar perturbations in
  inflation}}, \href{https://doi.org/10.1103/PhysRevD.85.041302}{\emph{Phys.
  Rev. D} {\bfseries 85} (2012) 041302}
  [\href{https://arxiv.org/abs/1108.0874}{{\ttfamily 1108.0874}}].

\bibitem{Bros:2009bz}
J.~Bros, H.~Epstein, M.~Gaudin, U.~Moschella and V.~Pasquier, \emph{{Triangular
  invariants, three-point functions and particle stability on the de Sitter
  universe}}, \href{https://doi.org/10.1007/s00220-009-0875-4}{\emph{Commun.
  Math. Phys.} {\bfseries 295} (2010) 261}
  [\href{https://arxiv.org/abs/0901.4223}{{\ttfamily 0901.4223}}].

\bibitem{Xianyu:2022jwk}
Z.-Z. Xianyu and H.~Zhang, \emph{{Bootstrapping one-loop inflation correlators
  with the spectral decomposition}},
  \href{https://doi.org/10.1007/JHEP04(2023)103}{\emph{JHEP} {\bfseries 04}
  (2023) 103} [\href{https://arxiv.org/abs/2211.03810}{{\ttfamily
  2211.03810}}].

\bibitem{Zhang:2025nzd}
H.~Zhang, \emph{{Dimensional regularization of bubble diagrams in de Sitter
  spacetime}}, \href{https://doi.org/10.1007/JHEP02(2026)119}{\emph{JHEP}
  {\bfseries 02} (2026) 119}
  [\href{https://arxiv.org/abs/2507.19318}{{\ttfamily 2507.19318}}].

\bibitem{Gelfand-Vilenkin}
I.~Gel'fand and N.~Vilenkin, \emph{{Generalized Functions, Vol. IV}}. Academic
  Press, New York, 1964,
  [\href{https://doi.org/https://doi.org/10.1016%2Fc2013-0-12221-0}].

\bibitem{Maurin}
K.~Maurin, \emph{General eigenfunction expansions and unitary representations
  of topological groups}, .

\bibitem{delaMadrid:2005qdg}
R.~de~la Madrid, \emph{{The role of the rigged Hilbert space in Quantum
  Mechanics}}, \href{https://doi.org/10.1088/0143-0807/26/2/008}{\emph{Eur. J.
  Phys.} {\bfseries 26} (2005) 287}
  [\href{https://arxiv.org/abs/quant-ph/0502053}{{\ttfamily
  quant-ph/0502053}}].

\bibitem{Kontorovich1938}
M.~Kontorovich and N.~Lebedev, \emph{On the method of solution of certain
  boundary value problems of the wave equation}, {\emph{Doklady Akademii Nauk
  SSSR} {\bfseries 19} (1938) 441}.

\bibitem{Lebedev1946}
N.~Lebedev, \emph{Sur une formule d'inversion}, {\emph{Doklady Akademii Nauk
  SSSR} {\bfseries 52} (1946) 655}.

\bibitem{Lebedev1949}
N.~Lebedev, \emph{Analog of the parseval theorem for the one integral
  transform}, {\emph{Doklady Akademii Nauk SSSR} {\bfseries 68} (1949) 653}.

\bibitem{Yakubovich1994}
S.~B. Yakubovich and B.~Fisher, \emph{On the theory of the kontorovich-lebedev
  transformation on distributions}, {\emph{Proceedings of the American
  Mathematical Society} {\bfseries 122} (1994) 773}.

\bibitem{Zemanian_1975}
A.~H. Zemanian, \emph{The kontorovich–lebedev transformation on distributions
  of compact support and its inversion},
  \href{https://doi.org/10.1017/S0305004100049471}{\emph{Mathematical
  Proceedings of the Cambridge Philosophical Society} {\bfseries 77} (1975)
  139–143}.

\bibitem{Erdelyi1954}
A.~Erd{\'e}lyi, W.~Magnus, F.~Oberhettinger and F.~Tricomi, \emph{Tables of
  Integral Transforms}, vol.~1-2. McGraw-Hill, New York, 1954.

\bibitem{Melville:2024ove}
S.~Melville and G.~L. Pimentel, \emph{{A de Sitter S-matrix from amputated
  cosmological correlators}},
  \href{https://doi.org/10.1007/JHEP08(2024)211}{\emph{JHEP} {\bfseries 08}
  (2024) 211} [\href{https://arxiv.org/abs/2404.05712}{{\ttfamily
  2404.05712}}].

\end{thebibliography}\endgroup
\end{document}